\documentclass[a4paper,11pt]{article}
\pdfoutput=1
\usepackage{jheppub}

\usepackage{amssymb,amsmath,amsfonts}
\usepackage[normalem]{ulem}
\usepackage[utf8x]{inputenc}
\usepackage{slashed}
\usepackage{graphicx}
\usepackage{tabularx}
\usepackage{here}
\usepackage{color}
\usepackage{csquotes} 
\usepackage{comment}
\usepackage{mathrsfs}
\usepackage{float}
\usepackage{ascmac}
\usepackage{multirow}
\usepackage{longtable}
\usepackage{bm}

\usepackage{booktabs,array}
\usepackage{microtype}
\usepackage{xcolor}
\usepackage{enumitem}
\usepackage{fancyhdr}
\usepackage[nameinlink,noabbrev]{cleveref}

\numberwithin{equation}{section}

\usepackage[italicdiff]{physics}
\usepackage{textcomp}
\usepackage{tikz}
\usetikzlibrary{arrows.meta,calc,shapes.geometric}

\newcommand{\eV}{\mbox{eV}}

\newcommand{\figref}[1]{Fig.~\ref{#1}}

\newcommand{\vevs}[1]{\langle #1 \rangle}

\makeatletter
\newcommand*\rel@kern[1]{\kern#1\dimexpr\macc@kerna}
\newcommand*\widebar[1]{%
  \begingroup
  \def\mathaccent##1##2{%
    \rel@kern{0.8}%
    \overline{\rel@kern{-0.8}\macc@nucleus\rel@kern{0.2}}%
    \rel@kern{-0.2}%
  }%
  \macc@depth\@ne
  \let\math@bgroup\@empty \let\math@egroup\macc@set@skewchar
  \mathsurround\z@ \frozen@everymath{\mathgroup\macc@group\relax}%
  \macc@set@skewchar\relax
  \let\mathaccentV\macc@nested@a
  \macc@nested@a\relax111{#1}%
  \endgroup
}
\makeatother

\preprint{
\begin{minipage}{5cm}
\small
\flushright
KYUSHU-HET-377
\end{minipage}}

\title{
Non-Invertible Froggatt-Nielsen Symmetry
}

\author{Hajime Otsuka$^{1,2}$ and}
\author{Ryo Yokokura$^{3}$}
\affiliation{
$^1$Department of Physics, Kyushu University, 744 Motooka, Nishi-ku, Fukuoka 819-0395, Japan
}
\affiliation{
$^2$Quantum and Spacetime Research Institute (QuaSR), Kyushu University, 744 Motooka, Nishi-ku, Fukuoka 819-0395, Japan}
\affiliation{
$^3$Department of Physics \& Research and Education Center for Natural Sciences, Keio University, Hiyoshi 4-1-1, Yokohama, Kanagawa 223-8521, Japan}

\emailAdd{otsuka.hajime@phys.kyushu-u.ac.jp}
\emailAdd{ryokokur@keio.jp}

\abstract{
We investigate non-invertible completions of anomalous flavor
symmetries and their application to Dirac neutrino masses.
We formulate symmetry defects dressed by  topological quantum field theories for flavor symmetries including
finite non-Abelian flavor transformations with rational anomaly
characters and modular flavor symmetries with the rational part
of modular anomaly cocycles, for gauge sectors with the
required exact higher-form symmetry. 
As a concrete example, we construct a Froggatt-Nielsen (FN) model in
which an order-four subgroup
$\mathbb Z_4^{\rm FN}\subset U(1)_{\rm FN}$ is completed into a
non-invertible defect through its mixed anomaly with an intermediate
hidden $U(1)_H$ gauge sector in the infrared monopole-decoupling limit. The non-invertible FN symmetry can be explicitly broken at the energy scale where
monopoles become dynamical. 
This breaking from the infrared viewpoint can be realized in an ultraviolet hidden 
$SU(2)_H$ theory as instanton effects, where a constrained instanton with topological charge one generates a fermion mixing term that violates FN charge by one unit. Its coefficient contains the nonperturbative factor
$\zeta_H\sim\exp[-8\pi^2/g_H^2(v_H)]$. 
When independent FN-breaking terms are subleading, integrating
out the heavy fermions gives
$(Y_\nu)_{ij}\propto\zeta_H\epsilon^{\ell_i+n_j}$ in a rotated basis for the right-handed neutrinos, to leading
order in the instanton-induced fermion mixing. 
The nonperturbative factor provides the exponential suppression of
the neutrino Yukawa couplings, while the FN charges determine the powers of $\epsilon$. 
For the chosen FN charges and generic order-one coefficients,
$m_2/m_3\sim\epsilon$, favoring normal ordering. 
An independent $\mathbb Z_8^D$ gauge symmetry forbids Majorana masses. 
The same FN charge assignment reproduces the leading
quark and charged-lepton Yukawa hierarchies.
}

\makeatletter
\gdef\@fpheader{}
\makeatother

\begin{document}

\maketitle

\section{Introduction}
\label{sec:introduction}

One of the basic questions in flavor physics is the origin of the
observed fermion masses and mixing angles. Flavor symmetries provide
a useful framework for organizing the allowed interactions and for
obtaining characteristic structures of Yukawa couplings. In the
Froggatt-Nielsen (FN) mechanism~\cite{Froggatt:1978nt}, a horizontal
$U(1)_{\rm FN}$ symmetry and a flavon expectation value generate
hierarchies in powers of
\begin{align}
\epsilon=\frac{\langle\varphi\rangle}{\Lambda_{\rm FN}}.
\end{align}
Here $\Lambda_{\rm FN}$ is the scale suppressing the flavon insertions.
For Dirac neutrinos, the overall Yukawa coupling is of order
$10^{-13}$. 
Abelian flavor symmetries have been used to account for quark and lepton mass textures while allowing small Dirac-neutrino masses~\cite{Gogoladze:2001kj}. Explicit constructions include models with an anomalous $U(1)$ symmetry and models with anomaly-free gauged family symmetries~\cite{Chen:2010uw,Chen:2011sb}. 
In this work, we address both the leading quark and
lepton flavor hierarchies and the smallness of Dirac-neutrino
masses within a single FN framework. The FN charges determine the powers of $\epsilon$ in the
Yukawa matrices, while hidden instantons provide
the common exponential suppression of the neutrino Yukawa couplings.

Nonperturbative mechanisms for small Dirac-neutrino masses have
been studied in intersecting D-brane compactifications, where
Euclidean D-brane instantons generate perturbatively forbidden
Dirac-neutrino Yukawa couplings
\cite{Cvetic:2008hi,Ibanez:2008my}; see also
Ref.~\cite{Blumenhagen:2009qh} for a review.
In a four-dimensional field-theoretic setting,
Ref.~\cite{Cordova:2022fhg} showed that nonperturbative breaking of
a non-invertible lepton-number symmetry can generate exponentially
small Dirac-neutrino Yukawa couplings. The present work combines
such a common suppression with a concrete FN messenger mechanism.
We derive the neutrino mass hierarchy from the same charge
assignment that gives the leading charged-fermion hierarchies,
rather than treating the neutrino texture as a separate input.

At the quantum level, a flavor transformation can be violated by
gauge-field configurations and may not remain an ordinary invertible
symmetry. Recent developments have shown that an anomalous chiral
transformation can instead survive as a non-invertible defect
\cite{Choi:2022jqy,Cordova:2022ieu}.
When a finite rotation produces a rational Fujikawa phase and the
gauge sector has the required exact higher-form symmetry, its wall
can be combined with a three-dimensional topological quantum field theory
(TQFT). The dressed wall is gauge invariant and topological, but
fusion with its orientation-reversed wall gives a condensation
defect rather than the identity. Dynamical monopoles break the
magnetic one-form symmetry supporting this construction and can
lead to nonperturbative violations of the associated selection rule.
The local operator and its coefficient must then be determined in
the ultraviolet gauge theory.

In our FN model, the Standard Model gauge sectors preserve an
order-four subgroup
\begin{align}
\mathbb Z_4^{\rm FN}
\subset
U(1)_{\rm FN}.
\end{align}
After the flavon acquires its expectation value, the residual
$\mathbb Z_2^{\rm FN}$ symmetry forbids the Dirac-neutrino
Yukawa interactions in perturbation theory. 
We then introduce a hidden gauge sector with the breaking sequence
\begin{align}
SU(2)_H
\longrightarrow
U(1)_H
\longrightarrow
1.
\end{align}
The intermediate Abelian theory has the
mixed anomaly needed to complete the order-four FN transformation
into a non-invertible defect in the formal monopole-decoupling
limit. Restoring finite-mass monopoles makes this defect
non-topological. In the ultraviolet theory, a unit constrained
instanton violates the FN charge by one unit and generates the
mixing $\mu_H(\eta N_D^c)$ between two hidden Weyl doublets.
The two components of $N_D^c$ correspond to right-handed neutrino fields in
left-handed Weyl notation; $\eta$ provides the second zero mode.

Two additional messengers transmit this mixing to the Standard Model
leptons, and give an additional mass for $\eta$. After they and $\eta$ are integrated out, the neutrino
Yukawa couplings have the form:
\begin{align}
(Y_\nu)_{ij}\propto\zeta_H C_{ij}\epsilon^{\ell_i+n_j}.
\end{align}
The coefficients $C_{ij}$ are determined by the messenger masses
and couplings, while the FN charges fix the powers of $\epsilon$. 
The nonperturbative factor is
\begin{align}
\zeta_H\simeq\mathcal K_H
\exp\!\left[-\frac{8\pi^2}{g_H^2(v_H)}+i\vartheta_{\rm UV}\right].
\label{eq:intro-central-result}
\end{align}
Here $g_H(v_H)$ is the hidden $SU(2)_H$ gauge coupling evaluated
at the adjoint expectation value $v_H$, which breaks
$SU(2)_H$ to $U(1)_H$. 
The factor $\mathcal K_H$ includes the instanton measure and
threshold effects, and $\vartheta_{\rm UV}$ is the hidden gauge
theta angle. Thus the hidden dynamics supplies the common
exponential suppression, whereas the FN interactions supply the flavor hierarchy.

Different flavon insertions in the two messenger couplings give
different powers of $\epsilon$ in the low-energy couplings,
although the components of $N_D^c$ have the same ultraviolet
FN charge. For these interactions, the
masses and left-handed mixing require no special vacuum alignment.
The tree-level Yukawa matrix has rank at most two, with
$m_2/m_3\sim\epsilon$ for generic order-one coefficients.
The mixing angles and CP phase depend on these coefficients. 
An independent $\mathbb Z_8^D$ gauge symmetry forbids Majorana
masses and the Weinberg operator, including after the instanton
is taken into account. The right-handed neutrinos remain in the
low-energy theory. After electroweak symmetry breaking, the
instanton-suppressed Yukawa couplings generate small Dirac
masses without invoking a Majorana seesaw. 
Here, the instanton determines a contribution to $\mu_H$, not every
possible ultraviolet contribution to the same interaction.
We assume independent FN-breaking masses and Yukawa couplings to
be subleading, and discuss a nonminimal $\mathbb Z_9$ gauge
extension that protects the chiral selection rule at renormalizable
order. The remaining higher-dimensional terms must still be
suppressed. 
The numerical example in Appendix~\ref{app:lepton-benchmark}
illustrates a choice of coefficients consistent with the lepton data under these assumptions.

Non-invertible symmetries have also been applied to other problems
in particle phenomenology
\cite{Putrov:2023jqi,Cordova:2023her,Craig:2024dnl,Kobayashi:2024yqq,Delgado:2024pcv,Hong:2025qbw,Choi:2026oqz,Sehayek:2026pvu,Delgado:2026lwj}
(see Refs.~\cite{Shao:2023gho,Brennan:2023mmt,Kaidi:2026urc}
for reviews). Nonperturbative breaking has been used to explain
exponential hierarchies in fermion masses and axion potentials
\cite{Cordova:2022ieu,Cordova:2022fhg}, and non-invertible
Peccei-Quinn symmetries have been studied in connection with the
strong CP problem~\cite{Cordova:2024ypu}.
On the flavor side, non-group-like fusion rules constrain Yukawa
textures, radiative neutrino masses, and dark-matter stability
\cite{Kobayashi:2024cvp,Kobayashi:2024yqq,Kobayashi:2025znw,
Kobayashi:2025ldi,Kobayashi:2025cwx}.
Applications to supersymmetric models also constrain baryon- and
lepton-number violating operators and flavor-changing neutral
currents~\cite{Kobayashi:2025lar,Nakai:2025thw}. 
We also formulate TQFT-dressed defects for finite non-Abelian
flavor transformations with rational anomaly characters and extend
the construction to the rational part of modular anomaly cocycles.

This paper is organized as follows. In Sec.~\ref{sec:noninv-review},
we review non-invertible chiral defects, their fusion, and their
breaking by dynamical monopoles. In Sec.~\ref{sec:application},
we specify the FN charges and hidden gauge sector and construct
the non-invertible FN defect. In Sec.~\ref{sec:instanton-breaking},
we derive the instanton-induced mixing and state the conditions
under which it dominates independent ultraviolet contributions.
Sec.~\ref{sec:messenger-z4} gives the messenger interactions,
identifies the light neutrinos, derives the neutrino mass hierarchy,
and explains how the discrete gauge symmetry forbids
Majorana masses. 
In Sec.~\ref{sec:phenomenology}, we discuss
the neutrino mass normalization, corrections to the zero mass,
and phenomenological consequences. The explicit lepton benchmark
is given in Appendix~\ref{app:lepton-benchmark}. In Sec.~\ref{sec:family}, we summarize
the extensions to finite non-Abelian flavor groups and modular
transformations. Sec.~\ref{sec:discussion} is devoted to the conclusions.
Appendix~\ref{app:flavor-completions} contains the detailed anomaly
and fusion analysis. The other appendices present the dyon calculation, an analysis
of vacuum stability in the scalar sector, the messenger mass
spectrum, and possible ultraviolet realizations of the discrete
symmetries.

\section{Non-invertible chiral symmetry in four dimensions}
\label{sec:noninv-review}

We review the construction of a non-invertible chiral symmetry in four
dimensions~\cite{Choi:2022jqy,Cordova:2022ieu}. 
A wall defined by the gauge-invariant anomalous current is
gauge invariant but not topological. Subtracting a Chern-Simons term
gives a conserved current, but the corresponding wall is not invariant
under large gauge transformations when the level is fractional. A
three-dimensional TQFT resolves this problem. We discuss the fusion of
the resulting defect and its breaking by dynamical monopoles.

\subsection{An anomalous chiral current}

Consider a four-dimensional theory with a compact Abelian gauge field $A$ and a classically conserved chiral current $j_5$. We normalize the anomaly as
\begin{align}
d\star j_5
=
\frac{\kappa}{8\pi^2}F\wedge F
\qquad
\mathrm{with}
\qquad
F=dA,
\label{eq:review-anomaly}
\end{align}
where $\kappa\in\mathbb Z$ is fixed by the charges and multiplicities of the chiral fermions \cite{Fujikawa}.
The simplified equation
\begin{align}
d\left(
\star j_5-\frac{\kappa}{8\pi^2}A\wedge F
\right)=0
\end{align}
suggests the codimension-one operator
\begin{align}
U_\alpha(\mathcal V)
=
\exp\left[
i\alpha
\int_{\mathcal V}
\left(
\star j_5-\frac{\kappa}{8\pi^2}A\wedge F
\right)
\right],
\label{eq:naive-wall}
\end{align}
where $\mathcal V$ is a closed three-manifold. The improved current is
conserved, but the Chern-Simons term is not invariant under large gauge
transformations for a generic angle $\alpha$. When
$\kappa\alpha/(2\pi)\in\mathbb Z$, the level is an allowed integer on a
spin manifold, and no nontrivial wall TQFT is required. In the following,
we consider a rational noninteger level
\cite{Choi:2022jqy,Cordova:2022ieu}.

\subsection{Gauge-invariant TQFT dressing, topologicality, and fusion}

For rational angles satisfying
\begin{align}
\kappa\alpha=2\pi\frac{p}{n},
\qquad
\gcd(p,n)=1,
\label{eq:rational-angle}
\end{align}
the anomalous wall can be completed by a three-dimensional minimal Abelian TQFT $\mathcal A^{n,p}$ coupled to the bulk magnetic one-form symmetry \cite{Choi:2022jqy,Cordova:2022ieu}. We write
\begin{align}
\mathcal D_{p/n}(M_3)
=
\exp\!\left[
 i\alpha\int_{M_3}\star j_5
\right]
Z_{\mathcal A^{n,p}}[M_3,F].
\label{eq:general-noninv-defect}
\end{align}
For $p=1$, the minimal TQFT has the explicit fractional-quantum-Hall presentation
\begin{align}
Z_{\mathcal A^{n,1}}[M_3,F]
=
\int\mathcal Dc\,
\exp\!\left\{
 i\int_{M_3}
\left[
\frac{n}{4\pi}c\wedge dc
+\frac{1}{2\pi}c\wedge F
\right]
\right\},
\label{eq:An1-action}
\end{align}
where $c$ is a compact $U(1)$ connection living only on the wall. The integer Chern-Simons level makes the complete path integral well defined under large gauge transformations. 
Integrating out $c$ locally gives the fractional response
\begin{align}
S_{\rm resp}
=
-\frac{1}{4\pi n}
\int_{M_3}A\wedge dA.
\end{align}
This expression captures the local response of the wall theory, but
it cannot by itself define a three-dimensional action for a
compact $U(1)$ gauge field. In particular, its fractional
Chern-Simons level is not well defined under general large gauge
transformations. The compact wall field $c$, including its
topological sectors, is therefore essential for the globally
well-defined TQFT $\mathcal A^{n,1}$.

We next check topologicality directly in correlation functions.
Let $M_3$ and $M_3'$ be two homologous wall supports such that
\begin{align}
M_3-M_3'=\partial V_4.
\end{align}
Deforming the bare chiral wall from $M_3$ to $M_3'$ changes its
contribution by
\begin{align}
\Delta S_{\rm chiral}
&=
\alpha\int_{V_4}d\star j_5
=
\frac{p}{4\pi n}
\int_{V_4}F\wedge F.
\label{eq:general-wall-deformation}
\end{align}
Thus, the bare chiral wall is not topological. Under the same
deformation, the wall TQFT $\mathcal A^{n,p}$ contributes the opposite
phase,
\begin{align}
\Delta S_{\mathcal A^{n,p}}
=
-\frac{p}{4\pi n}
\int_{V_4}F\wedge F,
\end{align}
so that the two contributions cancel. Therefore,
\begin{align}
\left\langle
\mathcal D_{p/n}(M_3)\,\mathcal X
\right\rangle
=
\left\langle
\mathcal D_{p/n}(M_3')\,\mathcal X
\right\rangle,
\label{eq:topological-correlation}
\end{align}
provided that the deformation does not cross any operator insertion
and that no magnetic source lies in $V_4$. 
Here $\mathcal X$ denotes other gauge-invariant operator insertions, such as $\mathcal O_1(x_1)\cdots\mathcal O_r(x_r)$, whose positions are kept fixed during the deformation.  
This invariance under smooth deformations of the defect support is the definition of topologicality for correlation functions.

On a local operator $\mathcal O_q$ of chiral charge $q$, the defect acts as
\begin{align}
\mathcal D_{p/n}:
\qquad
\mathcal O_q
\longmapsto
e^{i\alpha q}\mathcal O_q.
\label{eq:local-action}
\end{align}
The non-invertibility of the wall can be seen by 
the action on an 't Hooft loop.
The wall $\mathcal{D}_{p/n}$ acts on an 't Hooft loop 
$T(C_{\rm mon})$ 
on a temporally extended loop $C_{\rm mon}$ 
with the unit magnetic charge 
$\oint F = 2\pi $
as a 
non-invertible transformation for $|n| >1 $,
\begin{equation}
 \vevs{\mathcal{D}_{p/n} (S^2\times S^1)
T(C_{\rm mon})} = 0 ,
\label{eq:non-inv-mag}
\end{equation}
where $S^1$ is a temporally extended circle, 
and $S^2$ is a spatially extended sphere that encloses 
a time slice of the 't Hooft loop.
Since the magnetic flux on $S^2$ emitted from the 't Hooft loop 
generates an electric charge of the gauge field $c$ that violates 
the Gauss' law on $S^2$, the partition function 
in \eqref{eq:non-inv-mag} vanishes.
This is a manifestation of the non-invertible symmetry.

\subsection{Dynamical monopoles and explicit breaking}

In the absence of dynamical magnetic charges,
\begin{align}
dF=0,
\end{align}
and the two-form current $F/(2\pi)$ generates a magnetic one-form symmetry \cite{GaiottoKapustinSeibergWillett}. The construction in Eq.~\eqref{eq:general-noninv-defect} relies on this symmetry. A monopole worldline $C_{\rm mon}$
with the unit charge changes the Bianchi identity to
\begin{align}
dF=2\pi \,\delta_3(C_{\rm mon}),
\label{eq:Bianchi-monopole}
\end{align}
where 
$\delta_3(C_{\rm mon})$ is a 3-form such that 
$\int_{C_{\rm mon}} \omega_1 = \int \omega_1 
\wedge \delta_3(C_{\rm mon})$ for a 1-form $\omega_1$.
The magnetic one-form symmetry is then explicitly broken
by the dynamical monopoles, and the TQFT-dressed defect is no longer topological. Whether this breaking generates a given local symmetry-violating operator, and with what suppression, depends on the magnetic spectrum and the corresponding zero-mode structure. The FN model is treated explicitly in Secs.~\ref{sec:application}
and \ref{sec:instanton-breaking}.

\section{Anomalous FN symmetry and its non-invertible completion}
\label{sec:application}

We apply the construction of Sec.~\ref{sec:noninv-review} to an 
FN model in flat spacetime. In
Sec.~\ref{subsec:FN-review}, we specify the FN charges and determine
the subgroup preserved by the Standard Model gauge sectors. In
Sec.~\ref{subsec:UV-origin-anomaly}, we introduce the hidden
$SU(2)_H$ theory, its adjoint Higgsing to $U(1)_H$, and the
corresponding anomaly coefficients. In
Sec.~\ref{subsec:explicit-FN-defect}, we construct the non-invertible
FN defect in the monopole-decoupling limit and identify the subgroup
preserved by the flavon vacuum. Its breaking by finite-mass monopoles
and the resulting fermion interaction are discussed in
Sec.~\ref{sec:instanton-breaking}.

\subsection{FN charges and the residual subgroup}
\label{subsec:FN-review}

We introduce the global FN 
symmetry $U(1)_{\rm FN}$
and a complex flavon $\varphi$ \cite{Froggatt:1978nt}. For the Standard
Model fields, written as left-handed Weyl fields, FN charges are chosen as
\begin{align}
F(\varphi)&=-2,&F(H)&=0,\nonumber\\
F(Q_i)&=2q_i,&F(u_i^c)&=2u_i,&F(d_i^c)&=2d_i,\nonumber\\
F(L_i)&=2\ell_i,&F(e_i^c)&=2e_i.
\label{eq:FN-charge-convention}
\end{align}
The integers $q_i,u_i,d_i,\ell_i,e_i$ determine the number of flavon
insertions in the charged-fermion Yukawa interactions. For the
nonnegative charges used below, the FN-invariant interactions are
\begin{align}
\mathcal L_{\rm FN}^{\rm pert}\supset{}&
c^u_{ij}\left(\frac{\varphi}{\Lambda_{\rm FN}}\right)^{q_i+u_j}Q_iHu_j^c
+c^d_{ij}\left(\frac{\varphi}{\Lambda_{\rm FN}}\right)^{q_i+d_j}Q_iH^\dagger d_j^c
+c^e_{ij}\left(\frac{\varphi}{\Lambda_{\rm FN}}\right)^{\ell_i+e_j}
L_iH^\dagger e_j^c+\mathrm{h.c.}
\label{eq:FN-perturbative-lagrangian}
\end{align}
Here $c^{u,d,e}_{ij}$ are dimensionless coefficients of order unity,
and the generation indices are summed. Gauge and two-component
spinor contractions are implicit. Each term has vanishing total FN
charge. We define
\begin{align}
\epsilon=\frac{\langle\varphi\rangle}{\Lambda_{\rm FN}}>0,
\label{eq:FN-epsilon}
\end{align}
where $\Lambda_{\rm FN}$ is the FN scale. We work in a field basis
in which $\epsilon$ is real and positive. This does not impose CP
conservation. For the leading interactions considered below, the
constant FN phase does not give an independent CKM or PMNS
phase when the phases of FN-invariant combinations of scalar
expectation values are held fixed. 
Physical flavor phases remain in the Yukawa and messenger coefficients. 
The gauge theta angles and the instanton coefficient are understood
in this same field basis. A constant field rephasing does not remove
the dynamical FN phase, which remains in the spectrum unless an
interaction gives it a mass. The scalar vacuum is discussed in
Appendix~\ref{app:scalar-alignment}. 
Replacing $\varphi$ by its expectation value,
we obtain
\begin{align}
(Y_u)_{ij}=c^u_{ij}\epsilon^{q_i+u_j},
\qquad
(Y_d)_{ij}=c^d_{ij}\epsilon^{q_i+d_j},
\qquad
(Y_e)_{ij}=c^e_{ij}\epsilon^{\ell_i+e_j}.
\label{eq:charged-textures}
\end{align}

To make the anomaly calculation definite, we use the following charge
assignment at $\epsilon=0.2$:
\begin{align}
q&=(3,2,0),
&
u&=(4,2,0),
&
d&=(4,3,3),
\nonumber\\
\ell&=(1,0,0),
&
e&=(7,5,3).
\label{eq:benchmark-charges}
\end{align}
These charges fix the anomaly coefficients and hence the surviving FN
subgroup. The resulting flavor textures are derived in
Sec.~\ref{subsec:mass-hierarchy}.

An FN transformation can impose a quantum selection rule only if
its anomalous phases are accounted for. We first determine the
subgroup not broken by Standard Model gauge instantons. We normalize
the mixed anomaly coefficients as
\begin{align}
\mathcal A_{Faa}
=
\sum_f
F_f\,T_a(r_f)\,d_f^{(a)},
\end{align}
where $d_f^{(a)}$ denotes the multiplicity from the spectator gauge
quantum numbers and
\begin{align}
T(\mathbf N)=\frac12
\end{align}
for the fundamental representation of $SU(N)$. We then obtain
\begin{align}
\begin{aligned}
\mathcal A_{F33}
&=
\sum_i
\left[
F(Q_i)
+\frac12F(u_i^c)
+\frac12F(d_i^c)
\right],
\\
\mathcal A_{F22}
&=
\frac12
\sum_i
\left[
3F(Q_i)+F(L_i)
\right].
\end{aligned}
\label{eq:SM-FN-anomalies}
\end{align}

For a unit instanton, the index theorem gives
$2T_a(r_f)d_f^{(a)}$ zero modes for the Weyl fermion $f$.
Therefore, the corresponding 't Hooft vertex carries the FN charge:
\begin{align}
\Delta F_a
&=
\sum_f
F_f\,
2T_a(r_f)d_f^{(a)}
=
2\mathcal A_{Faa}.
\label{eq:instanton-FN-charge}
\end{align}
For these charges, we obtain
\begin{align}
\mathcal A_{F33}
=
26,
\qquad
\mathcal A_{F22}
=
16,
\end{align}
and hence
\begin{align}
\Delta F_{\rm QCD}
=
52,
\qquad
\Delta F_{\rm EW}
=
32.
\label{eq:SM-deltaF}
\end{align}
The subgroup preserved by both non-Abelian gauge sectors is
\begin{align}
\mathbb Z_{g_{\rm SM}}^{\rm FN}
\qquad
\mathrm{with}
\qquad
g_{\rm SM}
=
\gcd(52,32)
=
4.
\label{eq:SM-surviving-subgroup}
\end{align}
For a general FN charge
assignment, the surviving subgroup need not be $\mathbb Z_4^{\rm FN}$;
its order is determined by
$g_{\rm SM}=\gcd(\Delta F_{\rm QCD},\Delta F_{\rm EW})$.
Additional constraints from the remaining gauge sectors must then be
checked separately.
Note that the wall TQFT introduced below is coupled only to the hidden magnetic
one-form symmetry and cannot cancel a nontrivial QCD or electroweak
instanton phase. We therefore restrict ourselves to
$\mathbb Z_4^{\rm FN}$, which is the largest subgroup of
$U(1)_{\rm FN}$ with trivial QCD and electroweak instanton phases.

For the Abelian gauge factor, we take the direct-product global form:
\begin{align}
G_{\rm SM}
=
SU(3)_c
\times
SU(2)_L
\times
U(1)_{\widehat Y},
\label{eq:SM-global-form}
\end{align}
with $\widehat Y=6Y$. 
On a spin four-manifold, we take
\begin{align}
I_Y
=
\frac{1}{8\pi^2}
\int f_Y\wedge f_Y
\in
\mathbb Z.
\end{align}
The mixed FN--$[U(1)_{\widehat Y}]^2$ anomaly coefficient is
\begin{align}
\kappa_{FYY}
={}&
6\sum_iF(Q_i)
+
48\sum_iF(u_i^c)
+
12\sum_iF(d_i^c)
+
18\sum_iF(L_i)
+
36\sum_iF(e_i^c)
\nonumber\\
={}&
1992
(=
4\times498).
\label{eq:hypercharge-anomaly-value}
\end{align}
For the order-four FN transformation, the hypercharge contribution to
the Fujikawa phase is therefore
\begin{align}
\exp\!\left[
i\frac{\pi}{2}
\kappa_{FYY}I_Y
\right]
=
1,
\label{eq:hypercharge-trivial-phase}
\end{align}
indicating that hypercharge does not further reduce
$\mathbb Z_4^{\rm FN}$.

Thus the Standard Model gauge sectors preserve the order-four subgroup
$\mathbb Z_4^{\rm FN}$. To obtain a non-invertible completion of this
subgroup, we next introduce a hidden gauge sector for which the same
FN transformation has a nontrivial rational anomaly.

\subsection{Ultraviolet origin of the hidden Abelian theory and anomaly matching}
\label{subsec:UV-origin-anomaly}

We take the hidden gauge group to be $SU(2)_H$ and introduce the
left-handed Weyl doublets $N_D^c$ and $\eta$, with
\begin{align}
N_D^c=\begin{pmatrix}N_1^c\\N_2^c\end{pmatrix}\sim\mathbf2,
\qquad \eta\sim\mathbf2.
\end{align}
Their FN charges are $F(N_D^c)=2$ and $F(\eta)=-1$.
The components of $N_D^c$ are the charge conjugates of two right-handed
neutrinos. The additional doublet $\eta$ makes the number of hidden
Weyl doublets even, so that the Witten anomaly is absent
\cite{WittenSU2}.

Since the theory contains fundamental doublets, the gauge group is
$SU(2)_H$, not $SO(3)_H$. This global form fixes the Wilson and
't Hooft lines \cite{AharonySeibergTachikawa}. We introduce a real
adjoint scalar $\Phi=\Phi^aT^a$, with $T^a=\sigma^a/2$ and
$F(\Phi)=0$, to obtain an intermediate Abelian gauge theory. The
relevant kinetic terms are
\begin{align}
\mathcal L_H\supset{}&-\frac14G^a_{\mu\nu}G^{a\mu\nu}
+\frac12(D_\mu\Phi^a)(D^\mu\Phi^a)
+i\eta^\dagger\bar\sigma^\mu D_\mu\eta
+iN_D^{c\dagger}\bar\sigma^\mu D_\mu N_D^c.
\label{eq:hidden-lagrangian}
\end{align}
When we write
\begin{align}
(\eta N_D^c)=\eta^T\varepsilon_2N_D^c
\qquad\mathrm{with}\qquad
\varepsilon_2=\begin{pmatrix}0&1\\-1&0\end{pmatrix},
\label{eq:epsilon-hidden}
\end{align}
with the two-component spinor contraction implicit, 
the bilinear $(\eta N_D^c)$ and the adjoint Yukawa interaction
$(\eta\Phi N_D^c)$ both have FN charge one. Neither can be made
FN invariant by integer powers of the charge-minus-two flavon.
They are therefore absent from the FN-preserving perturbative action. 
To protect the Dirac nature of the neutrinos, we also impose
an independent discrete gauge symmetry $\mathbb Z_8^D$.
We assign $q_D(\eta)=2$ and $q_D(N_D^c)=6$ modulo eight.
These charges allow $(\eta N_D^c)$ but forbid the same-field
Majorana-type interactions. The full charge assignment and
anomaly cancellation are discussed in Sec.~\ref{subsec:diracness}.

The adjoint vacuum expectation value
\begin{align}
\langle\Phi\rangle
=
v_HT^3
\end{align}
breaks the hidden gauge group as
\begin{align}
SU(2)_H
\longrightarrow
U(1)_H.
\label{eq:first-Higgsing}
\end{align}
In Sec.~\ref{subsec:renormalizable-messenger-realization},
we introduce a scalar doublet $\Omega$ whose expectation value
breaks the remaining $U(1)_H$ gauge symmetry completely.
The scalar vacuum is also specified there, while the scalar
potential is given in Appendix~\ref{app:scalar-alignment}. 
The charged gauge bosons have mass
\begin{align}
M_{W_H}
=
g_Hv_H.
\label{eq:hidden-W-mass}
\end{align}
We normalize the compact Abelian connection as
\begin{align}
a_H
=
\frac{g_H}{2}A_H^3
\qquad
\mathrm{with}
\qquad
f_H
=
da_H,
\label{eq:Abelian-normalization}
\end{align}
so that
\begin{align}
\mathbf2
&\longrightarrow
(+1)\oplus(-1),
&
\mathbf3
&\longrightarrow
0\oplus(+2)\oplus(-2).
\end{align}

The vacuum manifold is
\begin{align}
\frac{SU(2)_H}{U(1)_H}
\simeq
S^2,
\end{align}
and
\begin{align}
\pi_2(S^2)
=
\mathbb Z.
\end{align}
The Abelian phase therefore contains finite-energy
't Hooft-Polyakov monopoles \cite{tHooft,Polyakov}. In the BPS
limit, the monopole solution and its mass are given in
Ref.~\cite{PrasadSommerfield}. With the normalization in
Eq.~\eqref{eq:Abelian-normalization}, the minimally charged monopole
satisfies
\begin{align}
\frac{1}{2\pi}
\int_{S^2_\infty}f_H
=
1.
\end{align}
Its mass in the BPS limit is
\begin{align}
M_{\rm mon}
=
\frac{4\pi v_H}{g_H}.
\label{eq:BPS-monopole-mass}
\end{align}
For a general scalar potential, this expression is modified by an
$\mathcal O(1)$ profile function.

Both hidden Weyl doublets contribute to the mixed FN anomaly. With
$F(\eta)=-1$, $F(N_D^c)=2$, and $T_H(\mathbf2)=1/2$, we obtain
\begin{align}
\mathcal A_{{\rm FN}\,HH}
=\sum_f F_fT_H(r_f)=\frac12.
\label{eq:hidden-nonabelian-anomaly}
\end{align}
An instanton of topological charge one has one chiral zero mode for
each doublet. Its two-fermion vertex therefore carries
\begin{align}
\Delta F_{\rm FN}=2\mathcal A_{{\rm FN}\,HH}=1.
\label{eq:UV-deltaF}
\end{align}
The sum of FN charges, rather than the individual charges, fixes this
violation. It is exactly one unit, so the instanton can generate
$(\eta N_D^c)$, which flavon insertions alone could not generate.
This is an anomaly of the global FN current, not an uncancelled
anomaly of a dynamical gauge symmetry.

After adjoint Higgsing, each doublet gives fields of $U(1)_H$ charge
$+1$ and $-1$. Thus the Abelian coefficient is
\begin{align}
\kappa_{{\rm FN}\,HH}
&=\sum_\lambda F_\lambda q_{H,\lambda}^{\,2}
=(-1)\bigl[(+1)^2+(-1)^2\bigr]
+2\bigl[(+1)^2+(-1)^2\bigr]=2,
\end{align}
and
\begin{align}
\left.d\star J_{\rm FN}\right|_{\rm hidden}
=\frac{2}{8\pi^2}f_H\wedge f_H.
\label{eq:IR-FN-anomaly}
\end{align}

The normalization in Eq.~\eqref{eq:Abelian-normalization} gives the
Cartan curvature
\begin{align}
\mathcal F_H
=
2f_HT^3.
\end{align}
Using $\operatorname{Tr}(T^3T^3)=1/2$, a purely Cartan curvature obeys
\begin{align}
\left.Q_{\rm inst}\right|_{\rm Cartan}
&=\frac{1}{8\pi^2}\int\operatorname{Tr}(\mathcal F_H\wedge\mathcal F_H)
=2I_H,
\label{eq:UV-IR-topological-number}
\end{align}
with 
\begin{align}
I_H&\equiv\frac{1}{8\pi^2}\int f_H\wedge f_H.     
\end{align}
This relation fixes the conversion between the
$SU(2)_H$ and $U(1)_H$ normalizations of the FN anomaly
and the theta terms. In particular,
\begin{align}
S_{E,\vartheta}=-i\vartheta_{\rm UV}Q_{\rm inst}
=-i\vartheta_{\rm Ab}I_H,
\label{eq:hidden-theta-normalization}
\end{align}
with $\vartheta_{\rm Ab}=2\vartheta_{\rm UV}$ in the Cartan sector. 
We use the Euclidean weight $e^{-S_E}$, so the sector
$Q_{\rm inst}=1$ carries $e^{i\vartheta_{\rm UV}}$.
Here $\vartheta_{\rm UV}$ appears in the unit-instanton factor, while
$\vartheta_{\rm Ab}$ appears in the Abelian wall construction. We choose
the instanton orientation such that $e^{i\vartheta_{\rm UV}}$ multiplies
the unbarred FN-charge-one vertex.\footnote{
Eq.~\eqref{eq:UV-IR-topological-number} relates the
non-Abelian topological charge $Q_{\rm inst}$ to the Abelian
integral $I_H$ for a field strength in the Cartan subalgebra. A unit $SU(2)_H$ instanton
cannot be represented everywhere by a smooth Abelian field.
Its FN-charge violation is determined by the fermion zero modes
in Eq.~\eqref{eq:UV-deltaF}.
}

We have therefore obtained the two ingredients required for the
non-invertible construction. The Standard Model gauge sectors
preserve $\mathbb Z_4^{\rm FN}$, while the intermediate hidden
$U(1)_H$ theory gives a nontrivial rational anomaly for the same
transformation. We now construct the corresponding defect.

\subsection{Non-invertible FN defect and the flavon vacuum}
\label{subsec:explicit-FN-defect}

The anomalous current by itself does not define a gauge-invariant
topological wall. For a rational Abelian anomaly, a
three-dimensional TQFT can cancel the phase produced by a wall
deformation while preserving invariance under large gauge
transformations~\cite{Choi:2022jqy,Cordova:2022ieu}.
We apply this construction to the FN transformation below.
The general construction and its fusion rule are reviewed in
Sec.~\ref{sec:noninv-review}.

We first take the formal monopole-decoupling limit of the intermediate $U(1)_H$ theory. In this limit, the monopole/dyon fugacity is set to zero,
\begin{align}
df_H=0,
\end{align}
and the magnetic one-form symmetry generated by $f_H/(2\pi)$ is exact. For the order-four FN rotation $\alpha=\pi/2$, the ratio of the Abelian anomaly coefficient to the order of the transformation is
\begin{align}
\frac{\kappa_{\rm FNHH}}{4}
=
\frac{2}{4}
=
\frac12.
\label{eq:hidden-anomaly-kappa}
\end{align}
Let us first include the Standard Model contributions to the FN
anomaly. For smooth backgrounds of the continuous gauge factors in flat spacetime, we
have\footnote{
The wall terms associated with $\mathbb Z_8^D$ are defined
by matching the continuous embedding in
Appendix~\ref{app:continuous-Dirac-lift}.
They are understood to be included in the FN wall and are
not written explicitly below. The $\mathcal A^{2,1}$ factor below
cancels the hidden $U(1)_H$ anomaly.
}
\begin{align}
d\star J_{\rm FN}
={}&\frac{1}{8\pi^2}\bigl[
52\operatorname{tr}(F_3\wedge F_3)
+32\operatorname{tr}(F_2\wedge F_2)
+1992 f_Y\wedge f_Y+2 f_H\wedge f_H
\bigr].
\label{eq:full-FN-anomaly}
\end{align}
Let $d\,\operatorname{CS}_3(A_a)=\operatorname{tr}(F_a\wedge F_a)$, with the fundamental trace normalized by $T(\mathbf N)=1/2$. For $\alpha=\pi/2$, define
\begin{align}
S_{\rm wall}^{\rm SM}
=-\frac{1}{4\pi}\int_{M_3}\bigl[
13\operatorname{CS}_3(A_3)+8\operatorname{CS}_3(A_2)
+498 a_Y\wedge da_Y
\bigr].
\label{eq:SM-integer-wall}
\end{align}
The three levels are integers for the direct-product gauge group used
here. These counterterms cancel the Standard Model contribution when
the wall is deformed. Although the finite Fujikawa phase is trivial in
a closed instanton sector, its integral over the four-volume swept by
the wall is not quantized, since this volume has a boundary.

The Standard Model contributions can therefore be removed by integer
wall counterterms, whereas the hidden contribution requires a
nontrivial wall theory. Following
Refs.~\cite{Choi:2022jqy,Cordova:2022ieu}, we attach the minimal TQFT $\mathcal A^{2,1}$, whose response cancels the hidden anomaly
with $\kappa_{\rm FNHH}/4=1/2$, and obtain
\begin{align}
\mathcal D_4^{\rm FN}(M_3)
=
\int\mathcal Dc\,
\exp\!\left\{
i\frac{\pi}{2}\int_{M_3}\star J_{\rm FN}
{+iS_{\rm wall}^{\rm SM}}
+i\int_{M_3}
\left[
\frac{2}{4\pi}c\wedge dc
+\frac{1}{2\pi}c\wedge f_H
\right]
\right\}.
\label{eq:FN-defect}
\end{align}
Here $c$ is a compact $U(1)$ gauge field on the wall.
$U_{\pi/2}^{\rm FN}$ denotes the FN rotation including
$\exp(iS_{\rm wall}^{\rm SM})$. Equivalently,
\begin{align}
\mathcal D_4^{\rm FN}
=
U_{\pi/2}^{\rm FN}\otimes\mathcal A^{2,1}.
\end{align}
The local response of the wall TQFT is
\begin{align}
S_{\rm resp}
=
-\frac{1}{8\pi}
\int_{M_3}a_H\wedge da_H,
\label{eq:FN-local-response}
\end{align}
understood as the response of the globally defined wall theory in
Eq.~\eqref{eq:FN-defect}, rather than as an independently defined
fractional Chern-Simons action.

\paragraph{Topologicality.}
The FN rotation and the Standard Model counterterms give the following
change when the wall is deformed across a four-volume $V_4$:
\begin{align}
\Delta S_{\rm FN}
=
\frac{\pi}{2}\int_{V_4}d\star J_{\rm FN}
+\Delta S_{\rm wall}^{\rm SM}
=
\frac{1}{8\pi}\int_{V_4}f_H\wedge f_H.
\label{eq:FN-wall-deformation}
\end{align}
The inflow of Eq.~\eqref{eq:FN-local-response} gives the opposite phase. Correlation functions containing $\mathcal D_4^{\rm FN}$ are therefore invariant under smooth wall deformations as long as the deformation crosses neither a magnetic worldline nor another operator insertion.

\paragraph{Non-invertible transformations.}
On a local operator of FN charge $F$,
\begin{align}
\mathcal D_4^{\rm FN}:
\qquad
\mathcal O_F
\longmapsto
e^{i\pi F/2}\mathcal O_F.
\end{align}
In particular, the charge-one bilinear transforms as
\begin{align}
(\eta N_D^c)
\longmapsto
i\,(\eta N_D^c).
\end{align}
The non-invertibility of the wall can be seen by 
the action on an 't Hooft loop.
The wall $\mathcal{D}_4^{\rm FN}$ acts on an 't Hooft loop 
 as a 
non-invertible transformation as in Eq.~\eqref{eq:non-inv-mag},
\begin{equation}
 \vevs{\mathcal{D}_4^{\rm FN} (S^2\times S^1)
T(C_{\rm mon})} = 0 ,
\label{noninvFN}
\end{equation}
where $S^1$ is a temporally extended circle, 
and $S^2$ is a spatially extended sphere that encloses 
a time slice of the 't Hooft loop.

\paragraph{Flavon vacuum.}

The parent wall is defined before choosing the flavon vacuum. Since
$F(\varphi)=-2$, the order-four FN transformation changes the sign of
$\langle\varphi\rangle$, whereas its order-two subgroup leaves this
expectation value invariant. Hence
\begin{align}
\mathbb Z_4^{\rm FN}\xrightarrow{\langle\varphi\rangle}
\mathbb Z_2^{\rm FN}
\end{align}
on local gauge-invariant operators in a fixed flavon vacuum. 
We call this residual $\mathbb Z_2^{\rm FN}$ transformation
FN parity; it acts as $(-1)^F$ on local gauge-invariant operators. 
The bilinear $(\eta N_D^c)$ is odd under this parity and remains
absent in perturbation theory. The realization of the same parity
after the remaining $U(1)_H$ is broken by $\langle\Omega\rangle$
is discussed in Sec.~\ref{subsec:renormalizable-messenger-realization}.

The FN selection rule therefore forbids $(\eta N_D^c)$
in the monopole-decoupled theory, despite the anomalous FN
current. The coefficient generated when dynamical monopoles
are included must be determined by the instanton calculation
below.

\section{Instanton-induced breaking of the non-invertible FN symmetry}
\label{sec:instanton-breaking}

The FN defect of Sec.~\ref{subsec:explicit-FN-defect} is topological
in the intermediate $U(1)_H$ theory when the monopole fugacity
vanishes. We now restore finite-mass monopoles and determine the
resulting FN-breaking interaction. In
Sec.~\ref{subsec:monopole-instanton-breaking}, we use the hidden
$SU(2)_H$ instanton and its fermion zero modes to derive the mixing
$\mu_H(\eta N_D^c)$ and explain its monopole interpretation. 
In Sec.~\ref{subsec:instanton-stability}, we discuss perturbative
stability and the suppression of independent ultraviolet
contributions. We also describe a nonminimal $\mathbb Z_9$
extension that protects the chiral mass at renormalizable order,
while higher-dimensional FN-breaking terms must remain subleading.

\subsection{The instanton-induced fermion mixing}
\label{subsec:monopole-instanton-breaking}
\label{sec:noninv-naturalness}

We now restore the finite-mass monopoles. A monopole worldline modifies the Bianchi identity as
\begin{align}
df_H=2\pi\,\delta_3(C_{\rm mon}),
\end{align}
so the magnetic one-form symmetry is no longer exact and $\mathcal D_4^{\rm FN}$ ceases to be topological when its wall crosses the worldline.

The loss of topologicality by itself does not determine the magnitude
or FN charge of a local symmetry-violating coupling. In the present
model, this information follows from the ultraviolet $SU(2)_H$
theory. As shown in Eq.~\eqref{eq:UV-deltaF}, the minimal topological
sector that violates the FN symmetry is an $SU(2)_H$ instanton of topological charge one
with $\Delta F_{\rm FN}=1$. The corresponding semiclassical factor is
\begin{align}
\zeta_H
\equiv
\mathcal K_H
\exp\!\left[
-\frac{8\pi^2}{g_H^2(v_H)}
+i\vartheta_{\rm UV}
\right].
\label{eq:monopole-spurion}
\end{align}
Here $\vartheta_{\rm UV}$ is the dimensionless coefficient of the
hidden $SU(2)_H$ topological term defined in
Eq.~\eqref{eq:hidden-theta-normalization}. It is distinct from the
flavon $\varphi$ and is not defined as a heavy-fermion mass phase. 
We regard $\zeta_H$ as an FN-breaking spurion with charge $-1$.
This assignment compensates the unit FN charge carried by the instanton-induced operator, so that, for example, $\zeta_H(\eta N_D^c)$ is neutral in the spurion analysis. 
The factor $\mathcal K_H$ contains the instanton-size integral,
fluctuation determinant, zero-mode normalization, hidden couplings,
and threshold factors.

\paragraph{Constrained instanton and zero modes.}
Since $\langle\Phi\rangle\neq0$, the scale-invariant BPST instanton is not an exact solution of the gauge-Higgs equations. The relevant semiclassical configuration is a constrained deformation of the BPST instanton \cite{BPST,Affleck}, for which
\begin{align}
S_{\rm inst}(\rho)
=
\frac{8\pi^2}{g_H^2(1/\rho)}
+c_\Phi v_H^2\rho^2+\cdots
\qquad
\mathrm{with}
\qquad
c_\Phi>0.
\label{eq:constrained-action}
\end{align}
The Higgs term suppresses large instantons and selects $\rho\sim v_H^{-1}$.

An $SU(2)_H$ instanton with topological charge one has one
fermion zero mode for each of $\eta$ and $N_D^c$. It therefore generates
a two-fermion interaction directly, without an additional mass insertion
\cite{tHooftInstanton,Cordova:2022ieu} as shown in Fig.~\ref{fig:instanton}:
\begin{align}
\Delta\mathcal L_{2f}
=\mu_H(\eta N_D^c)+\mathrm{h.c.}
\label{eq:general-hidden-mixing}
\end{align}
The mixing coefficient is
\begin{align}
\mu_H=v_H\zeta_H.
\label{eq:hidden-mixing}
\end{align}
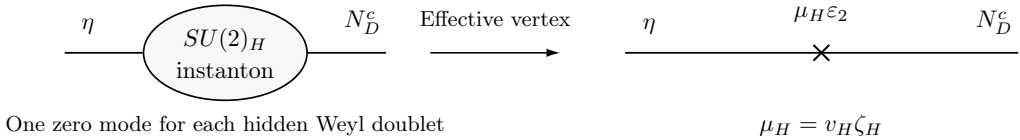
\begin{figure}[H]
\centering
\resizebox{\linewidth}{!}{%
  \begin{tikzpicture}[
  x=1cm,y=1cm,
  font=\fontsize{10}{12}\selectfont,
  fnhead/.style={anchor=west,font=\fontsize{11}{13}\selectfont},
  fnline/.style={line width=0.75pt},
  fnarrow/.style={-{Latex[length=2mm,width=1.25mm]},line width=0.65pt},
  fnnote/.style={font=\fontsize{9}{11}\selectfont,align=center},
  fncross/.style={line width=0.85pt}
]
\path[use as bounding box] (0,0) rectangle (16.6,3.78);
\begin{scope}[yshift=-4.72cm]

\node[ellipse,draw=black,fill=black!3,line width=0.7pt,
  minimum width=2.3cm,minimum height=1.20cm,align=center] (inst) at (3.35,6.70)
  {$SU(2)_H$\\instanton};
\draw[fnline] (1.00,6.70) -- (inst.west);
\draw[fnline] (inst.east) -- (5.70,6.70);
\node[above] at (1.34,6.84) {$\eta$};
\node[above] at (5.36,6.84) {$N_D^c$};
\draw[fnarrow] (6.35,6.70) -- (8.25,6.70)
  node[midway,above=7pt,fnnote] {Effective vertex};

\draw[fnline] (9.20,6.70) -- (14.95,6.70);
\draw[fncross] (11.955,6.585) -- (12.185,6.815);
\draw[fncross] (11.955,6.815) -- (12.185,6.585);
\node[above] at (9.55,6.84) {$\eta$};
\node[above] at (14.60,6.84) {$N_D^c$};
\node[above=9pt] at (12.07,6.70) {$\mu_H\varepsilon_2$};
\node[fnnote] at (3.35,5.64) {One zero mode for each hidden Weyl doublet};
\node at (12.07,5.64) {$\mu_H=v_H\zeta_H$};
\end{scope}
\end{tikzpicture}%
}
\caption{Instanton-induced two-fermion interaction.
A unit $SU(2)_H$ instanton has one zero mode for each hidden Weyl
 doublet and generates the mixing $\mu_H\varepsilon_2$, with
 $\mu_H=v_H\zeta_H$.}
\label{fig:instanton}
\end{figure}

The coefficient $\mu_H$ has mass dimension one, whereas $\zeta_H$ is
dimensionless. The factor $v_H$ follows from the dimension of the
bilinear and the instanton size $\rho\sim v_H^{-1}$. The mixing
connects $\eta$ and $N_D^c$. Schematically, its coefficient has the form
\begin{align}
\mu_H\sim e^{i\vartheta_{\rm UV}}
\int d\rho\,\rho^{-5}\rho^3\,\mathcal P_2(\rho)
\exp\!\left[-\frac{8\pi^2}{g_H^2(1/\rho)}
-c_\Phi v_H^2\rho^2\right],
\label{eq:two-mode-size-integral}
\end{align}
where $\mathcal P_2$ contains dimensionless measure and overlap factors.
After integration these factors are included in $\mathcal K_H$ in
Eq.~\eqref{eq:monopole-spurion}. Its value also depends on heavy
thresholds and is not calculated here. 
When the extension of
Appendix~\ref{app:UV-protection} is used, both the gauge coupling and
the prefactor are those of the effective theory obtained after the
additional heavy fermions are integrated out. Its two hidden Weyl doublets are $\eta$ and $N_D^c$. 
For more details on fermion zero modes and instanton-induced
interactions, see Refs.~\cite{tHooftInstanton,Schafer:1996wv}. 
In the direct one-instanton calculation discussed above, we use the zero modes
of the common gauge-covariant Weyl operator for $\eta$ and
$N_D^c$, before including corrections from the messenger
interactions. After factoring out their independent Grassmann
coordinates, the two modes have the same commuting wavefunction
$u_\alpha^{\,a}(x)$. The local Lorentz-scalar contraction
$\epsilon^{\alpha\beta}u_\alpha^{\,a}u_\beta^{\,b}$ is
antisymmetric in the $SU(2)_H$ indices $a,b$.
It therefore gives the singlet contraction
$\eta^T\varepsilon_2N_D^c$.
Since $\varepsilon_2\Phi$ is symmetric, the corresponding
adjoint contraction vanishes. No independent
$\eta^T\varepsilon_2\Phi N_D^c$ interaction is generated by
this direct contribution, including when $\Phi$ is replaced
by its expectation value.

Note that D-brane instantons also generate perturbatively forbidden
Dirac-neutrino Yukawa couplings
\cite{Cvetic:2008hi,Ibanez:2008my}.
Both mechanisms give exponentially suppressed couplings,
but the anomalous $U(1)$ is realized differently.
In the D-brane models, the gauge anomaly is cancelled by the
Green-Schwarz mechanism, and the gauge boson acquires a
St\"uckelberg mass. 
The corresponding $U(1)$ charges still impose selection rules
on perturbative matter couplings. The axionic dependence of the instanton factor compensates the matter-field charge, so that the
induced interaction remains gauge invariant. 
On the other hand, in the present construction, $U(1)_{\rm FN}$ is a global symmetry
with a mixed ABJ anomaly, and $\zeta_H$ describes its
breaking by a four-dimensional $SU(2)_H$ instanton.
The FN charge assigned to $\zeta_H$ keeps track of this breaking.
The non-invertible defect is defined using the magnetic
one-form symmetry of the intermediate $U(1)_H$ theory
in the monopole-decoupling limit. 
The two mechanisms can therefore lead to the same small
Dirac-neutrino masses. In the present model, the specified
messenger interactions also give a tree-level Yukawa matrix
of rank at most two and $m_2/m_3\sim\epsilon$ for generic
order-one coefficients. The same heavy fermions generate
the dimension-six interaction related to the neutrino Yukawa
matrix in Eq.~\eqref{eq:neutrino-dimension-six-correlation}.
These relations follow from the messenger interactions and are not generic consequences of D-brane instanton generation of a Yukawa coupling.

\paragraph{Infrared interpretation.}
The same semiclassical exponent can also be obtained from the monopole description in the intermediate $U(1)_H$ theory ~\cite{Fan:2021ntg}. Quantizing the electric collective coordinate of the monopole gives a dyon tower, and Poisson resummation gives the leading winding action
\begin{align}
S_{{\rm loop},1}
=
\frac{2\pi M_{\rm mon}}{M_{W_H}}
=
\frac{8\pi^2}{g_H^2(v_H)}
\end{align}
in the BPS limit. The leading $n=1$ dyon winding therefore reproduces the semiclassical exponent of the unit $SU(2)_H$ instanton. The complete dyon-worldline derivation is summarized in Appendix~\ref{app:dyon}.

\subsection{Perturbative stability and ultraviolet protection}
\label{subsec:instanton-stability}
\label{subsec:UV-protection-main}

We consider the regime in which the charge-one mixing is
dominated by the hidden instanton. 
Let us write its full coefficient as
\begin{align}
\mu_{\rm tot}=\mu_H+\delta\mu_{\rm UV}
\qquad\mathrm{with}\qquad |\delta\mu_{\rm UV}|\ll|\mu_H|,
\label{eq:instanton-dominance}
\end{align}
where $\delta\mu_{\rm UV}$ denotes an independent local contribution.
Other independent FN-breaking interactions are also taken to be
subleading. 
In the minimal model, the hidden gauge symmetry and
$\mathbb Z_8^D$ allow a bare $(\eta N_D^c)$ mass term, although
it violates the FN selection rule. These gauge symmetries alone
therefore do not enforce Eq.~\eqref{eq:instanton-dominance}.
The need to protect against a bare mass was also pointed out
for an $SU(2)$ theory with two Weyl doublets in
Ref.~\cite{Cordova:2022ieu}.

We address this problem with the nonminimal gauge
extension in Appendix~\ref{app:UV-protection}.
The gauge charges ensure that all renormalizable interactions
preserve an accidental chiral charge $p$.
The bilinear $(\eta N_D^c)$ carries nonzero $p$, whereas all
scalars are neutral. In particular, the scalar $S$ that gives
masses to the additional hidden fermions preserves $p$ when
it acquires an expectation value, although it breaks
$\mathbb Z_9$. Thus an independent $(\eta N_D^c)$ mass cannot
be generated perturbatively from these renormalizable
interactions. The hidden instanton violates $p$ and generates
the required mixing after the additional hidden fermions
are integrated out. The charge assignments and gauge-anomaly
cancellation are given in the appendix.

Independent higher-dimensional interactions can violate $p$.
For example,
$c_6(S^\dagger)^3(\eta N_D^c)/\Lambda_{\rm UV}^2$
gives $\delta\mu_{\rm UV}=c_6(f^*)^3/\Lambda_{\rm UV}^2$,
where $f=\langle S\rangle$.
Its suppression relative to $\mu_H$ requires
Eq.~\eqref{eq:Z9-quality-bound}.
With $|f|=v_H$, $|c_6|=1$,
$\Lambda_{\rm UV}=2.4\times10^{18}\,\mathrm{GeV}$,
and the benchmark scales in
Sec.~\ref{subsec:neutrino-normalization}, we obtain
$|\delta\mu_{\rm UV}|/|\mu_H|\simeq8.5\times10^{-4}$.
Thus this contribution is subleading without a small
dimensionless coefficient. Other independent operators must
be checked at the scales at which they are generated. 
For more details, see Appendix~\ref{app:UV-protection}.

With the independent FN-breaking coefficients set to zero, the
symmetry-restoring limit is
\begin{align}
\zeta_H\longrightarrow0.
\label{eq:zeta-zero}
\end{align}
The parent defect is topological in the monopole-decoupled
intermediate $U(1)_H$ theory. In the chosen flavon vacuum, the residual
FN parity forbids the charge-one bilinear. Its realization after
the remaining $U(1)_H$ is completely broken is described in
Sec.~\ref{subsec:renormalizable-messenger-realization}.

With FN-preserving perturbative vertices, loop corrections to the
mixing require an FN-breaking insertion. They therefore renormalize
$\mu_H$ or the independent breaking coefficients, rather than
generate an unsuppressed additive mass. For example, the perturbative
renormalization of the instanton contribution can be written as
\begin{align}
\mu_{\rm inst}^{\rm ren}
=v_H\zeta_H\left[1+\sum_{L\geq1}
\frac{c_L}{(16\pi^2)^L}+\cdots\right],
\label{eq:multiplicative-naturalness}
\end{align}
where $c_L$ depends on FN-preserving couplings. The prefactor
$\mathcal K_H$ and the subsequent running do not introduce
independent sources of FN violation. QCD and electroweak instantons
violate FN charge only in multiples of four, and the order-four
hypercharge phase is trivial.

The analysis below applies to ultraviolet realizations
satisfying Eq.~\eqref{eq:instanton-dominance}, with the
$\mathbb Z_9$ extension providing an explicit example of
protection at renormalizable order. 
Sec.~\ref{sec:messenger-z4} derives how
$\mu_H$ enters the neutrino Yukawa matrix, and
Sec.~\ref{sec:phenomenology} relates it to the observed neutrino
mass scale and discusses the phenomenological consequences.

\section{Neutrino mass hierarchy from messenger interactions}
\label{sec:messenger-z4}

In the previous section, we derived the instanton-induced mixing $\mu_H(\eta N_D^c)$, with $\mu_H=v_H\zeta_H$. 
We now specify the remaining fields and the interactions that
transmit this mixing to the Standard Model leptons.
Secs.~\ref{subsec:minimal-field-content} and
\ref{subsec:renormalizable-messenger-realization} give the field
content and the interactions responsible for hidden gauge-symmetry
breaking and the messenger masses. 
In Sec.~\ref{subsec:messenger-matching}, we integrate out the heavy
fermions and identify the light right-handed neutrinos. We then
derive the FN mass hierarchy in Sec.~\ref{subsec:mass-hierarchy} and verify anomaly cancellation for the discrete gauge symmetry
that forbids Majorana masses in Sec.~\ref{subsec:diracness}.

\subsection{Field contents}
\label{subsec:minimal-field-content}

The field content and charge assignments of the model
are summarized in Table~\ref{tab:matter-content}. We then explain
the roles of the additional neutrino, messenger, and scalar fields.

\begin{table}[htbp]
\centering
\caption{Field content of the model, grouped
by the roles of the fields. The visible block includes the three
right-handed neutrinos in $\mathbf2\oplus\mathbf1$ of $SU(2)_H$
and the FN flavon $\varphi$. The scalar $\Omega$ breaks the remaining
$U(1)_H$ gauge symmetry and generates the $\eta$--$X$ masses.
All fermions are left-handed Weyl fields. 
The symbol $F$ denotes the $U(1)_{\rm FN}$ charge. The exact
$\mathbb Z_8^D$ protects the Dirac nature of the neutrinos,
as discussed in Sec.~\ref{subsec:diracness}.}
\label{tab:matter-content}
\small
\setlength{\tabcolsep}{4pt}
\renewcommand{\arraystretch}{1.15}
\begin{tabular}{lcccccc}
\toprule
Sector & field & spin & $(SU(3)_c,SU(2)_L)_Y$
& $SU(2)_H$ & $F$ & $q_D$\\
\midrule
\multirow{9}{*}{Visible}
&$Q_i$&$1/2$&$(\mathbf3,\mathbf2)_{1/6}$
&$\mathbf1$&$2q_i$&2\\
&$u_i^c$&$1/2$&$(\bar{\mathbf3},\mathbf1)_{-2/3}$
&$\mathbf1$&$2u_i$&6\\
&$d_i^c$&$1/2$&$(\bar{\mathbf3},\mathbf1)_{1/3}$
&$\mathbf1$&$2d_i$&6\\
&$L_i$&$1/2$&$(\mathbf1,\mathbf2)_{-1/2}$
&$\mathbf1$&$2\ell_i$&2\\
&$e_i^c$&$1/2$&$(\mathbf1,\mathbf1)_1$
&$\mathbf1$&$2e_i$&6\\
&$N_D^c$&$1/2$&$(\mathbf1,\mathbf1)_0$
&$\mathbf2$&2&6\\
&$N_S^c$&$1/2$&$(\mathbf1,\mathbf1)_0$
&$\mathbf1$&3&6\\
&$H$&0&$(\mathbf1,\mathbf2)_{1/2}$
&$\mathbf1$&0&0\\
&$\varphi$&0&$(\mathbf1,\mathbf1)_0$
&$\mathbf1$&$-2$&0\\
\midrule
\multirow{2}{*}{Hidden}
&$\eta$&$1/2$&$(\mathbf1,\mathbf1)_0$
&$\mathbf2$&$-1$&2\\
&$\Phi$&0&$(\mathbf1,\mathbf1)_0$
&$\mathbf3$&0&0\\
\midrule
\multirow{3}{*}{Messenger}
&$\Omega$&0&$(\mathbf1,\mathbf1)_0$
&$\mathbf2$&1&0\\
&$X_1$&$1/2$&$(\mathbf1,\mathbf1)_0$
&$\mathbf1$&0&6\\
&$X_2$&$1/2$&$(\mathbf1,\mathbf1)_0$
&$\mathbf1$&2&6\\
\bottomrule
\end{tabular}
\end{table}

The $SU(2)_H$ doublet $N_D^c$ introduced in
Sec.~\ref{subsec:UV-origin-anomaly} contains two right-handed
neutrinos. We add one Standard Model and hidden-gauge singlet
$N_S^c$, so that the three right-handed neutrinos are assigned to
\begin{align}
N_D^c=\begin{pmatrix}N_1^c\\N_2^c\end{pmatrix}\sim\mathbf2,
\qquad N_S^c=N_3^c\sim\mathbf1.
\label{eq:neutrino-doublet-singlet}
\end{align}
Their FN charges are $F(N_D^c)=2$ and $F(N_S^c)=3$.
All are written as left-handed charge-conjugate fields and carry
$q_D=6$ under $\mathbb Z_8^D$. The singlet is needed to cancel the
discrete mixed gravitational anomaly for the chosen charges, as shown
in Sec.~\ref{subsec:diracness}. It does not enter the leading
mass-generation interactions below. The labels $1,2,3$ refer to this
gauge basis, not to the neutrino mass eigenstates.

We also introduce two Standard Model and hidden-gauge singlet Weyl
messengers $X_1,X_2$, with
\begin{align}
F(X_1)=0,\qquad F(X_2)=2.
\end{align}
Both carry $q_D=6$ under $\mathbb Z_8^D$.
They couple $L_iH$ to the doublet $\eta$. Two messengers are needed
both to give masses to the two components of $\eta$ and to transmit
two independent lepton flavor vectors. With only one singlet messenger $X$ coupled to $L_iH$,
the tree-level neutrino Yukawa matrix would have rank
at most one.

One complex scalar doublet $\Omega$, with $F(\Omega)=1$ and
$q_D(\Omega)=0$, breaks the remaining $U(1)_H$ gauge symmetry
completely and generates the $\eta$--$X$ masses.

\subsection{Hidden symmetry breaking and messenger interactions}
\label{subsec:renormalizable-messenger-realization}

We now construct the interactions that will convert the mixing $\mu_H$ into a neutrino Yukawa coupling. The expectation value of $\Omega$ can break the remaining
$U(1)_H$ and give masses to $\eta$ and $X_1,X_2$, while leaving
three light right-handed neutrino states. In the gauge where
$\langle\Phi\rangle=v_HT^3$, write
\begin{align}
\langle\Omega\rangle=\frac{w_H}{\sqrt2}
\begin{pmatrix}z_1\\z_2\end{pmatrix}
\qquad\mathrm{with}\qquad |z_1|^2+|z_2|^2=1.
\label{eq:scalar-VEVs}
\end{align}
A nonzero fundamental expectation value breaks the residual $U(1)_H$
completely. The dimensionless numbers $z_1,z_2$ specify its orientation
relative to the adjoint expectation value. 
We leave $z_1,z_2$
arbitrary, subject to the normalization above, and impose no alignment
condition in deriving the neutrino mass hierarchy. An explicit
potential and one locally stable vacuum are given in
Appendix~\ref{app:scalar-alignment}. The hierarchy of scalar scales
remains an input to the mass-generation model. The complete breaking
sequence is
\begin{align}
SU(2)_H\xrightarrow{v_H}U(1)_H\xrightarrow{w_H}1
\qquad\mathrm{with}\qquad w_H\ll v_H.
\label{eq:two-step-Higgsing}
\end{align}
At tree level the remaining gauge boson has
\begin{align}
M_{A_H}^2=\frac{g_H^2w_H^2}{4}.
\label{eq:minimal-photon-mass}
\end{align}
Thus there is no massless hidden photon. The intermediate $U(1)_H$
construction applies in the range $w_H\ll E\ll v_H$. Below $w_H$,
a monopole is attached to a vortex string \cite{TongMonopoleHiggs}. 
The breaking of $U(1)_H$ by $\langle\Omega\rangle$ gives
corrections suppressed by powers of $w_H/v_H$ to the
instanton-induced mixing calculated near the non-Abelian core.

The residual $\mathbb Z_2^{\rm FN}$ transformation, called FN parity
in Sec.~\ref{subsec:explicit-FN-defect}, is also preserved by the
$\Omega$ expectation value at the perturbative level. On the
elementary fields, it is represented by
\begin{align}
\mathcal P_{\rm FN}=(-1)^F z_H
\qquad\mathrm{with}\qquad
z_H=-\mathbf1_2\in SU(2)_H.
\label{eq:combined-FN-parity}
\end{align}
Both factors change the sign of $\Omega$, so their product preserves
its vacuum expectation value. It also preserves $\langle\varphi\rangle$ and
$\langle\Phi\rangle$. Under this transformation $\eta,X_1,X_2,L_i,H$
are even, while $N_D^c,N_S^c$ are odd. On gauge-invariant operators
it is the ordinary FN parity. It is distinct from the exact
$\mathbb Z_8^D$: the former forbids the perturbative Dirac neutrino Yukawa,
whereas the latter forbids Majorana masses even after the hidden
instanton is included.

Since $N_D^c$ is an $SU(2)_H$ doublet, while $L_i$ and $H$
are singlets, $L_iHN_D^c$ is not gauge invariant.
We therefore combine $N_D^c$ with $\Omega$ or its complex
conjugate to form an $SU(2)_H$ singlet. 
We define $\widetilde\Omega=\varepsilon_2\Omega^*$,
which has FN charge $-1$. This notation does not introduce
an additional scalar field. 
To identify the FN breaking required for the three
neutrino couplings, consider the gauge-invariant operators:
\begin{align}
\mathcal O^\nu_{i1}&=
\left(\frac{\varphi}{\Lambda_{\rm FN}}\right)^{\ell_i+1}
L_iH(\widetilde\Omega^\dagger N_D^c),\nonumber\\
\mathcal O^\nu_{i2}&=
\left(\frac{\varphi}{\Lambda_{\rm FN}}\right)^{\ell_i}
L_iH(\Omega^\dagger N_D^c),\nonumber\\
\mathcal O^\nu_{iS}&=
\left(\frac{\varphi}{\Lambda_{\rm FN}}\right)^{\ell_i+1}L_iHN_S^c.
\label{eq:nu-operator}
\end{align}
Before
replacing $\Omega$ by its expectation value, the first two require
coefficients of mass dimension $-1$, whereas the last has a
dimensionless coefficient. Here and below, $j=1,2,S$ labels their
right-handed fields. All three have the same FN charge:
\begin{align}
F(\mathcal O^\nu_{i1})=F(\mathcal O^\nu_{i2})
=F(\mathcal O^\nu_{iS})=1.
\label{eq:charge-one-neutrino-operator}
\end{align}
Additional integer powers of $\varphi$ and $\varphi^\dagger$ cannot
compensate this odd charge. More generally, a gauge singlet containing
one $N_D^c$ requires an odd number of fundamental scalar factors.
Their FN charge is odd, while $F(N_D^c)$ and $F(L_i)$ are even.
Hence no neutrino Yukawa is generated by FN-preserving scalar
insertions. 
We now give the FN-preserving messenger interactions. They do not
produce a neutrino Yukawa without the charge-one mixing derived in
Sec.~\ref{subsec:monopole-instanton-breaking}. 
As assumed in Sec.~\ref{subsec:instanton-stability},
we use the instanton-induced mixing $\mu_H$ as the source
of FN breaking in the following calculation and neglect
independent FN-breaking mass terms, Yukawa couplings and
other interactions.

The FN charges in Table~\ref{tab:matter-content} allow
\begin{align}
\mathcal L_{\eta X}\supset{}&
\lambda_1(\eta\Omega)X_1
+\lambda_2(\eta\widetilde\Omega)X_2\nonumber\\
&+\rho_1\frac{\varphi^\dagger}{\Lambda_{\rm FN}}
(\eta\widetilde\Omega)X_1
+\rho_2\frac{\varphi}{\Lambda_{\rm FN}}(\eta\Omega)X_2
+\mathrm{h.c.}
\label{eq:general-hidden-messenger-couplings}
\end{align}
Here $(\eta\Omega)=\eta^T\varepsilon_2\Omega$. The coefficients
$\lambda_{1,2}$ and $\rho_{1,2}$ are dimensionless hidden--messenger
couplings. The first line consists of renormalizable interactions. The second line contains the leading
FN-dressed cross couplings. Both $\varphi$ and $\varphi^\dagger$ must
be included in this non-supersymmetric theory. 
The singlet messengers $X_1$ and $X_2$ couple to
the Standard Model fields $L_i$ and $H$ through
\begin{align}
\mathcal L_{\rm portal}\supset
A_i\left(\frac{\varphi}{\Lambda_{\rm FN}}\right)^{\ell_i}L_iHX_1
+B_i\left(\frac{\varphi}{\Lambda_{\rm FN}}\right)^{\ell_i+1}L_iHX_2
+\mathrm{h.c.}
\label{eq:portal}
\end{align}
The dimensionless vectors $A=(A_1,A_2,A_3)^T$ and
$B=(B_1,B_2,B_3)^T$ describe the couplings of the three lepton
doublets to $X_1$ and $X_2$, respectively. They are independent inputs
to the messenger Lagrangian, not neutrino mixing angles. The extra
flavon in the second term follows from $F(X_2)-F(X_1)=2$. After the
flavon takes its expectation value, these interactions have the form
of renormalizable Yukawa vertices. 
We treat these flavon-dependent interactions as effective
operators below $\Lambda_{\rm FN}$, without specifying the
heavy fields that generate them.

The interactions in Eq.~\eqref{eq:general-hidden-messenger-couplings}
give the mass terms
\begin{align}
\mathcal L_{\eta X}^{\rm mass}\supset
X^TM_X\eta+\mathrm{h.c.}
\end{align}
Here $X^T=(X_1,X_2)$ and $\eta^T=(\eta^1,\eta^2)$. 
Substituting the expectation value
in Eq.~\eqref{eq:scalar-VEVs}, we obtain
\begin{align}
M_X
&=\frac{w_H}{\sqrt2}
\begin{pmatrix}
\lambda_1z_2-\epsilon\rho_1z_1^*
&
-\lambda_1z_1-\epsilon\rho_1z_2^*
\\
\epsilon\rho_2z_2-\lambda_2z_1^*
&
-\epsilon\rho_2z_1-\lambda_2z_2^*
\end{pmatrix}
=\widehat M_XU_\Omega.
\label{eq:MX-from-one-doublet}
\end{align}
Here, the vacuum orientation is contained in the matrix
\begin{align}
U_\Omega=\begin{pmatrix}z_2&-z_1\\z_1^*&z_2^*\end{pmatrix},
\label{eq:vacuum-orientation-matrix}
\end{align}
satisfying $U_\Omega^\dagger U_\Omega=\mathbf1_2$ and
$\det U_\Omega=1$, whereas the orientation-independent mass matrix is
\begin{align}
\widehat M_X
&=\frac{w_H}{\sqrt2}
\begin{pmatrix}
\lambda_1&-\epsilon\rho_1\\
\epsilon\rho_2&-\lambda_2
\end{pmatrix}
\equiv w_H\begin{pmatrix}a&b\epsilon\\c\epsilon&d\end{pmatrix}.
\label{eq:orientation-independent-mass}
\end{align}
This equality defines $a,b,c,d$ in terms of the original couplings;
they do not depend on $z_1,z_2$. The diagonal entries of
$\widehat M_X$ are of order $w_H$, whereas its off-diagonal entries
are of order $\epsilon w_H$ for generic order-one couplings.
Since $M_XM_X^\dagger=\widehat M_X\widehat M_X^\dagger$, the two
heavy masses are independent of the vacuum orientation. Thus both
components of $\eta$ acquire masses of order $w_H$ through the two
singlet messengers, without introducing a second scalar doublet. 
For the calculation of the neutrino Yukawa matrix below,
we define the dimensionless determinant:
\begin{align}
\Delta_X=\frac{\det M_X}{w_H^2}=ad-bc\epsilon^2.
\label{eq:messenger-determinant}
\end{align}
We take $\Delta_X$ to be of order unity, so that no small
messenger mass introduces an additional hierarchy.

After replacing $\varphi$ by its expectation value,
the coefficients of $L_iHX_1$ and $L_iHX_2$ in
Eq.~\eqref{eq:portal} are $A_i\epsilon^{\ell_i}$ and
$B_i\epsilon^{\ell_i+1}$, respectively.
Both couplings therefore contain the same factor
$\epsilon^{\ell_i}$ associated with $L_i$.
The charged-lepton Yukawa coupling in
Eq.~\eqref{eq:charged-textures} contains this factor as well,
together with $\epsilon^{e_j}$ associated with $e_j^c$.
We collect these factors in the diagonal matrices:
\begin{align}
\begin{aligned}
D_L&=\operatorname{diag}
\left(\epsilon^{\ell_1},\epsilon^{\ell_2},\epsilon^{\ell_3}\right),\\
D_e&=\operatorname{diag}
\left(\epsilon^{e_1},\epsilon^{e_2},\epsilon^{e_3}\right).
\end{aligned}
\label{eq:FN-diagonal-factors}
\end{align}

The interactions of $L_i,H$ with the singlet messengers
can then be written as
\begin{align}
\mathcal L_{\rm portal}\supset
(LH)^TPX+\mathrm{h.c.},
\label{eq:portal-matrix-form}
\end{align}
where $P$ is a dimensionless $3\times2$ matrix.
Factoring $\epsilon^{\ell_i}$ out of its $i$th row gives
\begin{align}
P
&=D_L
\begin{pmatrix}
A_1&\epsilon B_1\\
A_2&\epsilon B_2\\
A_3&\epsilon B_3
\end{pmatrix}
=
\begin{pmatrix}
\epsilon^{\ell_1}A_1&\epsilon^{\ell_1+1}B_1\\
\epsilon^{\ell_2}A_2&\epsilon^{\ell_2+1}B_2\\
\epsilon^{\ell_3}A_3&\epsilon^{\ell_3+1}B_3
\end{pmatrix}.
\label{eq:visible-portal-matrix}
\end{align}
Similarly, factoring $\epsilon^{\ell_i}$ and
$\epsilon^{e_j}$ out of the rows and columns of the
charged-lepton Yukawa matrix gives
\begin{align}
Y_e=D_LC_eD_e
\qquad\mathrm{with}\qquad
(C_e)_{ij}=c^e_{ij}.
\label{eq:charged-lepton-factorization}
\end{align}
The same $D_L$ appears in $P$ and $Y_e$ because both
interactions contain the lepton doublets $L_i$.
This factorization does not rotate the fermion fields,
and $Y_e$ is generally nondiagonal in this basis.

Note that we use canonically normalized fermion fields in the Lagrangian above.
The two heavy Dirac masses remain of order $w_H$ as
$\epsilon\to0$, so their inverse masses do not introduce a factor
of $1/\epsilon$. On the other hand, the coupling of each $L_i$
to $X_2$ contains one more power of $\epsilon$ than its coupling
to $X_1$. 
In Sec.~\ref{subsec:messenger-matching}, we integrate out
the heavy fermions and show that a unitary rotation of the
right-handed neutrinos separates the vacuum orientation from the
Yukawa matrix. In that basis, the two nonzero columns differ by
one power of $\epsilon$.

\subsection{Light right-handed states and tree-level matching}
\label{subsec:messenger-matching}

We now derive the Dirac-neutrino Yukawa coupling by integrating out
$\eta$ and the two singlet messengers $X_1,X_2$. The Standard Model fields and the three right-handed neutrinos
described by $N_D^c$ and $N_S^c$ remain in the low-energy
effective theory. 
After $\Phi$ and $\Omega$ acquire their vacuum expectation values,
the relevant interactions are
\begin{align}
\mathcal L\supset
(LH)^TPX+X^TM_X\eta+\mu_H\eta^T\varepsilon_2N_D^c
+\mathrm{h.c.}
\label{eq:two-doublet-mass-terms}
\end{align}
Here $\mu_H=v_H\zeta_H$ is the mixing coefficient derived in
Eq.~\eqref{eq:hidden-mixing}. We use $(LH)_i=L_iH$ and
$X^T=(X_1,X_2)$; $N_S^c$ does not enter these interactions.
At $\mu_H=0$, the two components of $\eta$ pair with $X_1,X_2$
to form two heavy Dirac fermions. In the order $(X,\eta)$, their
symmetric Weyl mass matrix is
\begin{align}
\mathbb M_H=\begin{pmatrix}0&M_X\\M_X^T&0\end{pmatrix}.
\label{eq:heavy-matrix-main}
\end{align}
We denote the two heavy Dirac masses at $\mu_H=0$ by
$\sigma_-\leq\sigma_+$. 
The two masses are the nonnegative diagonal entries obtained by
biunitary diagonalization of $M_X$. We take both masses to be of
order $w_H$ and much larger than $|\mu_H|$ and the electroweak
scale $v$. Their explicit expressions are given in
Appendix~\ref{app:heavy-fermion-stability}.

Before eliminating the heavy fields, we identify the light states
from the fermion bilinear terms. Before electroweak symmetry breaking,
$L_iHX_\alpha$ is an interaction rather than a mass term and does not
enter this mass matrix. The mass terms couple the two fields
$\eta^1,\eta^2$, each with $q_D=2$, to the five fields
$X_1,X_2,N_1^c,N_2^c,N_S^c$, each with $q_D=6$:
\begin{align}
\mathcal L_{\rm mass}\supset
\begin{pmatrix}\eta^1&\eta^2\end{pmatrix}
\mathcal M_D
\begin{pmatrix}
X_1\\X_2\\N_1^c\\N_2^c\\N_S^c
\end{pmatrix}
+\mathrm{h.c.}
\label{eq:mass-bilinear-field-basis}
\end{align}
Here the $2\times5$ mass matrix is
\begin{align}
\mathcal M_D=
\left(M_X^T\quad\mu_H\varepsilon_2\quad0_{2\times1}\right).
\label{eq:rectangular-dirac-matrix}
\end{align}
All these
fields are written as left-handed Weyl fields; they have been grouped
by their discrete charges, not by chirality or mass.
Since $M_X$ is nonsingular, $\mathcal M_D$ has rank two. Two linear
combinations of the five $q_D=6$ fields pair with $\eta$ to form heavy
Dirac fermions. The other three combinations are massless before
electroweak symmetry breaking. At $\mu_H=0$ they are exactly
$N_D^c,N_S^c$. For nonzero $\mu_H$, these massless directions satisfy
$M_X^TX+\mu_H\varepsilon_2N_D^c=0$ and have no $\eta$ component:
\begin{align}
X=-\mu_H M_X^{-T}\varepsilon_2N_D^c,\qquad \eta=0.
\label{eq:light-right-handed-embedding}
\end{align}
Thus the two light states descending from $N_D^c$ contain $X$
components of order $\mu_H/w_H$, whereas $N_S^c$ remains unmixed
in the mass terms of Eq.~\eqref{eq:two-doublet-mass-terms}.

To determine the normalization of these light states, we start from
the canonical kinetic terms of the seven fermion components,
\begin{align}
\mathcal L_{\rm kin}=
iX^\dagger\bar\sigma^\mu\partial_\mu X
+i\eta^\dagger\bar\sigma^\mu\partial_\mu\eta
+iN_D^{c\dagger}\bar\sigma^\mu\partial_\mu N_D^c
+iN_S^{c\dagger}\bar\sigma^\mu\partial_\mu N_S^c.
\label{eq:canonical-fermion-kinetic-terms}
\end{align}
Here we omit gauge interactions and terms involving scalar
fluctuations for simplicity. Substituting
Eq.~\eqref{eq:light-right-handed-embedding}
into the $X$ kinetic term gives
\begin{align}
\left.iX^\dagger\bar\sigma^\mu\partial_\mu X\right|_{\rm light}
={}&i|\mu_H|^2N_D^{c\dagger}\varepsilon_2^\dagger
(M_X^{-T})^\dagger M_X^{-T}\varepsilon_2
\bar\sigma^\mu\partial_\mu N_D^c.
\label{eq:induced-light-kinetic-term}
\end{align}
The $\eta$ kinetic term vanishes on these light directions. Adding
the original kinetic terms of $N_D^c$ and $N_S^c$, we obtain
\begin{align}
\mathcal L_{\rm kin}^{\rm light}
=i\begin{pmatrix}N_D^c\\N_S^c\end{pmatrix}^{\!\dagger}
Z_R\bar\sigma^\mu\partial_\mu
\begin{pmatrix}N_D^c\\N_S^c\end{pmatrix}.
\label{eq:light-kinetic-Lagrangian}
\end{align}
The kinetic matrix is then given by
\begin{align}
Z_R=\begin{pmatrix}
\mathbf1_2+|\mu_H|^2\varepsilon_2^\dagger
(M_X^{-T})^\dagger M_X^{-T}\varepsilon_2&0\\
0&1
\end{pmatrix}.
\label{eq:light-kinetic-matrix}
\end{align}
The identity terms come from the original right-handed-neutrino
kinetic terms, while the additional term comes from their $X$
components. 

Since $Z_R$ is positive Hermitian, we use its positive inverse square
root to define the canonically normalized light fields,
\begin{align}
\begin{pmatrix}N_D^c\\N_S^c\end{pmatrix}
=Z_R^{-1/2}n_{\rm light}^c.
\label{eq:minimal-right-normalization}
\end{align}
The kinetic term then becomes
\begin{align}
\mathcal L_{\rm kin}^{\rm light}
=i n_{\rm light}^{c\dagger}\bar\sigma^\mu
\partial_\mu n_{\rm light}^c.
\label{eq:canonical-light-kinetic-term}
\end{align}
The vector $n_{\rm light}^c$ contains the three canonically normalized
light Weyl fields. 
Since $Z_R-\mathbf1_3=\mathcal O(|\mu_H|^2/w_H^2)$,
the factor $Z_R^{-1/2}$ multiplying the Yukawa matrix is
$\mathbf1_3+\mathcal O(|\mu_H|^2/w_H^2)$.

\begin{figure}[htbp]
\centering
\resizebox{\linewidth}{!}{%
\begin{tikzpicture}[
  x=1cm,y=1cm,
  font=\fontsize{10}{12}\selectfont,
  fnhead/.style={anchor=west,font=\fontsize{11}{13}\selectfont},
  fnline/.style={line width=0.75pt},
  fnarrow/.style={-{Latex[length=2mm,width=1.25mm]},line width=0.65pt},
  fnnote/.style={font=\fontsize{9}{11}\selectfont,align=center},
  fncross/.style={line width=0.85pt}
]
\path[use as bounding box] (0,0) rectangle (16.6,4.53);

\coordinate (portal) at (2.90,2.80);
\draw[fnline] (0.80,2.80) -- (portal);
\draw[fnline,dashed] (2.90,3.70) -- (portal);
\fill (portal) circle (1.4pt);
\node[above] at (1.03,2.99) {$L_i$};
\node[right] at (3.01,3.66) {$H$};

\coordinate (mixing) at (11.67,2.80);
\draw[fnline] (portal) -- (mixing);
\draw[fnline] (mixing) -- (15.70,2.80);
\node[above=8pt] at (7.20,2.80) {$(M_X^{-T})_{\alpha a}$};
\node[fnnote] at (7.20,2.35) {Heavy Dirac propagator};
\node[above] at (3.83,2.99) {$X_\alpha$};
\node[above] at (9.56,2.99) {$\eta^a$};
\node[above] at (15.02,2.99) {$(N_D^c)^b$};
\draw[fncross] (11.555,2.685) -- (11.785,2.915);
\draw[fncross] (11.555,2.915) -- (11.785,2.685);
\node[above=10pt] at (11.67,2.80) {$\mu_H(\varepsilon_2)_{ab}$};

\node[align=center] at (2.90,1.66)
  {$P_{i1}=A_i\epsilon^{\ell_i}$\\[5pt]
   $P_{i2}=B_i\epsilon^{\ell_i+1}$};
\end{tikzpicture}%
}
\caption{Dirac Yukawa coupling $Y_\nu\simeq
   \left(-\mu_H P M_X^{-T}\varepsilon_2\quad 0_{3\times1}\right)$ from heavy-fermion exchange.
The solid internal line joins the portal $L_iHX_\alpha$ to the
 instanton-induced mixing. It represents the chirality-changing
 propagator of the heavy Dirac fermions formed from $X$ and $\eta$,
 with $M_X$ included in the quadratic action. Its low-momentum
 flavor factor is $(M_X^{-T})_{\alpha a}$, not a separate vertex
 or a single insertion of $M_X$. The cross denotes one insertion
 of $\mu_H(\varepsilon_2)_{ab}$; no additional instanton factor
 multiplies it. Here $\alpha=1,2$ labels the singlet messengers
 and $a,b=1,2$ label hidden-doublet components. 
 The light right-handed neutrinos remain in the low-energy
effective theory. Since $N_S^c$ does not enter these interactions,
the corresponding column of the Yukawa matrix vanishes. 
The expression uses $Z_R^{-1/2}\simeq\mathbf1_3$ for
the normalization of the light right-handed neutrinos.
The correction to this factor is
$\mathcal O(|\mu_H|^2/w_H^2)$.}
\label{fig:FN-messenger-generation}
\end{figure}
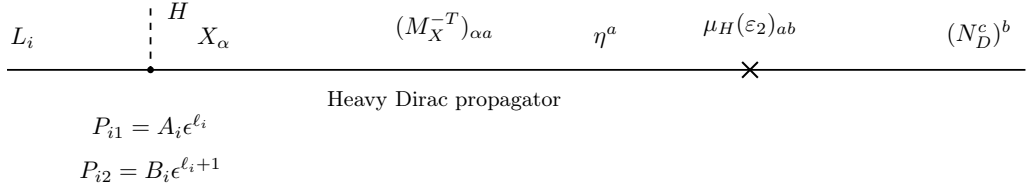

We next integrate out $\eta$ and $X_1,X_2$ to obtain the interaction
between $L_iH$ and the light right-handed neutrinos. At energies
well below their masses, the leading Yukawa coupling follows by
neglecting derivatives in the heavy-field equations of motion (see Fig.~\ref{fig:FN-messenger-generation}). 
Unlike the bilinear mass analysis above, this calculation includes
the couplings $L_iHX_\alpha$. From Eq.~\eqref{eq:two-doublet-mass-terms}, the equations for $X$ and $\eta$, respectively, are given by
\begin{align}
M_X\eta+P^T(LH)=0,\qquad
M_X^TX+\mu_H\varepsilon_2N_D^c=0.
\end{align}
Solving for the fields to be eliminated gives
\begin{align}
\eta=-M_X^{-1}P^T(LH),\qquad
X=-\mu_HM_X^{-T}\varepsilon_2N_D^c.
\label{eq:eta_EOM}
\end{align}
We substitute these expressions into 
Eq.~\eqref{eq:two-doublet-mass-terms}, including $X^TM_X\eta$,
and use Eq.~\eqref{eq:minimal-right-normalization}. The effective
interaction among the light fields is
\begin{align}
\mathcal L_{\rm eff}\supset
(LH)^TY_\nu n_{\rm light}^c+\mathrm{h.c.}
\label{eq:effective-neutrino-Lagrangian}
\end{align}
As shown in~\figref{fig:FN-messenger-generation}, its Yukawa matrix is given by
\begin{align}
Y_\nu=\left(-\mu_HP M_X^{-T}\varepsilon_2\quad0_{3\times1}\right)
Z_R^{-1/2}.
\label{eq:general-messenger-matching}
\end{align}

Every nonzero entry therefore contains the instanton-induced mixing
$\mu_H$. The overall sign can be absorbed into a field rephasing.
Terms with additional derivatives are not included in this Yukawa
interaction. After electroweak symmetry breaking, corrections to the
left-handed field normalization are of order $v^2/w_H^2$, with
$\langle H^0\rangle=v/\sqrt2$.

The factorization in Eq.~\eqref{eq:MX-from-one-doublet} makes the dependence on the vacuum orientation explicit. Since
$U_\Omega^*\varepsilon_2=\varepsilon_2U_\Omega$, we have
\begin{align}
M_X^{-T}\varepsilon_2=
\widehat M_X^{-T}\varepsilon_2U_\Omega.
\label{eq:inverse-orientation-factorization}
\end{align}
Let $\mathcal U_\Omega=\operatorname{diag}(U_\Omega,1)$, acting on
the three light right-handed fields, and let $\widehat Z_R$ be
Eq.~\eqref{eq:light-kinetic-matrix} with $M_X$ replaced by
$\widehat M_X$. Then
\begin{align}
Z_R=\mathcal U_\Omega^\dagger\widehat Z_R\mathcal U_\Omega.
\label{eq:kinetic-orientation-factorization}
\end{align}
Its positive inverse square root transforms in the same way. Thus
\begin{align}
Y_\nu=\widehat Y_\nu\mathcal U_\Omega,
\label{eq:Ynu-orientation-factorization}
\end{align}
where the Yukawa matrix in the rotated right-handed basis is
\begin{align}
\widehat Y_\nu=
\left(-\mu_HP\widehat M_X^{-T}\varepsilon_2\quad0_{3\times1}\right)
\widehat Z_R^{-1/2}.
\label{eq:orientation-independent-matching}
\end{align}
The field redefinition is
$\widehat n_{\rm light}^c=\mathcal U_\Omega n_{\rm light}^c$.
It is unitary and preserves the canonical kinetic terms. In particular,
\begin{align}
Y_\nu Y_\nu^\dagger=\widehat Y_\nu\widehat Y_\nu^\dagger.
\label{eq:orientation-independent-observables}
\end{align}
For the interactions in Eq.~\eqref{eq:two-doublet-mass-terms}, the
neutrino masses and left-handed mixing are therefore independent of $z_1,z_2$ at fixed $w_H,\mu_H$
and dimensionless couplings. This statement includes the light-field
normalization. It does not require the full scalar or gauge sector,
or the instanton prefactor, to be orientation independent.

At leading order in $\mu_H/w_H$, we may set
$\widehat Z_R=\mathbf1_3$ in
Eq.~\eqref{eq:orientation-independent-matching}. The inverse transpose
of Eq.~\eqref{eq:orientation-independent-mass} is
\begin{align}
\widehat M_X^{-T}=\frac{1}{w_H\Delta_X}
\begin{pmatrix}d&-c\epsilon\\-b\epsilon&a\end{pmatrix}.
\end{align}
Together with the portal matrix in
Eq.~\eqref{eq:visible-portal-matrix}, this gives
\begin{align}
\widehat Y_\nu
&=\frac{\mu_H}{w_H\Delta_X}D_L
\left(\epsilon(aB-cA),\ b\epsilon^2B-dA,\ 0\right)\nonumber\\
&\equiv\zeta_\nu D_L(\epsilon C_1,C_2,0),
\label{eq:factorized-Ynu}
\end{align}
where we define the common factor
\begin{align}
\zeta_\nu=\frac{\mu_H}{w_H}=\frac{v_H}{w_H}\zeta_H,
\label{eq:matched-yukawa-spurion}
\end{align}
and the two flavor vectors
\begin{align}
C_1=\frac{aB-cA}{\Delta_X},\qquad
C_2=\frac{b\epsilon^2B-dA}{\Delta_X}.
\label{eq:effective-flavor-vectors}
\end{align}
The vectors $C_1,C_2$ are fixed by the original portal couplings
$A,B$ and $\widehat M_X$; they are not additional independent
couplings and do not depend on $z_1,z_2$. 
For generic order-one inputs and $\Delta_X=\mathcal O(1)$,
the two vectors are linearly independent, with components
of order unity. 
The factor $\zeta_\nu$ contains the same instanton
suppression as $\mu_H$. 
In Eq.~\eqref{eq:factorized-Ynu}, the first nonzero
column contains one more power of $\epsilon$ than the second.
This difference follows from the FN factors in the messenger
interactions, not from a vacuum alignment condition.

At this order, the first two columns in Eq.~\eqref{eq:factorized-Ynu}
couple to the components of $U_\Omega N_D^c$, and the third to
$N_S^c$, up to the small field normalization described above.
In the original $N_D^c$ component basis, both columns can contain
order-one terms. Those leading terms are proportional to the same
vector $D_LC_2$; the second independent flavor direction is
suppressed by $\epsilon$. The zero column follows from the
specified portal structure, not from $\mathbb Z_8^D$. Corrections
involving $N_S^c$ are discussed in
Sec.~\ref{subsec:singlet-corrections}.
In the rotated basis, the notation
$(\widehat Y_\nu)_{ij}=\zeta_\nu C_{ij}\epsilon^{\ell_i+n_j}$
allows $n=(1,0,1)$ and $C=(C_1,C_2,0)$. 
These $n_j$ specify the column-dependent powers of $\epsilon$
in the low-energy Yukawa matrix, not different ultraviolet charges
for the two components of $N_D^c$.

The rank bound follows before choosing any numerical coefficients:
\begin{align}
\operatorname{rank}Y_\nu\leq\operatorname{rank}P\leq2.
\label{eq:general-rank-bound}
\end{align}
A single $X$ would supply one column in $P$ and could produce only
one nonzero mass through this tree-level path. It would also leave
one component of $\eta$ unpaired at $\mu_H=0$. Two $X$ fields are
therefore the smallest number for this mechanism. 
These results apply to the specified messenger interactions
at tree level. 
Additional FN-preserving higher-dimensional interactions can
modify the Yukawa matrix and its dependence on the vacuum
orientation. An example and the condition for its contribution
to be subleading are given in
Appendix~\ref{app:heavy-fermion-stability}.

\subsection{FN textures and the neutrino mass hierarchy}
\label{subsec:mass-hierarchy}

The same Standard Model charge assignment is used in both sectors.
From the Yukawa couplings in Eq.~\eqref{eq:charged-textures}
and the charge assignment in Eq.~\eqref{eq:benchmark-charges}, the
charged-fermion textures are
\begin{align}
\begin{aligned}
Y_u&\sim\begin{pmatrix}
\epsilon^7&\epsilon^5&\epsilon^3\\
\epsilon^6&\epsilon^4&\epsilon^2\\
\epsilon^4&\epsilon^2&1
\end{pmatrix},\qquad
Y_d\sim\begin{pmatrix}
\epsilon^7&\epsilon^6&\epsilon^6\\
\epsilon^6&\epsilon^5&\epsilon^5\\
\epsilon^4&\epsilon^3&\epsilon^3
\end{pmatrix},\qquad
Y_e&\sim\begin{pmatrix}
\epsilon^8&\epsilon^6&\epsilon^4\\
\epsilon^7&\epsilon^5&\epsilon^3\\
\epsilon^7&\epsilon^5&\epsilon^3
\end{pmatrix}.
\end{aligned}
\label{eq:benchmark-textures}
\end{align}
For generic order-one coefficients, biunitary diagonalization
gives the following hierarchy of physical Yukawa couplings:
\begin{align}
\begin{aligned}
y_u:y_c:y_t&\sim\epsilon^7:\epsilon^4:1,\\
y_d:y_s:y_b&\sim\epsilon^7:\epsilon^5:\epsilon^3,\\
y_e:y_\mu:y_\tau&\sim\epsilon^8:\epsilon^5:\epsilon^3.
\end{aligned}
\label{eq:charged-mass-hierarchies}
\end{align}
The left-handed quark charges $q=(3,2,0)$ give the CKM structure:
\begin{align}
|V_{\rm CKM}|\sim
\begin{pmatrix}
1&\epsilon&\epsilon^3\\
\epsilon&1&\epsilon^2\\
\epsilon^3&\epsilon^2&1
\end{pmatrix}.
\label{eq:CKM-FN-structure}
\end{align}
Here $\sim$ indicates the leading powers of $\epsilon$, with
coefficients of order unity left implicit and no accidental
cancellations assumed. The coefficients and CP phase are
not fixed by the charge assignment, but $\epsilon=0.2$ gives the leading charged-fermion hierarchies.

The powers of $\epsilon$ in the quark mixing angles also
determine the suppression of the Jarlskog invariant. 
In the standard parameterization of the unitary CKM matrix, the Jarlskog invariant is
\begin{align}
J_{\rm CKM}
=s^q_{12}c^q_{12}s^q_{23}c^q_{23}s^q_{13}(c^q_{13})^2
\sin\delta_{\rm CKM},
\label{eq:quark-Jarlskog-angles}
\end{align}
where $\theta^q_{ij}$ are the quark mixing angles,
$s^q_{ij}=\sin\theta^q_{ij}$,
$c^q_{ij}=\cos\theta^q_{ij}$, and $\delta_{\rm CKM}$
is the CP-violating phase. 
The estimates $s^q_{12}=\mathcal O(\epsilon)$,
$s^q_{23}=\mathcal O(\epsilon^2)$ and
$s^q_{13}=\mathcal O(\epsilon^3)$ therefore give
\begin{align}
J_{\rm CKM}=\mathcal O(\epsilon^6)\sin\delta_{\rm CKM}.
\label{eq:quark-CP-FN-order}
\end{align}
For $|\sin\delta_{\rm CKM}|=\mathcal O(1)$, this gives
$|J_{\rm CKM}|=\mathcal O(\epsilon^6)$.
At $\epsilon=0.2$, $\epsilon^6=6.4\times10^{-5}$, consistent
with the observed magnitude $|J_{\rm CKM}|\simeq3.1\times10^{-5}$
for order-one mixing coefficients~\cite{ParticleDataGroup:2026mpi}.
Thus the smallness of $J_{\rm CKM}$ can be explained by the
quark mixing hierarchy without requiring a small CP-violating
phase. The value of $\delta_{\rm CKM}$ and the sign of
$J_{\rm CKM}$ depend on the Yukawa coefficients.

For neutrinos, we use the rotated right-handed basis of
Sec.~\ref{subsec:messenger-matching}. 
In the following flavor analysis, we denote $\widehat Y_\nu$ and $\widehat n_{\rm light}^c$ by $Y_\nu$
and $n_{\rm light}^c$, respectively. This does not change any mass
or left-handed mixing. The neutrino texture is
\begin{align}
\frac{Y_\nu}{\zeta_\nu}
=D_L(\epsilon C_1,C_2,0)
\sim\begin{pmatrix}
\epsilon^2&\epsilon&0\\
\epsilon&1&0\\
\epsilon&1&0
\end{pmatrix}
\qquad\mathrm{with}\qquad D_L=\operatorname{diag}(\epsilon,1,1).
\label{eq:neutrino-FN-texture}
\end{align}
Here $C_1,C_2$ are fixed by the messenger masses and portal couplings
through Eq.~\eqref{eq:effective-flavor-vectors}. The first two columns
couple to the two combinations $U_\Omega N_D^c$, and the third to
$N_S^c$. The zero column gives one zero
mass at this order. The additional factor $\epsilon$ in the first
column can account for the hierarchy between the two nonzero masses.

To relate these statements to physical masses, we use the following
biunitary diagonalization for $f=e,\nu$:
\begin{align}
Y_f^{\rm diag}=W_f^\dagger Y_fV_f
=\operatorname{diag}(y_{f1},y_{f2},y_{f3}),
\label{eq:yukawa-diagonal-form}
\end{align}
where $W_f,V_f$ are unitary and the diagonal entries are real,
nonnegative and ordered from smallest to largest. In the two-component
convention of Eq.~\eqref{eq:FN-perturbative-lagrangian}, the
left-handed field rotation is $U_{fL}=W_f^*$, so that
$U_{fL}^TY_fV_f=Y_f^{\rm diag}$. After
$\langle H^0\rangle=v/\sqrt2$, the neutrino mass term is
$\nu_L^TM_\nu n_{\rm light}^c+\mathrm{h.c.}$, with
\begin{align}
M_\nu=\frac{v}{\sqrt2}Y_\nu,
\label{eq:Dirac-mass-from-Yukawa}
\end{align}
where $v=246\,\mathrm{GeV}$. The mass eigenvalues are
\begin{align}
m_i=\frac{v}{\sqrt2}y_i
\qquad\mathrm{with}\qquad y_i\equiv y_{\nu i}.
\label{eq:neutrino-mass-eigenvalues}
\end{align}

The two nonzero masses can be obtained without diagonalizing a
general $3\times3$ matrix. Let $u=D_LC_1$ and $v_\nu=D_LC_2$
denote the two flavor vectors including the left-handed FN factors.
The nonzero block of $Y_\nu^\dagger Y_\nu$ is
$|\zeta_\nu|^2G_\nu$, where
\begin{align}
G_\nu=\begin{pmatrix}
\epsilon^2u^\dagger u&\epsilon u^\dagger v_\nu\\
\epsilon v_\nu^\dagger u&v_\nu^\dagger v_\nu
\end{pmatrix}.
\label{eq:neutrino-Gram-matrix}
\end{align}
The diagonal Yukawa entries therefore satisfy
\begin{align}
y_1=0,\qquad
y_{2,3}^{\,2}=\frac{|\zeta_\nu|^2}{2}
\left[\operatorname{tr}G_\nu
\mp\sqrt{(\operatorname{tr}G_\nu)^2-4\det G_\nu}\right].
\label{eq:exact-two-neutrino-eigenvalues}
\end{align}
In particular,
$\det G_\nu=\epsilon^2[\|u\|^2\|v_\nu\|^2-|u^\dagger v_\nu|^2]$.
It is nonzero if the two flavor vectors are independent. The smaller
mass is controlled by the component of $u$ perpendicular to $v_\nu$,
\begin{align}
u_\perp=u-v_\nu\frac{v_\nu^\dagger u}{v_\nu^\dagger v_\nu}.
\label{eq:orthogonal-flavor-component}
\end{align}
The two nonzero diagonal Yukawa couplings are
\begin{align}
y_2&=|\zeta_\nu|\,\epsilon\|u_\perp\|[1+\mathcal O(\epsilon^2)],
\label{eq:two-mass-expansion}
\\
y_3&=|\zeta_\nu|\,\|v_\nu\|[1+\mathcal O(\epsilon^2)].
\label{eq:leading-largest-Yukawa}
\end{align}
For generic coefficients, the second- and third-generation
components of $u$ and $v_\nu$ are of order unity and are not
nearly proportional. Both $\|u_\perp\|$ and $\|v_\nu\|$ are
then of order unity. Consequently,
\begin{align}
Y_\nu^{\rm diag}\sim|\zeta_\nu|\operatorname{diag}(0,\epsilon,1).
\label{eq:neutrino-diagonal-scaling}
\end{align}
Thus $m_1=0$ at this order. The two nonzero masses obey
\begin{align}
\frac{m_2}{m_3}\simeq\epsilon\frac{\|u_\perp\|}{\|v_\nu\|}
\sim\epsilon,
\label{eq:neutrino-mass-scaling}
\end{align}
whereas their overall scale is
\begin{align}
m_3\sim\frac{v}{\sqrt2}\frac{v_H}{w_H}|\zeta_H|.
\label{eq:largest-neutrino-mass-scaling}
\end{align}
If the two Yukawa vectors are nearly proportional, $m_2$
receives a further suppression. This concerns their lepton-flavor
components, not the scalar vacuum orientation $z_1,z_2$.
No condition on that vacuum orientation was used. The mass ratio is independent of the common instanton
factor at this order. This is the separation between the origin of
the hierarchy and the origin of the overall small mass.

The charged-current mixing depends on the left-handed
rotations,
\begin{align}
U_{\rm PMNS}=U_{eL}^\dagger U_{\nu L}.
\label{eq:PMNS-minimal}
\end{align}
It must therefore be evaluated with $Y_e$ and $Y_\nu$ in the same
lepton-doublet basis.

For the charge assignment $\ell=(1,0,0)$, mixing of the first
generation with the other two is suppressed by $\epsilon$ in
both the charged-lepton and neutrino left-handed rotations,
whereas the $2$--$3$ rotations are generally of order unity.
For generic nondegenerate flavor coefficients and no accidental
cancellation between these rotations, we obtain
\begin{align}
|U_{\rm PMNS}|\sim
\begin{pmatrix}
1&\epsilon&\epsilon\\
\epsilon&1&1\\
\epsilon&1&1
\end{pmatrix}.
\label{eq:PMNS-FN-structure}
\end{align}
Here $\sim$ indicates the leading powers of $\epsilon$,
with coefficients of order unity left implicit. 
Thus $\sin\theta_{12},\sin\theta_{13}=\mathcal O(\epsilon)$,
while $\sin\theta_{23}=\mathcal O(1)$.
The leading powers of $\epsilon$ in these mixing angles agree
with those of the BL2 pattern studied in Abelian flavor models~\cite{Ding:2012wh}. 
The two suppressed angles have independent order-one coefficients
and need not be numerically equal. 
In particular, the observed value of $\sin\theta_{12}$
can be reproduced by a coefficient of a few multiplying
$\epsilon$. 
The common instanton factor cancels from the mixing matrix at
this order. Neither the numerical mixing angles nor the CP phase
are fixed by the FN charges.

For a unitary PMNS matrix in the standard angle convention,
the lepton Jarlskog invariant is
\begin{align}
J_{\rm CP}
=s_{12}c_{12}s_{23}c_{23}s_{13}c_{13}^2
\sin\delta_{\rm CP},
\label{eq:lepton-Jarlskog-angles}
\end{align}
where $s_{ij}=\sin\theta_{ij}$ and $c_{ij}=\cos\theta_{ij}$
refer to the lepton mixing angles.
Using the mixing-angle estimates from
Eq.~\eqref{eq:PMNS-FN-structure}, we obtain
\begin{align}
J_{\rm CP}=\mathcal O(\epsilon^2)\sin\delta_{\rm CP}.
\label{eq:lepton-CP-FN-order}
\end{align}
As in the quark sector, this estimate assumes generic mixing
coefficients and does not fix the sign of the invariant or require
a nonzero CP phase. 
These estimates use Yukawa matrices defined at the same
renormalization scale. We do not include their
renormalization-group evolution or threshold corrections
and leave these effects for future work.

\subsection{Discrete gauge symmetry and Dirac neutrinos}
\label{subsec:diracness}

Integrating out $\eta$ and $X_1,X_2$ gives the Dirac-neutrino
Yukawa coupling in Eq.~\eqref{eq:general-messenger-matching}.
Its smallness follows from the instanton-induced mixing $\mu_H$.
To forbid independent Majorana mass terms, we use the
$\mathbb Z_8^D$ gauge symmetry specified in
Table~\ref{tab:matter-content}. 
It is independent of the non-invertible FN defect. In particular,
\begin{align}
q_D(LHn_{\rm light}^c)&=0,&
q_D(n_{\rm light}^cn_{\rm light}^c)&=4,&
q_D((LH)(LH))&=4\pmod8.
\end{align}
All scalar expectation values are neutral under this symmetry.
Thus Dirac masses are allowed, while Majorana masses and the
Weinberg operator are forbidden. More generally, on Standard Model
fields $q_D=6(B-L)\pmod8$, so baryon-number-preserving local
$\Delta L=2$ operators are excluded. Hidden gauge instantons also
preserve $\mathbb Z_8^D$, since
$q_D(\eta)+q_D(N_D^c)=0\pmod8$.

We now check the anomaly-cancellation conditions for the
$\mathbb Z_8^D$ gauge symmetry. We take the spin structure and the $\mathbb Z_8^D$
gauge field to be independent, corresponding to
$\mathrm{Spin}\times\mathbb Z_8^D$.
The order-two element of $\mathbb Z_8^D$ is not identified
with fermion parity. 
For left-handed Weyl fields, let
\begin{align}
\mathcal A_1=\sum_f d_fq_{D,f},\qquad
\mathcal A_3=\sum_f d_fq_{D,f}^{\,3}.
\end{align}
The pure discrete and mixed gravitational conditions for
$\mathrm{Spin}\times\mathbb Z_8$ are given by~\cite{Hsieh:2018ifc}
\begin{align}
\mathcal A_1=0\pmod4,\qquad \mathcal A_3=0\pmod8.
\label{eq:Z8-anomaly-conditions}
\end{align}
Taking $6=-2\pmod8$, the separate contributions are
\begin{align}
(\mathcal A_1,\mathcal A_3)_{\rm SM}&=(6,24),\nonumber\\
(\mathcal A_1,\mathcal A_3)_{\eta+N_D^c}&=(0,0),\nonumber\\
(\mathcal A_1,\mathcal A_3)_{X_1+X_2}&=(-4,-16),\nonumber\\
(\mathcal A_1,\mathcal A_3)_{N_S^c}&=(-2,-8).
\label{eq:Z8-anomaly-sums}
\end{align}
The total vanishes. Here the SM contribution does not include any
right-handed neutrino. Removing $N_S^c$ would leave
$\mathcal A_1=2\pmod4$. 
We therefore include $N_S^c$ to cancel the discrete
mixed gravitational anomaly, even though it does not enter
the leading interactions that generate the neutrino masses.

With $\mathcal A_{DG^2}=\sum_f q_{D,f}2T_G(r_f)d_f^{(G)}$, the
mixed non-Abelian coefficients are
\begin{align}
(\mathcal A_{D33},\mathcal A_{D22},\mathcal A_{DHH})=(0,24,0).
\end{align}
They are multiples of eight. For integer hypercharge
$\widehat Y=6Y$, we also obtain
\begin{align}
\sum_f d_fq_{D,f}\widehat Y_f^{\,2}=-432,\qquad
\sum_f d_fq_{D,f}^{\,2}\widehat Y_f=0.
\end{align}
Appendix~\ref{app:continuous-Dirac-lift} gives a continuous
realization of these discrete charges using an anomaly-free
$U(1)_X$ gauge symmetry. This additional gauge symmetry is
distinct from the hidden $U(1)_H$. We use that construction
to specify the finite gauge theory and the wall terms inherited
after $U(1)_X$ breaking. 
The $U(1)_X$ gauge boson is not included as a dynamical
field in the effective theory studied here.

All listed matter charges are even. The faithful action on these
fields is consequently $\mathbb Z_8^D/\mathbb Z_2\simeq\mathbb Z_4$.
Nevertheless, we define the gauge theory with group
$\mathbb Z_8^D$, including its line operators, rather than
replacing it by a $\mathbb Z_4$ gauge theory. 
The Majorana operators remain charged under the faithful action.
Neither this discrete symmetry nor anomaly cancellation makes the
Dirac Yukawa small. Its smallness follows from the instanton-dominated
FN breaking assumed in Sec.~\ref{subsec:instanton-stability}.

\section{Phenomenological implications}
\label{sec:phenomenology}

The preceding section gives the FN hierarchy and separates it from
the common factor $\mu_H/w_H$. In
Sec.~\ref{subsec:neutrino-normalization}, we use the atmospheric
mass splitting to constrain this factor, keeping its dependence on
the flavor coefficients explicit, and relate it to illustrative
hidden-sector scales. Sec.~\ref{subsec:singlet-corrections}
discusses how an additional interaction involving $N_S^c$ can
lift the zero mass. Finally, Sec.~\ref{subsec:predictions}
compares the mass spectrum with experimental constraints and
gives the relation to the dimension-six interaction induced by
the same heavy fermions.

An explicit lepton example with real, order-one charged-lepton and
messenger coefficients is given in
Appendix~\ref{app:lepton-benchmark}. It reproduces the
charged-lepton masses, neutrino mass ratio and three mixing angles
in a common flavor basis. This example conserves CP in lepton
mixing. The general model allows complex coefficients; the mixing
angles and CP phase are not predictions of the FN charges.

\subsection{Neutrino mass scale and hidden-sector scales}
\label{subsec:neutrino-normalization}

We now relate the normal-ordering, rank-two spectrum of
Sec.~\ref{subsec:mass-hierarchy} to the required common suppression
of the neutrino Yukawa couplings. 
We use the NuFIT 6.0 normal-ordering best-fit mass-squared
differences, including the Super-Kamiokande atmospheric-neutrino
data, listed in Table~\ref{tab:NuFIT} in
Appendix~\ref{app:lepton-benchmark}~\cite{Esteban:2024eli}. 
Setting $m_1=0$, we obtain
\begin{align}
m_2&=\sqrt{\Delta m_{21}^2}\simeq8.65\times10^{-3}\,\mathrm{eV},\nonumber\\
m_3&=\sqrt{\Delta m_{31}^2}\simeq5.01\times10^{-2}\,\mathrm{eV}.
\label{eq:neutrino-masses-numerical}
\end{align}
Their ratio is
\begin{align}
\frac{m_2}{m_3}=\sqrt{\frac{\Delta m_{21}^2}{\Delta m_{31}^2}}
\simeq0.173.
\label{eq:mass-ratio-numerical}
\end{align}
Through Eq.~\eqref{eq:neutrino-mass-eigenvalues}, the corresponding
Yukawa values are
\begin{align}
y_2=\frac{\sqrt2m_2}{v}\simeq4.98\times10^{-14},\qquad
y_3=\frac{\sqrt2m_3}{v}\simeq2.88\times10^{-13}.
\label{eq:required-yukawas}
\end{align}
At leading order in $|\mu_H|/w_H$, we write
$y_i=|\zeta_\nu|s_i$ for $i=2,3$, where $y_i$ are the nonzero
diagonal Yukawa couplings defined in
Eq.~\eqref{eq:neutrino-mass-eigenvalues}. The dimensionless
$s_2,s_3$ depend on the FN factors and the input coefficients,
not on the common factor $|\zeta_\nu|$. For generic order-one
coefficients, $s_3=\mathcal O(1)$ and $s_2/s_3\sim\epsilon$.
The atmospheric mass splitting fixes the combination
\begin{align}
\frac{|\mu_H|}{w_H}s_3=|\zeta_\nu|s_3
=\frac{\sqrt{2\Delta m_{31}^2}}{v}
\simeq2.88\times10^{-13}.
\label{eq:neutrino-normalization-condition}
\end{align}
Thus $|\zeta_\nu|\simeq2.88\times10^{-13}/s_3$;
$|\mu_H|/w_H$ is determined only after the flavor coefficients
are specified. The numerical example in
Appendix~\ref{app:lepton-benchmark} gives $s_3\simeq1.00$.
Note that the Yukawa couplings obtained after fixing $|\zeta_\nu|$ and
the resulting solar mass splitting are given there as a check of
compatibility, not as additional predictions of the FN charges.

We can now relate the overall neutrino mass scale to the hidden instanton. 
Using $m_3=(v/\sqrt2)|\zeta_\nu|s_3$ and
Eqs.~\eqref{eq:monopole-spurion} and
\eqref{eq:matched-yukawa-spurion} gives
\begin{align}
m_3=\frac{v}{\sqrt2}\frac{v_H}{w_H}
|\mathcal K_H|s_3\exp\!\left[-\frac{8\pi^2}{g_H^2(v_H)}\right].
\label{eq:quantitative-matching}
\end{align}
Here $s_3$ is the flavor factor defined above.
The factor $v_H/w_H$ follows from
$\zeta_\nu=(v_H/w_H)\zeta_H$, while $\mathcal K_H$ contains
the instanton prefactor. 
For specified values of $v_H/w_H$, $s_3$ and the prefactor, the required
gauge coupling satisfies
\begin{align}
g_H(v_H)=\left[
\frac{8\pi^2}{\displaystyle
\ln\!\left(\frac{v}{\sqrt2m_3}\frac{v_H}{w_H}|\mathcal K_H|s_3\right)}
\right]^{1/2}.
\label{eq:hidden-coupling-inversion}
\end{align}
This is not an independent prediction of $g_H$: the prefactor has
not been calculated and, in general, also depends on the gauge
coupling and the heavy spectrum.

For the scalar benchmark in Appendix~\ref{app:scalar-alignment}, take
\begin{align}
v_H=10^{10}\,\mathrm{GeV},\qquad
\frac{w_H^2}{v_H^2}=5.0\times10^{-3}.
\label{eq:hidden-benchmark-inputs}
\end{align}
We set $|\mathcal K_H|=1$ for this illustrative estimate and use
$s_3\simeq1.00$ from the lepton example in
Appendix~\ref{app:lepton-benchmark}.
This gives $w_H\simeq7.07\times10^8\,\mathrm{GeV}$. The two scales,
prefactor and flavor coefficients are illustrative inputs. With
$|\zeta_\nu|$ fixed by Eq.~\eqref{eq:neutrino-normalization-condition},
they give $g_H(v_H)\simeq1.58$ and
\begin{align}
|\zeta_H|=\frac{w_H}{v_H}|\zeta_\nu|
\simeq2.04\times10^{-14},
\qquad
|\mu_H|=w_H|\zeta_\nu|\simeq0.204\,\mathrm{MeV}.
\label{eq:two-doublet-scale-values}
\end{align}
The values in Table~\ref{tab:hidden-benchmark} distinguish these
assumptions from the masses and mixing inferred from them. The
monopole mass is the BPS estimate, not a full calculation in the
scalar potential of Appendix~\ref{app:scalar-alignment}.

\begin{table}[htbp]
\centering
\caption{An illustrative hidden-sector benchmark using the atmospheric
splitting and the lepton coefficients of
Appendix~\ref{app:lepton-benchmark}. The first three entries are
scale and prefactor inputs; the remaining entries follow from
the specified mass formulas and flavor coefficients. 
The neutrino data therefore constrain a combination of
parameters, rather than fixing the individual hidden-sector
scales.}
\label{tab:hidden-benchmark}
\small
\renewcommand{\arraystretch}{1.15}
\begin{tabular}{llp{7.0cm}}
\toprule
Quantity & Value & Role in the calculation\\
\midrule
$v_H$ & $10^{10}\,\mathrm{GeV}$ & Input adjoint-Higgs scale\\
$w_H$ & $7.07\times10^8\,\mathrm{GeV}$ & Input doublet-Higgs scale\\
$|\mathcal K_H|$ & $1$ & Assumed instanton prefactor\\
\midrule
$g_H(v_H)$ & $1.58$ & Inferred from Eq.~\eqref{eq:hidden-coupling-inversion}\\
$|\mu_H|$ & $0.204\,\mathrm{MeV}$ & Mixing required by Eq.~\eqref{eq:two-doublet-scale-values}\\
$M_{W_H}$ & $1.58\times10^{10}\,\mathrm{GeV}$ & $g_Hv_H$, Eq.~\eqref{eq:hidden-W-mass}\\
$M_{\rm mon}$ & $7.94\times10^{10}\,\mathrm{GeV}$ & BPS estimate $4\pi v_H/g_H$\\
$M_{A_H}$ & $5.60\times10^8\,\mathrm{GeV}$ & $g_Hw_H/2$, Eq.~\eqref{eq:minimal-photon-mass}\\
$m_{X\eta,1}=m_{X\eta,2}$ & $1.02w_H$ & Heavy Dirac masses at $\mu_H=0$\\
\bottomrule
\end{tabular}
\end{table}

This benchmark has $w_H^2/v_H^2=5.0\times10^{-3}$, so
$\Omega$ breaks the remaining $U(1)_H$ at a scale well below
the adjoint expectation value $v_H$.
It also gives $|\mu_H|/w_H\simeq2.88\times10^{-13}$.
The $X_1,X_2$ components of the light right-handed neutrinos
in Eq.~\eqref{eq:light-right-handed-embedding} are therefore
very small. Their contribution to the kinetic normalization
is of order $|\mu_H|^2/w_H^2$ and can be neglected when
deriving the Yukawa coupling to first order in $\mu_H$.
We evaluate the instanton exponential using the value of $g_H(v_H)$. The one-loop renormalization-group
equation for the hidden $SU(2)_H$ gauge coupling $g_H$
is given in Appendix~\ref{app:technical-checks}.

\subsection{Corrections to the massless-neutrino limit}
\label{subsec:singlet-corrections}

The tree-level Yukawa matrix derived by integrating out $\eta$ and
$X_1,X_2$ in Sec.~\ref{subsec:messenger-matching} has rank at most
two. The singlet $N_S^c$, introduced for discrete anomaly
cancellation, does not enter the portal interactions in
Eq.~\eqref{eq:portal}, so its column in
Eq.~\eqref{eq:factorized-Ynu} vanishes. For the rank-two solutions
considered above, one neutrino is therefore massless. This zero
mass is not required by $\mathbb Z_8^D$: the discrete symmetry
forbids Majorana masses and the Weinberg operator, but allows
nonzero Dirac masses for all three neutrinos. We now examine
additional interactions involving $N_S^c$.

First, FN-preserving higher-dimensional kinetic terms can mix
$N_S^c$ with $\widetilde\Omega^\dagger N_D^c$. We denote their
suppression scale by $\Lambda_{\rm kin}$. This notation distinguishes
it from $\Lambda$, the suppression scale of the additional portal
in Appendix~\ref{app:heavy-fermion-stability}. After $\Omega$
acquires its expectation value, the kinetic mixing is of order
$w_H/\Lambda_{\rm kin}$ for order-one coefficients. Canonical
normalization then requires a nonsingular field redefinition.
It can give a nonzero third column in the Yukawa matrix, but
cannot change its rank. Thus these kinetic terms do not
generate a third nonzero neutrino mass, although their effects
must be included when $w_H/\Lambda_{\rm kin}$ is not negligible.

Second, an additional interaction with an independent lepton
flavor vector can generate a third nonzero mass. We denote the
suppression scale of the following interaction by $\Lambda_7$
and its dimensionless coefficients by $d_i$:
\begin{align}
\Delta\mathcal L_7=
\frac{d_i}{\Lambda_7^3}
\left(\frac{\varphi}{\Lambda_{\rm FN}}\right)^{\ell_i+1}
(L_iHN_S^c)(\eta N_D^c)^\dagger+\mathrm{h.c.}
\label{eq:dimension-seven-rank-lifting}
\end{align}
The product $(L_iHN_S^c)(\eta N_D^c)^\dagger$ has mass dimension
seven, which accounts for the factor $\Lambda_7^{-3}$. The flavon
insertions are instead suppressed by the FN scale
$\Lambda_{\rm FN}$ and give $\epsilon^{\ell_i+1}$ after replacing
$\varphi$ by its expectation value. No equality is assumed among
$\Lambda_7$, $\Lambda_{\rm kin}$, $\Lambda$ and $\Lambda_{\rm FN}$;
they parameterize the coefficients of different interactions.

The full operator in Eq.~\eqref{eq:dimension-seven-rank-lifting}
has zero FN charge and preserves $\mathbb Z_8^D$.
As discussed in Sec.~\ref{subsec:monopole-instanton-breaking},
a unit $SU(2)_H$ instanton has one fermion zero mode for each
of $\eta$ and $N_D^c$. The corresponding two-fermion vertex
is given in Eq.~\eqref{eq:general-hidden-mixing}.
The conjugate bilinear $(\eta N_D^c)^\dagger$ in
Eq.~\eqref{eq:dimension-seven-rank-lifting} can contract with
both fermion legs of this vertex, leaving $L_iHN_S^c$ as
the external fields. In the path integral, this supplies the
two fermion factors required to saturate the Grassmann
integrals over the zero modes. No additional hidden mass
insertion is required. The opposite instanton orientation
gives the Hermitian-conjugate interaction. 
Only one instanton is involved, so the induced Yukawa coupling
contains the same exponential suppression as $\mu_H$.
The zero-mode overlaps and heavy-field contributions need not
be the same as those in the two-fermion vertex; their difference
is included in $\mathcal R_7$ below.
For the dominant instanton size $\rho\sim v_H^{-1}$,
dimensional analysis gives
\begin{align}
(\delta Y_\nu)_{iS}\sim
\zeta_H\mathcal R_7d_i\epsilon^{\ell_i+1}
\left(\frac{v_H}{\Lambda_7}\right)^3.
\label{eq:rank-lifting-estimate}
\end{align}
The dimensionless factor $\mathcal R_7$ compares the zero-mode
overlap and heavy-field contributions for this operator with
those included in $\mathcal K_H$, the prefactor of $\zeta_H$.
It has not been evaluated, so
Eq.~\eqref{eq:rank-lifting-estimate} is a parametric estimate,
not a numerical prediction.

Let us write $(\delta y_S)_i=(\delta Y_\nu)_{iS}$ for the additional
Yukawa couplings to $N_S^c$. If this column is not a linear
combination of the two existing columns, the corrected mass
matrix has rank three. Here we keep the first two columns
unchanged and neglect the additional kinetic mixing discussed
above. We take $\|\delta y_S\|\ll y_2$, where $y_2$ is the
smaller nonzero diagonal Yukawa coupling before this correction,
so that the new mass can be calculated perturbatively.

Let $w_{\nu1}$ be the normalized eigenvector of the uncorrected
$Y_\nu Y_\nu^\dagger$ with zero eigenvalue. It specifies the flavor
composition of the left-handed neutrino that is massless before
this correction. It is orthogonal to the two existing Yukawa
columns, which span the same space as $D_LA,D_LB$; hence
$P^\dagger w_{\nu1}=0$. The inner product
$w_{\nu1}^\dagger\delta y_S$ is the induced Yukawa coupling of
this state to $N_S^c$.

Since the original third column vanishes, adding $\delta y_S$
changes $Y_\nu Y_\nu^\dagger$ by $\delta y_S\delta y_S^\dagger$,
without cross terms with the first two columns. First-order
perturbation theory for this Hermitian matrix gives the leading
shift of its zero eigenvalue,
\begin{align*}
\delta(y_1^2)
=w_{\nu1}^\dagger(\delta y_S\delta y_S^\dagger)w_{\nu1}
=|w_{\nu1}^\dagger\delta y_S|^2.
\end{align*}
Here first order refers to the perturbation of $Y_\nu Y_\nu^\dagger$,
which is quadratic in $\delta y_S$. Taking the square root and
multiplying by $v/\sqrt2$ gives the lightest Dirac mass:
\begin{align}
m_{\rm lightest}
=\frac{v}{\sqrt2}|w_{\nu1}^\dagger\delta y_S|
\left[1+\mathcal O\!\left(\frac{\|\delta y_S\|^2}{y_2^2}\right)\right].
\label{eq:lightest-mass-projection}
\end{align}
The inequality $|w_{\nu1}^\dagger\delta y_S|\leq\|\delta y_S\|$
gives an upper estimate without using the direction of the
additional column. Combining Eq.~\eqref{eq:rank-lifting-estimate}
with $y_3=|\zeta_\nu|s_3$ and
$\zeta_\nu=(v_H/w_H)\zeta_H$, we obtain, for order-one flavor
coefficients and $s_3=\mathcal O(1)$,
\begin{align}
\frac{m_{\rm lightest}}{m_3}\lesssim
\mathcal O\!\left[|\mathcal R_7|\epsilon\frac{w_H}{v_H}
\left(\frac{v_H}{\Lambda_7}\right)^3\right].
\label{eq:lightest-mass-estimate}
\end{align}
For the charges used here, the coupling to the previously massless
neutrino contains one additional power of $\epsilon$.
With $D_L=\operatorname{diag}(\epsilon,1,1)$, the two conditions
$P^\dagger w_{\nu1}=0$ determine the second and third components
of $w_{\nu1}$ in terms of the first. After dividing the second
condition by its common factor $\epsilon$, their coefficient
matrix has determinant $(A_2B_3-A_3B_2)^*$. If this determinant and the individual
coefficients are of order unity, the second and third components
are suppressed by $\epsilon$ relative to the first. Normalization
then gives $|(w_{\nu1})_1|=\mathcal O(1)$ and
$|(w_{\nu1})_{2,3}|=\mathcal O(\epsilon)$.
For the same charges, Eq.~\eqref{eq:rank-lifting-estimate} gives
\begin{align*}
\delta y_S\sim\zeta_H\mathcal R_7
\left(\frac{v_H}{\Lambda_7}\right)^3
\begin{pmatrix}
\epsilon^2d_1\\
\epsilon d_2\\
\epsilon d_3
\end{pmatrix}.
\end{align*}
Each term in $w_{\nu1}^\dagger\delta y_S$ therefore contains two
powers of $\epsilon$ times the common factor
$\zeta_H\mathcal R_7(v_H/\Lambda_7)^3$.
In the absence of a further cancellation in this coupling,
we obtain
\begin{align}
\frac{m_{\rm lightest}}{m_3}
=\mathcal O\!\left[
|\mathcal R_7|\epsilon^2\frac{w_H}{v_H}
\left(\frac{v_H}{\Lambda_7}\right)^3\right].
\label{eq:lightest-mass-projected-scaling}
\end{align}
This estimate assumes generic flavor coefficients and
$\|\delta y_S\|\ll y_2$. A further cancellation can reduce the
mass; if $\delta y_S$ lies entirely in the span of the existing
columns, the neutrino remains massless.
Eq.~\eqref{eq:lightest-mass-estimate} remains a conservative
bound because it does not use the additional suppression in
the coupling to the previously massless state. 
The common factor $\zeta_H$ cancels in this ratio. A small nonzero
lightest mass generated in this way therefore remains consistent
with the instanton origin of the neutrino Yukawa couplings.
Majorana masses remain forbidden by $\mathbb Z_8^D$. This estimate applies only to the operator in
Eq.~\eqref{eq:dimension-seven-rank-lifting} and does not classify all
higher-dimensional corrections.

Independent local FN-breaking masses or Yukawa couplings must
be treated separately. Their coefficients need not contain
$\zeta_H$, since they need not be generated by the hidden
instanton. As assumed in
Sec.~\ref{subsec:instanton-stability}, we require their
contributions to be subleading. In particular, an independent
contribution $\delta\mu_{\rm UV}$ to the coefficient of
$(\eta N_D^c)$ must satisfy
$|\delta\mu_{\rm UV}|\ll|\mu_H|$, as in
Eq.~\eqref{eq:instanton-dominance}. Independent FN-breaking
Yukawa couplings are likewise assumed to give subleading
contributions to the neutrino mass matrix. These are assumptions
about the ultraviolet theory, not consequences of the minimal
field content or of $\mathbb Z_8^D$.
Appendix~\ref{app:UV-protection} gives a possible nonminimal
ultraviolet extension whose accidental chiral symmetry forbids
the independent mass at renormalizable order. The remaining
higher-dimensional FN-breaking terms must still be sufficiently
suppressed for the instanton contribution to dominate.

\subsection{Phenomenological consequences}
\label{subsec:predictions}

We now discuss the consequences of the neutrino mass matrix and the
additional constraints on the lepton and scalar sectors. 
For the interactions of $\eta$ and the two singlet messengers
$X_1,X_2$ specified in Sec.~\ref{subsec:messenger-matching}, the tree-level neutrino Yukawa matrix has rank at most two. 
For the chosen FN charges and generic order-one coefficients,
the two nonzero neutrino masses have a ratio of order $\epsilon$,
favoring normal ordering. 
The numerical values of the PMNS angles and CP phase depend on the chosen coefficients and are not fixed by the FN charge assignment.

In the normal-ordering, rank-two limit, $m_1=0$. Using the NuFIT 6.0 best-fit oscillation parameters listed
in Table~\ref{tab:NuFIT}~\cite{Esteban:2024eli}, we obtain
\begin{align}
\sum_i m_i=\sqrt{\Delta m_{21}^2}+\sqrt{\Delta m_{31}^2}
\simeq5.88\times10^{-2}\,\mathrm{eV}.
\label{eq:neutrino-mass-sum}
\end{align}
The effective mass in beta decay is
\begin{align}
m_\beta=\left(\sum_i|U_{ei}|^2m_i^2\right)^{1/2}
\simeq8.84\times10^{-3}\,\mathrm{eV}.
\label{eq:absolute-mass-predictions}
\end{align}
Here $U=U_{\rm PMNS}$ denotes the unitary lepton mixing matrix defined in Eq.~\eqref{eq:PMNS-minimal}. 
For comparison, the lepton benchmark in
Appendix~\ref{app:lepton-benchmark} gives
$m_\beta\simeq8.90\times10^{-3}\,\mathrm{eV}$. 
These values of $\sum_i m_i$ and $m_\beta$ follow from the assumed rank-two normal spectrum and the oscillation inputs, rather than from an independent determination of the overall
neutrino-mass scale. Other models with the same spectrum and oscillation inputs give the same values. 
Both values of $m_\beta$ lie well below the direct KATRIN limit
\begin{align}
m_\beta<0.45\,\mathrm{eV}\qquad(90\%\ \mathrm{C.L.})
\label{eq:KATRIN-limit}
\end{align}
from 259 days of tritium data~\cite{KATRIN:2024cdt}.

The mass sum can also be constrained by cosmological observations, but these bounds depend on the data and cosmological assumptions.
For example, DESI DR2 BAO combined with the baseline CMB data gives
$\sum_i m_i<0.064\,\mathrm{eV}$ in flat $\Lambda$CDM and
$\sum_i m_i<0.16\,\mathrm{eV}$ when the dark-energy equation of state
is allowed to vary as $w(a)=w_0+(1-a)w_a$, where $a$ is the cosmological
scale factor, both at 95\%~\cite{DESI:2025zgx}.
The 2026 combination including DESI Lyman-$\alpha$ full-shape information
and updated CMB data gives the corresponding limits
$0.0592\,\mathrm{eV}$ and $0.166\,\mathrm{eV}$~\cite{DESI:2026lnd}. 
The flat $\Lambda$CDM limit from this latter combination is already close to the value in
Eq.~\eqref{eq:neutrino-mass-sum}. 
These cosmological bounds have not been evaluated for the
full particle content of the model. 
In particular, the abundances of the additional light states must be checked before using these bounds.

For the messenger interactions considered in
Sec.~\ref{subsec:messenger-matching}, $N_S^c$ remains massless
and has no Yukawa coupling to the active neutrinos.
Since it is also a Standard Model gauge singlet, it does not
induce active--sterile oscillations in this approximation.
Its contribution to the cosmological radiation density depends
on its abundance. The higher-dimensional interactions discussed
in Sec.~\ref{subsec:singlet-corrections}, or gauge interactions
in an ultraviolet completion such as that in
Appendix~\ref{app:continuous-Dirac-lift}, can produce $N_S^c$
at high temperatures even when its low-energy Yukawa coupling
is negligible. The abundances of $N_S^c$, the light states
from $N_D^c$, and the flavon phase must therefore be determined
for a specified thermal history. If they contribute appreciably
to the radiation density, the cosmological mass limits must
be reassessed with these contributions included.
This thermal analysis is left for future work.

Independently of these cosmological considerations, the exact $\mathbb Z_8^D$ forbids baryon-number-preserving $\Delta L=2$ processes, including neutrinoless double-beta decay. 
This remains true when an additional Dirac interaction gives
a small mass to the previously massless neutrino, as discussed
in Sec.~\ref{subsec:singlet-corrections}, provided that
$\mathbb Z_8^D$ remains unbroken. The absence of Majorana masses is therefore an exact
selection rule within the stated model, whereas $m_1=0$ is a result of
the specified leading interactions.

The absence of these $\Delta L=2$ processes does not exclude charged-lepton flavor violation that conserves total lepton number.
In particular, the heavy fermions that generate $Y_\nu$ also induce a dimension-six interaction.
Substituting the solution $\eta=-M_X^{-1}P^T(LH)$ obtained in Sec.~\ref{subsec:messenger-matching} into the kinetic term of $\eta$ gives
\begin{align}
\Delta\mathcal L^{(6)}=
i(LH)^\dagger C^{(6)}\bar\sigma^\mu\partial_\mu(LH).
\label{eq:LFV-dimension-six-operator}
\end{align}
The coefficient, of mass dimension minus two, is
\begin{align}
C^{(6)}=P^*(M_X^{-1})^\dagger M_X^{-1}P^T.
\label{eq:LFV-dimension-six-coefficient}
\end{align}
It remains nonzero at $\mu_H=0$ and modifies the weak interactions
of the light neutrinos~\cite{Abada:2007ux}. 
We compare $C^{(6)}$ and $Y_\nu$ at the scale where the heavy fermions are integrated out, at tree level and to lowest order in external momenta, $|\mu_H|/w_H$ and $v/w_H$. 
Eq.~\eqref{eq:general-messenger-matching}
and $\varepsilon_2\varepsilon_2^\dagger=\mathbf1_2$ then imply
\begin{align}
Y_\nu Y_\nu^\dagger=|\mu_H|^2\bigl(C^{(6)}\bigr)^*.
\label{eq:neutrino-dimension-six-correlation}
\end{align}
The complex conjugation follows from our left-handed Weyl convention.
The relation is due to the mixing $\mu_H\varepsilon_2$ and requires
neither degenerate messenger masses nor a particular vacuum
orientation. Similar relations occur in Dirac-neutrino models with
flavor-universal mixing~\cite{CentellesChulia:2025eck}.

After electroweak symmetry breaking, the dimension-six interaction changes the active-neutrino kinetic matrix by $v^2C^{(6)}/2$. Canonical normalization then changes the
charged-current couplings. In the charged-lepton mass basis, we write the resulting mixing matrix as
\begin{align}
N_{\rm CC}=(\mathbf1_3-\eta_{\rm NU})U_{\rm PMNS}.
\label{eq:charged-current-nonunitarity}
\end{align}
The Hermitian matrix $\eta_{\rm NU}$ describes the leading deviation of $N_{\rm CC}$ from unitarity.
Using $U_{eL}=W_e^*$ from Eq.~\eqref{eq:yukawa-diagonal-form} and the relation above, we obtain
\begin{align}
\eta_{\rm NU}
&=\frac{v^2}{4}U_{eL}^\dagger C^{(6)}U_{eL}
=\frac{1}{2|\mu_H|^2}
U_{\rm PMNS}\operatorname{diag}(0,m_2^2,m_3^2)U_{\rm PMNS}^\dagger.
\label{eq:neutrino-nonunitarity-correlation}
\end{align}
Its flavor structure is therefore fixed by the neutrino masses and
mixing, up to the common factor $|\mu_H|^{-2}$. In particular, its two nonzero eigenvalues have ratio $m_2^2/m_3^2\sim\epsilon^2$. 
For heavy masses much larger than $m_W$, the off-diagonal
entries of $\eta_{\rm NU}$ also control the heavy-neutral-fermion
contribution to $\ell_\alpha\to\ell_\beta\gamma$
\cite{CentellesChulia:2025eck}. 
For the benchmark $|\mu_H|\simeq0.204\,\mathrm{MeV}$, the largest
eigenvalue of $\eta_{\rm NU}$ is about $3.0\times10^{-14}$, so this
LFV contribution is too small to observe.

Eq.~\eqref{eq:neutrino-dimension-six-correlation} continues to hold when $P$ or $M_X$ is changed, provided that the same matrices are used to calculate both $Y_\nu$ and $C^{(6)}$.
In contrast, independent hidden mixing or kinetic terms, and an additional $N_S^c$ Yukawa coupling, can modify this relation. 
A contribution $\delta\mu_{\rm UV}(\eta N_D^c)$ with the same singlet contraction replaces $\mu_H$ by $\mu_H+\delta\mu_{\rm UV}$ and preserves the form of the relation
with this replacement. The flavor relation itself therefore does not establish an instanton origin. 
Loop corrections and renormalization-group evolution must also be
included in a precision test~\cite{Wang:2023bdw,Ardu:2024tzb}.

In addition to the heavy-fermion contribution, a dynamical flavon can induce charged-lepton flavor violation and is subject to astrophysical constraints \cite{Ema:2016ops,Calibbi:2016hwq,Calibbi:2020jvd}. In particular, a light phase mode dominated by $\varphi$ would give an electron coupling above the red-giant bound for the
present benchmark~\cite{Capozzi:2020cbu}. 
A viable realization of this benchmark therefore requires
the mass and couplings of the light FN phase mode to be
consistent with this bound.
Assessing possible domain walls requires an analysis of the
effective potential for the FN phase. In particular, one must
determine whether the instanton-induced fermion interaction
produces an energy difference between FN-related vacua.
We leave the calculation of this potential and of the thermal
abundances discussed above, together with an analysis of
possible domain-wall evolution, for future work.

\section{Extensions to finite and modular flavor symmetries}
\label{sec:family}

The FN model above is an Abelian example of the TQFT dressing of
an anomalous flavor transformation. We now discuss the extension
to finite flavor groups and modular transformations.
The relevant data are a rational anomaly, a gauge sector with
the required exact higher-form symmetry and a fixed global form,
and a wall theory that cancels the anomaly. Contributions from
other gauge sectors must be trivial or cancelled separately.
As in the FN example, we work in flat spacetime. An extension to
general curved backgrounds would require the mixed gravitational
contribution to be absent or cancelled.
The detailed anomaly and fusion analysis is given in
Appendix~\ref{app:flavor-completions}.

\subsection{Finite flavor groups}
\label{subsec:finite-flavor-summary}

For a finite flavor group $G_f$, the local Fujikawa Jacobian
depends on the determinants of the fermion representation
matrices~\cite{Araki:2008ek}. Each determinant is a
one-dimensional character, so the anomaly detects only the
Abelianization $G_f/[G_f,G_f]$. In a single compact Abelian gauge
sector $U(1)_V$, write the anomalous phase as
$J_V(g)=\exp[2\pi i\alpha_V(g)\mathcal I_V]$, where
$\mathcal I_V$ is the properly quantized topological number and
$\alpha_V(g)\in\mathbb Q/\mathbb Z$.
Multiplicativity of the Jacobian makes $\alpha_V$ a rational
character. Its kernel $G_0$ is normal, and
$G_f/G_0\simeq\operatorname{Im}\alpha_V$ is cyclic.
Several independent gauge sectors can instead give a finite
Abelian quotient. These statements do not require the full
flavor group to be Abelian~\cite{Kobayashi:2021xfs}. 
Thus $\exp[2\pi i\alpha_V(g)]$ is the anomalous phase in a
unit topological sector, and $G_0$ consists of the transformations
for which this phase is unity. 
Here $U(1)_V$ denotes the gauge sector used for the flavor anomaly,
not the continuous lift of $\mathbb Z_8^D$ in Appendix~\ref{app:continuous-Dirac-lift}.

If $\alpha_V(g)=p/n$ in lowest terms, the minimal Abelian wall
TQFT $\mathcal A^{n,p}$ supplies the required anomaly
\cite{Cordova:2022ieu,Choi:2021kmx,Choi:2022zal}.
More generally, for a wall theory $\mathcal B_g$ with that anomaly,
the dressed defect is
\begin{align}
\mathfrak D(g,\mathcal B_g;M_3)
=U_g(M_3)Z_{\mathcal B_g}[M_3,F_V].
\label{eq:flavor-extension-defect}
\end{align}
Here $U_g$ implements the original flavor transformation.
Group multiplication and TQFT stacking give
\begin{align}
\mathfrak D(g,\mathcal B_g)\times\mathfrak D(h,\mathcal B_h)
=\mathfrak D(gh,\mathcal B_g\boxtimes\mathcal B_h).
\label{eq:flavor-extension-fusion}
\end{align}
The symbol $\boxtimes$ denotes placing the two wall theories
on the same three-dimensional interface, with both coupled
to the bulk gauge field. 
The additive anomaly character ensures that the stacked theory
has the anomaly required for $gh$. This gives associative
TQFT-valued fusion data under the assumptions above. 
For $h\in G_0$, the wall theory can be taken to be trivial and
$U_h$ is an ordinary invertible wall. When the wall TQFT carries
no nontrivial action of $G_f$, as assumed here, we have
\begin{align}
\mathfrak D(g,\mathcal B_g)U_h
=U_{ghg^{-1}}\mathfrak D(g,\mathcal B_g).
\label{eq:flavor-extension-crossed-relation}
\end{align}
Thus the anomaly character is Abelian, but the defect
relations also involve the conjugation $h\mapsto ghg^{-1}$ for $h\in G_0$.
For $\alpha_V(g)=p/n\neq0\pmod1$, 
the minimally dressed wall obeys
the non-invertible transformation on an 't Hooft loop as in 
Eq.~\eqref{eq:non-inv-mag}, 
\begin{align}
\vevs{\mathfrak D(g,\mathcal A^{n,p}) T(C_{\rm mon})}
 =0 .
\label{eq:flavor-extension-noninvertibility}
\end{align}

As an example, consider the three left-handed Weyl fermions
$\psi,\eta,\lambda$ of Appendix~\ref{app:flavor-completions},
with $U(1)_V$ charges $(+1,-1,0)$ and $A_4$ representations
$(\mathbf1',\mathbf1,\mathbf1'')$, respectively. These fields
are separate from those of the FN model. The gauge anomalies
cancel, and the neutral fermion $\lambda$ cancels the purely
gravitational determinant character. For the order-three generator
$T$, we use $\rho_{\mathbf1'}(T)=e^{2\pi i/3}$. The charged
fermion $\psi$ then gives
\begin{align}
J_V(T)=\exp\!\left(\frac{2\pi i}{3} \mathcal I_V\right).
\label{eq:A4-extension-anomaly}
\end{align}
Thus $\alpha_V(T)=1/3$ and $G_0=\ker\alpha_V=V_4$, the Klein
subgroup, with $A_4/V_4\simeq\mathbb Z_3$. The $T$ wall is dressed by
$\mathcal A^{3,1}$ and satisfies
Eq.~\eqref{eq:flavor-extension-noninvertibility} with $(n,p)=(3,1)$.
Since $V_4$ acts trivially on these one-dimensional representations,
this fermion sector realizes only the quotient. A faithful
$A_4$ extension can include an uncondensed gauge-neutral scalar
triplet with an $A_4$-invariant potential, without changing the
chiral anomaly. 
The construction identifies the anomalous flavor directions;
it does not replace the ordinary representation selection rules
by new non-Abelian texture predictions.

\subsection{Modular transformations and stabilized vacua}
\label{subsec:modular-flavor-summary}

A modular flavor transformation acts simultaneously on the
modulus, the matter fields, and the coupling functions
\cite{Feruglio:2017spp}. For $\gamma\in SL(2,\mathbb Z)$, the
anomaly receives contributions from the automorphy factors and
from the determinants of the finite modular representations.
In a string-derived theory, Green-Schwarz and threshold terms,
as well as the remaining fermionic and supergravity
contributions, must be included~\cite{Ibanez:1992hc}. 
We restrict attention to transformations for which the complete gauge-field-dependent anomaly has a rational phase specified by $\alpha_V(\gamma,\tau)$. 
Its composition
law is the additive cocycle condition:
\begin{align}
\alpha_V(\gamma_1\gamma_2,\tau)
=\alpha_V(\gamma_1,\gamma_2\tau)
+\alpha_V(\gamma_2,\tau)\pmod1.
\label{eq:modular-extension-cocycle}
\end{align}
This condition permits a compatible TQFT dressing of the
modular walls. For a dynamical modulus, the transformations
must be closed under composition and inversion, with rational
anomaly data defined on a common domain. Rationality at an
isolated value of $\tau$ is not sufficient. 
On a connected domain, a continuous rational-valued cocycle is
locally constant, so the wall TQFT data must also be fixed there.

After modulus stabilization at $\langle\tau\rangle=\tau_\star$,
we can instead restrict to the subgroup that preserves the full
vacuum, including the expectation values of the other fields:
\begin{align}
H_{\rm vac}\subseteq H_{\tau_\star}
\equiv\{\gamma\in SL(2,\mathbb Z)\mid\gamma\tau_\star=\tau_\star\}.
\label{eq:modular-extension-stabilizer}
\end{align}
For transformations in $H_{\rm vac}$, the modulus remains at
$\tau_\star$. The cocycle condition therefore reduces to the usual
multiplication law for the anomaly phases at this fixed vacuum.
Equivalently, the restricted anomaly is a character of $H_{\rm vac}$. This construction does not require its continuation
to generic values of the modulus. If
$\alpha_V(\gamma,\tau_\star)=0\pmod1$, the wall theory can be
chosen trivial and the residual defect is invertible. For
$\alpha_V(\gamma,\tau_\star)=p/n\neq0\pmod1$ in lowest terms,
the minimally dressed residual defect is
\begin{align}
\mathfrak D_\gamma
=U_\gamma(\tau_\star)Z_{\mathcal A^{n,p}}[F_V].
\label{eq:modular-extension-residual-defect}
\end{align}
Under the gauge-sector assumptions stated above, fusion with its
orientation reverse gives
\begin{align}
\vevs{\mathfrak D_\gamma T(C_{\rm mon})}
 =0 .
\label{eq:modular-extension-noninvertibility}
\end{align}
Modulus stabilization selects the transformations that act within
the chosen vacuum; their non-invertibility follows from the
nontrivial TQFT dressing. Transformations that change the vacuum
are not residual internal symmetries of the fixed-vacuum theory. 
The complete expressions and their assumptions are collected
in Appendix~\ref{subsec:modular-flavor}.
These extensions concern the anomaly sector; the preceding FN
model provides the explicit messenger realization of a small
Dirac-neutrino Yukawa coupling.

\section{Conclusions}
\label{sec:discussion}

We have studied non-invertible completions of anomalous flavor
symmetries and their application to Dirac neutrino masses.
As a concrete phenomenological application, we considered an FN model with a hidden $SU(2)_H$ gauge symmetry and two hidden Weyl doublets $\eta$ and $N_D^c$ in the effective theory.
For our charge assignment, the Standard Model gauge sectors
preserve a $\mathbb Z_4^{\rm FN}$ subgroup of the anomalous
$U(1)_{\rm FN}$ symmetry. The adjoint expectation value breaks
$SU(2)_H$ to $U(1)_H$. The resulting mixed
$U(1)_{\rm FN}$--$[U(1)_H]^2$ anomaly allows the corresponding
FN transformation to be completed as the non-invertible defect
$\mathcal D_4^{\rm FN}$ in the monopole-decoupled Abelian theory. 
Finite-mass monopoles break the magnetic one-form symmetry,
so this defect is no longer topological.
In the ultraviolet $SU(2)_H$ theory, a constrained instanton
with topological charge one generates the mixing
$\mu_H(\eta N_D^c)$, with $\mu_H=v_H\zeta_H$ and
$\zeta_H\sim\exp[-8\pi^2/g_H^2(v_H)]$.
The bilinear $(\eta N_D^c)$ carries one unit of FN charge.

We then introduced the two singlet messengers $X_1,X_2$
with interactions containing the flavon factors required by
the FN charges. These interactions transmit the nonperturbative
FN breaking to the neutrino sector. 
After integrating out the heavy fields, we obtain
$(\widehat Y_\nu)_{ij}\simeq\zeta_\nu C_{ij}\epsilon^{\ell_i+n_j}$ in a rotated basis for the right-handed neutrinos at leading order in $|\mu_H|/w_H$, with $\zeta_\nu=(v_H/w_H)\zeta_H$. 
In the instanton-dominated regime, the common factor $\zeta_H$ provides the exponential suppression of the
Dirac-neutrino Yukawa couplings, while the FN charges fix the powers of $\epsilon$ in the flavor matrices. 
The same charge assignment gives
the leading quark and charged-lepton hierarchies.
Although the two components of $N_D^c$ have the same ultraviolet FN charge, the different flavon insertions in the
messenger interactions give different powers of $\epsilon$ in the two nonzero columns of $\widehat Y_\nu$. 
For these messenger interactions, the dependence on the orientation of $\langle\Omega\rangle$ with respect to
$\langle\Phi\rangle$ can be absorbed into a unitary rotation
of the right-handed neutrinos. At fixed $w_H,\mu_H$ and
dimensionless couplings, the neutrino masses and left-handed
mixing are therefore independent of this orientation and require no special vacuum alignment.
The resulting tree-level neutrino Yukawa matrix has rank at most two. 
For the chosen FN charges and generic order-one coefficients, the powers of $\epsilon$ give
$m_2/m_3\sim\epsilon$ and suggest normal ordering.
The FN charges also give the leading powers of $\epsilon$ in
the mixing angles and Jarlskog invariants, while their numerical values depend on the flavor coefficients. 
A numerical example with real, order-one lepton and messenger coefficients is given in
Appendix~\ref{app:lepton-benchmark}. 
In the normal-ordering rank-two limit, $m_1=0$ at tree level.
With the best-fit oscillation parameters used here, the mass sum is
$\sum_i m_i\simeq0.059~\mathrm{eV}$.

When combined with the instanton-induced bilinear,
the FN-preserving dimension-seven interaction generates an
additional Yukawa coupling to the singlet right-handed neutrino
$N_S^c$, proportional to the same instanton factor $\zeta_H$. 
When this interaction couples $N_S^c$ to the left-handed neutrino combination that was previously massless, it
generates a small Dirac mass for that state.
Thus a nonzero lightest mass is compatible with the same
nonperturbative origin of neutrino masses.
We also presented a nonminimal gauge extension with a $\mathbb Z_9$ remnant in Appendix~\ref{app:UV-protection}. 
The allowed renormalizable interactions preserve an accidental chiral symmetry, which prevents an independent $(\eta N_D^c)$ mass term from being generated perturbatively.
The hidden instanton violates this symmetry and generates the required mixing.

The Dirac nature of the neutrinos is protected independently by the
exact discrete gauge symmetry $\mathbb Z_8^D$.
It forbids Majorana masses for the right-handed neutrinos, the Weinberg operator, and more generally baryon-number-preserving local $\Delta L=2$ operators.
Neutrinoless double-beta decay is therefore absent as long as
$\mathbb Z_8^D$ remains exact.
For the leading messenger interactions, the same heavy fermions also generate a dimension-six interaction whose coefficient $C^{(6)}$ satisfies 
$Y_\nu Y_\nu^\dagger=|\mu_H|^2(C^{(6)})^*$
at tree level and to leading order in the heavy-mass expansion. 
The resulting heavy-fermion contribution to charged-lepton flavor violation is negligible in the high-scale benchmark. 

For finite non-Abelian flavor symmetries, we constructed
TQFT-dressed defects associated with rational anomaly characters and
showed that their fusion is determined by flavor-group multiplication
and TQFT stacking. We also extended the construction to the rational
part of modular anomaly cocycles. After modulus stabilization, a
modular transformation that leaves the full vacuum invariant remains
as a residual symmetry. An anomaly-free residual transformation admits
an invertible defect, whereas a transformation with a nontrivial
rational anomaly admits a non-invertible TQFT-dressed defect under
the gauge-sector conditions specified in Sec.~\ref{sec:family}.
The resulting gauge-invariant topological defects have non-invertible fusion and realize these anomalous transformations
as symmetries of the quantum theory. 
Exploring the phenomenological consequences of these completions
in finite-group and modular flavor models is an interesting
direction for future work.

The neutrino-mass calculation can be refined by evaluating the
constrained-instanton prefactor and threshold corrections.
A more detailed comparison with the observed fermion masses and
mixing parameters should include the renormalization-group evolution
of the Yukawa couplings. 
It would also be interesting to extend the FN construction to Majorana neutrinos, with symmetry assignments that allow
lepton-number violation. 
We leave these studies, together with a calculation of the
effective potential for the FN phase and an analysis of the
thermal history, for future work.

\section*{Acknowledgments}

This work was supported by JSPS KAKENHI Grant Numbers JP25H01539 (H.O.), JP26K07087 (H.O.), and JP25K17394 (R.Y.).

\appendix
\section{Semiclassical suppression from dyon worldlines}
\label{app:dyon}

This appendix gives the infrared worldline derivation of the exponent~\cite{Fan:2021ntg} used in Sec.~\ref{subsec:monopole-instanton-breaking}. It is complementary to the constrained-instanton calculation: the dyon description illustrates the breaking of the magnetic one-form symmetry, while the instanton zero modes determine the induced local fermion operator.

When monopoles of finite mass are included as dynamical particles,
\begin{align}
df_H
=
2\pi\,\delta_3(C_{\rm mon}),
\end{align}
and the magnetic one-form symmetry is no longer exact. The defect in
Eq.~\eqref{eq:FN-defect} consequently ceases to be topological.
In the ultraviolet theory, a hidden instanton with topological charge
one gives $\Delta F_{\rm FN}=1$ and generates the bilinear
$(\eta N_D^c)$. This bilinear is odd under the combined FN parity in
Eq.~\eqref{eq:combined-FN-parity} and is forbidden in perturbation
theory. Its coefficient is proportional to the spurion $\zeta_H$
defined in Eq.~\eqref{eq:monopole-spurion}.

The monopole has a periodic electric collective coordinate. Quantizing this coordinate produces a tower of dyons labeled by $r\in\mathbb Z$ \cite{WittenEffect,Cordova:2022ieu,Cordova:2022fhg}. 
In estimating the dyon masses from the electric collective coordinate,
we use $\vartheta=\vartheta_{\rm UV}$, not
$\vartheta_{\rm Ab}=2\vartheta_{\rm UV}$. The formula below does not
include excitations of the matter fields.
\begin{align}
M_r^2(\vartheta)
\simeq
M_{\rm mon}^2
+
M_{W_H}^2
\left(
r-\frac{\vartheta}{2\pi}
\right)^2.
\label{eq:dyon-mass}
\end{align}
Using the proper-time parameter $\tau$, the bosonic worldline
contribution is schematically
\begin{align}
\Gamma(\vartheta)
\sim
\sum_{r\in\mathbb Z}
\int_0^\infty
\frac{d\tau}{\tau^\alpha}
\exp\left[
-\frac{\tau}{2}M_r^2(\vartheta)
\right],
\end{align}
where $\alpha$ is the exponent of the proper-time measure. We focus on
the exponential suppression below.
After Poisson resummation,
\begin{align}
\Gamma(\vartheta)
\sim
\sum_{n\in\mathbb Z}
\int_0^\infty
\frac{d\tau}{\tau^\beta}
\exp\left[
-\frac{M_{\rm mon}^2\tau}{2}
-\frac{2\pi^2n^2}{M_{W_H}^2\tau}
+in\vartheta
\right],
\end{align}
where $\beta=\alpha+1/2$, since Poisson resummation gives an additional
factor $\tau^{-1/2}$.
For $n\neq0$, the saddle point is
$\tau_*=2\pi|n|/(M_{\rm mon}M_{W_H})$. The corresponding action is
\begin{align}
S_{{\rm loop},n}
=
\frac{2\pi |n|M_{\rm mon}}{M_{W_H}}
=
\frac{8\pi^2|n|}{g_H^2(v_H)}
\label{eq:dyon-action}
\end{align}
in the BPS limit. The contributions with positive and negative $n$
are complex conjugates. A general scalar potential changes the
monopole profile and the moment of inertia associated with its
electric collective coordinate. We therefore use the BPS spectrum
to estimate the exponent, not as the exact spectrum of the model
with two stages of hidden gauge-symmetry breaking. The local fermion
vertex follows from the ultraviolet constrained instanton; the
fluctuation determinant and heavy-particle threshold factors have
not been evaluated. For $|n|=1$, the BPS estimate is
\begin{align}
\exp[-S_{{\rm loop},1}]
=
\exp\!\left[
-\frac{8\pi^2}{g_H^2(v_H)}
\right].
\label{eq:dyon-suppression}
\end{align}
The worldline calculation reproduces the leading instanton exponent in the BPS limit. The induced fermion operator is fixed by the ultraviolet instanton zero modes.

\section{Consistency checks for the minimal hidden and messenger sectors}
\label{app:technical-checks}

The derivation of the neutrino Yukawa matrix in
Sec.~\ref{subsec:messenger-matching} uses a vacuum with $w_H\ll v_H$
and two heavy messenger masses of order $w_H$.
Here we give a scalar potential and messenger couplings that realize
these assumptions without introducing an additional scalar doublet. 
We also examine the instanton-size dependence of the fermion
mixing derived in Sec.~\ref{subsec:monopole-instanton-breaking},
including the effect of the running hidden gauge coupling.

\subsection{Scalar potential and vacuum stability}
\label{app:scalar-alignment}

The mass hierarchy derived in the main text does not require a
particular orientation of $\langle\Omega\rangle$. Here we give one
locally stable scalar vacuum realizing the two hidden Higgs scales.
Since $F(\Omega)=1$, the interaction
$\varphi\Omega^T\varepsilon_2\Phi\Omega$ is allowed and must be included
when determining this vacuum. We use the illustrative potential
\begin{align}
V={}&\frac{\lambda_\Phi}{4}(\Phi^a\Phi^a-v_0^2)^2
+m_\Omega^2\Omega^\dagger\Omega
+\lambda_\Omega(\Omega^\dagger\Omega)^2
+\kappa\Phi^a\Phi^a\Omega^\dagger\Omega\nonumber\\
&+\mu_\Omega\Omega^\dagger\Phi\Omega
+\lambda_f(|\varphi|^2-f_0^2)^2
+\left[\lambda_m\varphi\Omega^T\varepsilon_2\Phi\Omega
+\mathrm{h.c.}\right],
\label{eq:scalar-potential}
\end{align}
where we choose real scalar couplings and real vacuum
expectation values in the field basis and gauge specified below. 
This is an
assumption about the scalar sector, not a consequence of taking
$\epsilon>0$. The coefficients coupling $L_iH$ to $X_1,X_2$ are
independent of this choice and are taken real in
Appendix~\ref{app:lepton-benchmark}.
We set the other FN-preserving interactions between the scalars,
including those involving the Standard Model Higgs field, to zero. This is a choice for the
example, not a consequence of the FN symmetry.

We analyze the scalar vacuum for $\lambda_m\neq0$. 
In the basis
$\langle\varphi\rangle=v_\varphi>0$ and the gauge
$\langle\Phi\rangle=v_HT^3$, parameterize the doublet vacuum by
\begin{align}
\langle\Omega\rangle=\frac{w_H}{\sqrt2}
\begin{pmatrix}\sin\beta\\\cos\beta\end{pmatrix}.
\label{eq:nearby-alignment}
\end{align}
Here $\beta$ specifies the direction of $\langle\Omega\rangle$ in this gauge. 
It gives $z_1=\sin\beta$ and $z_2=\cos\beta$ in
Eq.~\eqref{eq:scalar-VEVs}.
The orientation-dependent potential is
\begin{align}
V_{\rm orient}=-\frac{v_Hw_H^2}{4}
\left[\mu_\Omega\cos2\beta
+2\lambda_m v_\varphi\sin2\beta\right].
\end{align}
For $\mu_\Omega,\lambda_m>0$, its minimum satisfies
\begin{align}
\tan2\beta=\frac{2\lambda_m v_\varphi}{\mu_\Omega}.
\label{eq:scalar-tilt}
\end{align}
This equation fixes the vacuum angle for the scalar.
For the messenger interactions used in the main text, the neutrino
mass hierarchy does not require a particular value of $\beta$.
For the stationary conditions with respect to the magnitudes of the
scalar expectation values, it is convenient to define the positive
quantity
\begin{align}
R=\sqrt{\mu_\Omega^2+4\lambda_m^2v_\varphi^2}.
\label{eq:scalar-radial-parameter}
\end{align}
The remaining stationary conditions can be satisfied by choosing
\begin{align}
m_\Omega^2&=\frac{v_HR}{2}-\kappa v_H^2-\lambda_\Omega w_H^2,\nonumber\\
v_0^2&=v_H^2+
\frac{w_H^2(\kappa v_H-R/4)}{\lambda_\Phi v_H},\nonumber\\
f_0^2&=v_\varphi^2-
\frac{v_Hw_H^2\lambda_m^2}{4\lambda_f R}.
\label{eq:complete-scalar-stationarity}
\end{align}

To give an explicit locally stable vacuum, we choose the dimensionless
couplings
\begin{align}
\lambda_\Phi=\lambda_\Omega=\lambda_f=1,\qquad
\kappa=\frac14,\qquad\lambda_m=\frac12.
\label{eq:stable-scalar-slice}
\end{align}
The mass scales and expectation values are also chosen as
\begin{align}
\mu_\Omega=\Lambda_{\rm FN}=v_H,\qquad
v_\varphi=0.2v_H,\qquad w_H^2=0.005v_H^2.
\label{eq:scalar-benchmark-scales}
\end{align}
Eqs.~\eqref{eq:scalar-tilt} and
\eqref{eq:complete-scalar-stationarity} then determine the vacuum angle and the quadratic parameters of the potential. 
They give $\beta\simeq0.0987$,
$m_\Omega^2/v_H^2\simeq0.255$, and
$f_0^2/v_H^2\simeq0.0397$.

The quartic part is bounded below for this choice. One sufficient
condition follows from
\begin{align}
|\Omega^T\varepsilon_2\Phi\Omega|
\leq\frac{\sqrt{\Phi^a\Phi^a}}{2}\Omega^\dagger\Omega.
\end{align}
For $\kappa>0$, completing the square in the mixed terms gives a
lower bound with the possible negative term
$-\lambda_m^2|\varphi|^2(\Omega^\dagger\Omega)/(4\kappa)$.
For $\lambda_\Phi,\lambda_\Omega,\lambda_f,\kappa>0$, the positive
quartics dominate if
\begin{align}
\frac{\lambda_m^2}{4\kappa}<2\sqrt{\lambda_\Omega\lambda_f}.
\end{align}
The benchmark satisfies these inequalities.

We also evaluate the Hessian with respect to all nine real scalar
components, writing
$\Omega_j=(o_{jR}+io_{jI})/\sqrt2$ and
$\varphi=(t_R+it_I)/\sqrt2$. At the classical level, the Hessian has
three zero eigenvalues associated with gauge transformations and
one associated with the global FN phase. Denoting the five
nonzero eigenvalues by $m_{s,k}^2$, we obtain
\begin{align}
\frac{1}{v_H^2}\operatorname{diag}(m_{s,1}^2,\ldots,m_{s,5}^2)
\simeq\operatorname{diag}(0.00984,0.160,1.02,1.02,2.00).
\label{eq:minimal-scalar-spectrum}
\end{align}
The five nonzero scalar squared masses are positive, establishing
local stability in these directions. The global FN phase remains
massless at the classical level. Its anomaly-induced potential
and cosmological evolution have not been evaluated here.

\subsection{Heavy-fermion spectrum and messenger realization}
\label{app:heavy-fermion-stability}

We first give the heavy masses for general messenger couplings in
Eq.~\eqref{eq:two-doublet-mass-terms}. At $\mu_H=0$, the two Dirac
masses $\sigma_-\leq\sigma_+$ are the nonnegative diagonal entries obtained
by biunitary diagonalization of $M_X$. They equal those of
$\widehat M_X$. Let us define
\begin{align}
T_X=|a|^2+|d|^2+\epsilon^2(|b|^2+|c|^2).
\label{eq:heavy-trace-invariant}
\end{align}
The two masses at $\mu_H=0$ satisfy
\begin{align}
\sigma_\pm^2=\frac{w_H^2}{2}
\left[T_X\pm\sqrt{T_X^2-4|\Delta_X|^2}\right].
\label{eq:general-heavy-singular-values}
\end{align}
For generic $a,b,c,d=\mathcal O(1)$ and
$|\Delta_X|=\mathcal O(1)$, both masses are of order $w_H$,
independently of $z_1,z_2$.
At nonzero $\mu_H$, the rectangular matrix in
Eq.~\eqref{eq:rectangular-dirac-matrix} obeys
\begin{align}
\mathcal M_D\mathcal M_D^\dagger
=M_X^TM_X^*+|\mu_H|^2\mathbf1_2.
\label{eq:heavy-mass-gram-general}
\end{align}
The two heavy Dirac masses before electroweak symmetry breaking
are given exactly by
\begin{align}
M_{H,\pm}^2=\sigma_\pm^2+|\mu_H|^2.
\label{eq:heavy-masses-finite-mu}
\end{align}

Including the canonical normalization of the light right-handed
neutrino kinetic terms in
Eqs.~\eqref{eq:light-kinetic-matrix} and
\eqref{eq:general-messenger-matching}, we also obtain
\begin{align}
Y_\nu Y_\nu^\dagger
=|\mu_H|^2P
\left(M_X^*M_X^T+|\mu_H|^2\mathbf1_2\right)^{-1}P^\dagger.
\label{eq:finite-mu-Yukawa-product}
\end{align}
This tree-level expression keeps the full dependence on $\mu_H$
for the mass terms in Eq.~\eqref{eq:two-doublet-mass-terms}, with
external momenta neglected. It does not include the additional
interactions of Sec.~\ref{subsec:singlet-corrections} or the
corrections from canonical normalization of the active-neutrino
kinetic terms after electroweak symmetry breaking.
For $|\mu_H|\ll\sigma_-$, it reduces to the leading result used
in the main text. The dimension-six relation in
Eq.~\eqref{eq:neutrino-dimension-six-correlation} is used only
at the stated leading order in $|\mu_H|/w_H$ and $v/w_H$.

For the lepton example in Appendix~\ref{app:lepton-benchmark},
we choose the couplings in
Eq.~\eqref{eq:general-hidden-messenger-couplings} as
\begin{align}
\lambda_1=\sqrt2
\qquad\mathrm{with}\qquad \lambda_2=-\lambda_1.
\label{eq:minimal-scalar-messenger-couplings}
\end{align}
For the terms containing $\varphi$ or $\varphi^\dagger$ in the
same equation, we take
\begin{align}
\rho_1=\rho_2=-\sqrt2.
\label{eq:scalar-messenger-cross-couplings}
\end{align}
These couplings give $a=b=d=1,c=-1$ in
Eq.~\eqref{eq:orientation-independent-mass}. For any $z_1,z_2$
satisfying $|z_1|^2+|z_2|^2=1$, the original mass matrix is therefore
\begin{align}
M_X=w_H\begin{pmatrix}1&\epsilon\\-\epsilon&1\end{pmatrix}U_\Omega.
\end{align}
All four couplings have ${\cal O}(1)$ magnitude and can be
used with the locally stable vacuum of the preceding subsection.
At $\mu_H=0$, the two heavy Dirac masses are
\begin{align}
m_{X\eta,1}=m_{X\eta,2}=w_H\sqrt{1+\epsilon^2}\simeq1.02w_H.
\label{eq:two-doublet-heavy-spectrum}
\end{align}
Neither heavy Dirac mass is small compared with $w_H$, so the
neutrino hierarchy is not enhanced by a small messenger mass. 
The equality of the two heavy masses is a choice of benchmark;
the rank bound and the vacuum-orientation independence in
Sec.~\ref{subsec:messenger-matching} do not require this degeneracy.

The independence of the neutrino masses and left-handed mixing from
$z_1,z_2$ applies to the interactions used in the calculation, not to
every FN-preserving higher-dimensional operator. For example, the
charges allow the additional interaction
\begin{align}
\Delta\mathcal L_{\rm portal}\supset
\frac{\xi_i}{\Lambda^3}
\left(\frac{\varphi}{\Lambda_{\rm FN}}\right)^{\ell_i}
\left(\Omega^T\varepsilon_2\Phi\Omega\right)^\dagger
L_iHX_2+\mathrm{h.c.}
\label{eq:orientation-dependent-portal}
\end{align}
Here $\xi_i$ is dimensionless and $\Lambda$ is the suppression scale
of this operator, not necessarily $\Lambda_{\rm FN}$. Since
$\langle\Omega\rangle^T\varepsilon_2\langle\Phi\rangle
\langle\Omega\rangle=-v_Hw_H^2z_1z_2/2$, its correction
$\delta P_{i2}$ to the coefficient $B_i\epsilon^{\ell_i+1}$ of
$L_iHX_2$ obeys
\begin{align}
\left|\frac{\delta P_{i2}}{B_i\epsilon^{\ell_i+1}}\right|
\leq\left|\frac{\xi_i}{B_i}\right|
\frac{v_Hw_H^2}{4\epsilon\Lambda^3}.
\label{eq:orientation-portal-bound}
\end{align}
We require $|\delta P_{i2}|\ll|B_i\epsilon^{\ell_i+1}|$.
For the benchmark $w_H^2/v_H^2=5.0\times10^{-3}$ and $\epsilon=0.2$,
the additional assumption $\Lambda\geq v_H$ makes the ratio in
Eq.~\eqref{eq:orientation-portal-bound} no larger than
$6.3\times10^{-3}|\xi_i/B_i|$ for any vacuum orientation.
This operator changes $P$ but leaves the mixing
$\mu_H\varepsilon_2$ unchanged. It can therefore introduce dependence on the vacuum orientation
while preserving
Eq.~\eqref{eq:neutrino-dimension-six-correlation} at the stated order,
provided that the modified $P$ is used in calculating both $Y_\nu$
and $C^{(6)}$. The bound above applies only to
Eq.~\eqref{eq:orientation-dependent-portal}. Other operators containing
$\Phi$ and loop corrections that depend on $z_1,z_2$ must be checked
separately; they are not included in the leading calculation.

\subsection{Hidden gauge running and the instanton measure}
\label{app:hidden-gauge-running}

In Sec.~\ref{subsec:monopole-instanton-breaking}, we estimate
the instanton-induced mixing $\mu_H$ using instantons of size
$\rho\sim v_H^{-1}$. Here we examine the size dependence of
the integral in Eq.~\eqref{eq:two-mode-size-integral}.
The running of the hidden gauge coupling determines its
small-size behavior, while the adjoint expectation value
suppresses large instantons.

We first consider energies above the masses generated by hidden
gauge-symmetry breaking and below any additional heavy-particle
thresholds. In this range, the $SU(2)_H$-charged matter consists
of the two Weyl doublets $\eta,N_D^c$, the real adjoint scalar
$\Phi$, and the complex scalar doublet $\Omega$.
The one-loop coefficient is
\begin{align}
b_0
&=\frac{11}{3}C_2(SU(2))
-\frac23\sum_{\rm Weyl}T(r_f)
-\frac16\sum_{\rm real}T(r_s)\nonumber\\
&=\frac{22}{3}-\frac23-\frac13-\frac16
=\frac{37}{6}.
\label{eq:hidden-beta-coefficient}
\end{align}
The four terms are the contributions from the gauge fields,
the two Weyl doublets, $\Phi$, and $\Omega$, respectively.
The complex doublet is counted as two real scalar multiplets.
The gauge coupling obeys
\begin{align}
\mu\frac{dg_H}{d\mu}
=-\frac{b_0}{16\pi^2}g_H^3+\cdots.
\label{eq:hidden-one-loop-running}
\end{align}

Let $\mu_0$ be a reference renormalization scale in this range.
As long as $1/\rho$ lies in the same range, the one-loop
solution gives
\begin{align}
\exp\!\left[-\frac{8\pi^2}{g_H^2(1/\rho)}\right]
=
\exp\!\left[-\frac{8\pi^2}{g_H^2(\mu_0)}\right]
(\mu_0\rho)^{b_0}.
\end{align}
Thus the running supplies a factor $\rho^{b_0}$ to the
instanton weight. Before the external fermion zero modes
are inserted, the measure has the size dependence
$d\rho\,\rho^{b_0-5}=d\rho\,\rho^{7/6}$, up to normalization
and logarithmic factors.
The two external fermion zero modes supply a further factor
$\rho^3$~\cite{tHooftInstanton,Schafer:1996wv}.
The corresponding small-size dependence of the bilinear
coefficient in Eq.~\eqref{eq:two-mode-size-integral} is therefore
\begin{align}
d\rho\,\rho^{-5}(\mu_0\rho)^{b_0}\rho^3
\propto d\rho\,\rho^{25/6}.
\end{align}
This positive power suppresses smaller instantons within
the range where the stated field content is applicable.

At large sizes, the adjoint expectation value contributes
$c_\Phi v_H^2\rho^2$ to the action, as in
Eq.~\eqref{eq:constrained-action}.
The resulting factor $\exp[-c_\Phi v_H^2\rho^2]$ suppresses
large instantons~\cite{Affleck,tHooftInstanton}.
Together, these small- and large-size behaviors support
the estimate $\rho\sim v_H^{-1}$ for the minimal-sector
contribution used in Sec.~\ref{subsec:monopole-instanton-breaking}.

This estimate does not determine the full instanton prefactor
$\mathcal K_H$. If $1/\rho$ crosses a threshold of additional
ultraviolet particles, both the running coupling and the
instanton measure must be matched to the theory containing
those particles. The power law above is not continued across
such thresholds and does not by itself establish that
the mixing is insensitive to the ultraviolet theory.
The remaining prefactor and threshold dependence are included
in $\mathcal K_H$ and are not evaluated here.

\section{Possible ultraviolet realizations of the discrete symmetries}
\label{app:UV-protection}

The main text uses $\mathbb Z_8^D$ as an exact finite gauge symmetry
and assumes that the FN-breaking mass is dominated by the instanton.
We treat these two issues separately. The first subsection embeds
$\mathbb Z_8^D$ in an anomaly-free continuous gauge theory. The
remaining subsections give a nonminimal gauge extension with a
$\mathbb Z_9$ remnant that forbids an independent FN-breaking mass
at renormalizable order. The additional gauge and matter fields are
not counted among the fields required for the leading messenger
mechanism. The extension protects a chiral selection rule at
renormalizable order; it does not determine all FN flavor couplings
or forbid every higher-dimensional breaking term.

\subsection{Continuous gauge embedding of $\mathbb Z_8^D$}
\label{app:continuous-Dirac-lift}

We give an embedding of the discrete gauge symmetry
in an anomaly-free continuous gauge theory. We include $U(1)_X$
as a separate factor of the gauge group and use the integer
hypercharge normalization $\widehat Y=6Y$. For Standard Model
fields, we define
\begin{align}
X=6(B-L)-\widehat Y.
\end{align}
Their charges are
\begin{align}
X(Q_i,u_i^c,d_i^c,L_i,e_i^c,H)=(1,2,-4,-3,0,-3).
\end{align}
For the additional fields, we assign
\begin{align}
X(\eta)=-6,\qquad
X(N_D^c)=X(N_S^c)=X(X_1)=X(X_2)=6,\nonumber\\
X(\Phi)=X(\Omega)=X(\varphi)=0.
\label{eq:optional-X-charges}
\end{align}
The unit charge of $Q_i$ fixes a faithful integer normalization.
All messenger interactions specified in the main text are invariant
under $U(1)_X$. This continuous theory specifies the discrete gauge
group and the terms in the FN wall action used in
Sec.~\ref{subsec:diracness}. 
The $U(1)_X$ gauge boson is not included as a dynamical
field in the effective theory used in the main text.
The group $U(1)_X$ is distinct from the intermediate hidden
gauge group $U(1)_H$.

The Standard Model fields without right-handed neutrinos give
$\mathcal A_{{\rm grav}-X}=-18$ and $\mathcal A_{X^3}=-648$.
The doublets $\eta,N_D^c$ cancel each other, and the three singlet
fields $X_1,X_2,N_S^c$ contribute $(18,648)$. Thus the full theory,
rather than the visible sector, is anomaly free. The mixed
$SU(2)_H^2-X$ anomaly is proportional to $-6+6=0$.
All other mixed gauge anomalies also vanish. No additional chiral
fermion is needed for this $U(1)_X$ lift.

A scalar $S_8$ of charges $(X,\widehat Y)=(8,0)$ can break
$U(1)_X$ to $\mathbb Z_8^X$. The transformation
\begin{align}
g_D=\exp\left[\frac{2\pi i}{8}(X+\widehat Y)\right]
\end{align}
leaves the Higgs expectation value invariant, since
$X(H)+\widehat Y(H)=0$.
It gives precisely the $q_D$ assignments of Table~\ref{tab:matter-content}.
The masses of the gauge boson and the radial mode of $S_8$ can be
chosen above the cutoff of the effective theory used in the main text.
Alternatively, the continuous theory can be used only to define the
finite gauge theory and its wall action.

For the FN charges in Table~\ref{tab:matter-content}, we obtain
\begin{align}
\kappa_{FXX}=1452,\qquad
\kappa_{FX\widehat Y}=-672,\qquad
\kappa_{F\widehat Y\widehat Y}=1992.
\end{align}
These coefficients are multiples of four. If the $U(1)_X$ gauge
field is present, the additional Chern-Simons terms with integer
coefficients on the FN wall are
\begin{align}
S_{\rm wall}^{X}=-\frac{1}{4\pi}\int_{M_3}
\left[363\,a_X\wedge da_X
-168(a_X\wedge da_Y+a_Y\wedge da_X)\right].
\label{eq:X-integer-wall}
\end{align}
The mixed anomaly convention is
$d\star J_{\rm FN}\supset(2\kappa_{FX\widehat Y}/8\pi^2)f_X\wedge f_Y$.
These counterterms cancel the $U(1)_X$ and mixed
$U(1)_X$--hypercharge contributions when the FN wall is deformed.
After $U(1)_X$ is broken to its discrete subgroup, they determine
the corresponding terms in the discrete-gauge contribution to
the FN wall action. The hidden wall theory $\mathcal A^{2,1}$
is unchanged.

\subsection{A nonminimal $\mathbb Z_9$ extension}

We now use a further compact $U(1)_A$ with integer charge $a$ and
its $\mathbb Z_9$ remnant to protect the chiral mass selection rule. 
The ultraviolet gauge group in this construction also
contains $U(1)_X$, whose gauge invariance restricts the
interactions of the additional fermions. 
Replacing the continuous gauge symmetries by their discrete
subgroups can change the allowed interactions. We therefore use
the continuous charges to specify this ultraviolet Lagrangian.

The charges of the light fields and the three additional hidden
Weyl pairs are listed in Table~\ref{tab:Z9-hidden}.
\begin{table}[htbp]
\centering
\caption{Charges for the ultraviolet extension that forbids an
independent fermion mass at renormalizable order. Tilded fermions are independent left-handed
fields. All fields are Standard Model singlets. The quantity $p$ is the accidental chiral charge discussed
below. The charges $F$ of the light fields are those of Table~\ref{tab:matter-content}.}
\label{tab:Z9-hidden}
\small\renewcommand{\arraystretch}{1.1}
\begin{tabular}{lccrrrr}
\toprule
field&copies&$SU(2)_H$&$X$&$a$&$p$&$F$\\
\midrule
$\eta$&1&$\mathbf2$&$-6$&0&0&$-1$\\
$N_D^c$&1&$\mathbf2$&6&3&1&2\\
$N_S^c$&1&$\mathbf1$&6&3&1&3\\
$X_1$&1&$\mathbf1$&6&0&0&0\\
$X_2$&1&$\mathbf1$&6&0&0&2\\
$P_I$&3&$\mathbf2$&$-6$&2&0&0\\
$\widetilde P_I$&3&$\mathbf2$&6&$-3$&0&0\\
$S$ (scalar)&1&$\mathbf1$&0&1&0&0\\
\bottomrule
\end{tabular}
\end{table}
All Standard Model fields and $\Phi,\Omega,\varphi,S_8$ have $a=p=0$.
The new scalar $S$ has no FN charge. When $S$ acquires an expectation
value, the additional hidden doublets have the mass matrix
\begin{align}
M_P=\lambda f
\qquad\mathrm{with}\qquad f=\langle S\rangle.
\label{eq:Z9-heavy-masses}
\end{align}
We take the dimensionless matrix $\lambda$ to be nonsingular. 
The bilinear $(\eta N_D^c)$ has $U(1)_A$ charge three,
while $S$ has charge one. Three factors of $S^\dagger$
therefore make the bilinear gauge invariant:
\begin{align}
(S^\dagger)^3(\eta N_D^c).
\label{eq:Z9-first-dressed-mass}
\end{align}
This operator has mass dimension six. Since $S$ is neutral
under $p$ and $(\eta N_D^c)$ has $p=1$, the operator still
violates the accidental chiral symmetry. 
The bare bilinear and $S^{(\dagger)}(\eta N_D^c)$ are forbidden.
The $S$ expectation value breaks $\mathbb Z_9$ completely;
the perturbative selection rule after this breaking follows
from an accidental symmetry, not an unbroken $\mathbb Z_9$
acting on the light bilinear.

\subsection{Anomaly cancellation and the accidental chiral charge}

A scalar $T_9$, of $a=9$ and $X=0$, gives
\begin{align}
U(1)_A\xrightarrow{\langle T_9\rangle}\mathbb Z_9
\xrightarrow{\langle S\rangle}1.
\label{eq:Z9-gauge-breaking}
\end{align}
We assume the hierarchy $M_*\sim|\langle T_9\rangle|\gg|f|$.
To cancel the remaining Abelian anomalies, introduce the singlets
in Table~\ref{tab:Z9-singlets}. Their FN and $p$ charges vanish except
for $F(E_3)=p(E_3)=3$ and
$F(\widetilde E_3)=p(\widetilde E_3)=-3$.
\begin{table}[htbp]
\centering
\caption{Additional Standard Model and hidden singlet fermions
introduced to cancel the remaining Abelian anomalies in the
ultraviolet extension. Each pair contains two left-handed Weyl
fields. These spectators belong to the UV example, not to the
minimal messenger sector.}
\label{tab:Z9-singlets}
\small\renewcommand{\arraystretch}{1.1}
\begin{tabular}{lccc}
\toprule
pair&number of pairs&$(X,\widetilde X)$&$(a,\widetilde a)$\\
\midrule
$(n_1,\widetilde n_1)$&1&$(0,0)$&$(2,-3)$\\
$(n_{2,3},\widetilde n_{2,3})$&2&$(0,0)$&$(0,-1)$\\
$(E_{1,2},\widetilde E_{1,2})$&2&$(2,-2)$&$(-1,-8)$\\
$(E_3,\widetilde E_3)$&1&$(2,-2)$&$(-2,-7)$\\
$(R_1,\widetilde R_1)$&1&$(0,0)$&$(4,5)$\\
$(R_2,\widetilde R_2)$&1&$(0,0)$&$(1,8)$\\
$(R_3,\widetilde R_3)$&1&$(0,0)$&$(0,9)$\\
\bottomrule
\end{tabular}
\end{table}
The Yukawa interactions $Sn_i\widetilde n_i$, $T_9E_i\widetilde E_i$
and $T_9^\dagger R_i\widetilde R_i$ are allowed. We also include all
other gauge-invariant renormalizable interactions. The neutral-fermion mass matrix can be nonsingular,
but making all these fermions heavy requires a choice of couplings
and mass scales.

To verify anomaly cancellation in the $U(1)_A$ extension,
we define the following coefficients, including the multiplicities $d_f$ of each gauge representation:
\begin{align}
\begin{aligned}
\mathcal A_A&=\sum_fd_fa_f,&
\mathcal A_{AAA}&=\sum_fd_fa_f^3,\\
\mathcal A_{AXX}&=\sum_fd_fa_fX_f^2,&
\mathcal A_{AAX}&=\sum_fd_fa_f^2X_f.
\end{aligned}
\label{eq:UV-anomaly-definitions}
\end{align}
The first is the mixed gravitational coefficient; the other three
are Abelian gauge-anomaly coefficients.
Table~\ref{tab:Z9-hidden} and the $n$ pairs give
\begin{align}
(\mathcal A_A,\mathcal A_{AAA},\mathcal A_{AXX},\mathcal A_{AAX})_{\rm lower}
=(0,-54,108,342),
\end{align}
while the $E,R$ fields give $(0,54,-108,-342)$. Thus all four
anomalies cancel. The hidden mixed coefficient is
\begin{align}
2\mathcal A_{SU(2)_H^2-A}=0+3+3(2-3)=0.
\label{eq:Z9-hidden-gauge-anomaly}
\end{align}
There are eight hidden Weyl doublets, so the Witten anomaly is absent.
The additional $X$-charged singlets are vectorlike under $U(1)_X$.
All other mixed continuous gauge anomalies vanish. The threshold
terms in the discrete gauge theory are fixed by matching to this
anomaly-free continuous theory.

Every gauge-invariant renormalizable fermion mass term and Yukawa interaction allowed by this ultraviolet field content preserves the charge $p$.
This statement includes interactions containing either a scalar or
its complex conjugate. In
particular,
\begin{align}
p(N_D^c)=p(N_S^c)=1,\qquad
p(\eta)=p(X_1)=p(X_2)=0,
\end{align}
and all scalars have $p=0$. The interactions
$(\eta N_D^c)$, $(\eta\Phi N_D^c)$, $LHN_S^c$, and the couplings
$(\eta\Omega)N_S^c$, $(\eta\widetilde\Omega)N_S^c$ violate $p$.
They cannot be generated by perturbative vertices preserving this
charge, even after $S$ and $T_9$ condense. On the other hand,
\begin{align}
\sum_{\text{hidden doublets}}p_f\,2T_H(\mathbf2)=1,
\end{align}
so the hidden instanton violates this accidental chiral symmetry.
The gauge symmetries remain nonanomalous.

For the main-text fields, the parity $(-1)^p$ agrees with the
combination $(-1)^Fz_H$ in Eq.~\eqref{eq:combined-FN-parity}.
Because the central element $z_H$ also acts on an $SU(2)_H$ doublet,
$p$ and $F$ need not agree modulo two for that field. The extension protects
the absence of an independent chiral mass at renormalizable order,
but does not impose the complete FN hierarchy.
For example, $LHX_2$ is permitted by the additional gauge symmetries;
its extra $\varphi$ factor still follows from the separate FN
selection rule. Any ultraviolet theory that generates the
flavon-dependent messenger interactions must preserve both the
FN selection rule and the accidental chiral symmetry.

The new FN anomalies are
\begin{align}
\kappa_{FAA}=-72,\qquad \kappa_{FAX}=72,\qquad
\kappa_{FA\widehat Y}=0.
\label{eq:Z9-new-FN-anomalies}
\end{align}
The FN anomaly coefficients listed in
Appendix~\ref{app:continuous-Dirac-lift} are unchanged.
The hidden instanton still gives $\Delta F_{\rm FN}=1$. 
The Abelian coefficients are multiples of four and give integer
wall counterterms for the order-four FN rotation. These terms must
be matched at the ultraviolet thresholds. A trivial anomaly phase
on a closed four-manifold does not by itself guarantee invariance
under local deformations of the wall.

\subsection{Instanton matching and phase conventions}

We now match the nonperturbative interaction of this UV example onto
the coefficient $\mu_H$ used in the main text. In the enlarged theory
an instanton has eight fermion zero modes.
Three insertions of the conjugate heavy-fermion mass terms absorb
the six $P,\widetilde P$ zero modes, leaving $\eta,N_D^c$ as external
fields. At leading order in $M_P\rho$, the interaction has the form
\begin{align}
\Delta\mathcal L_{\rm inst}\sim
\frac{\mathcal K_{\rm UV}\det\lambda^*}{v_H^2}
 e^{-8\pi^2/g_{H,\rm UV}^2(v_H)+i\vartheta_{\rm UV}}
(S^\dagger)^3(\eta N_D^c)+\mathrm{h.c.},
\label{eq:Z9-instanton-operator}
\end{align}
and hence
\begin{align}
\mu_{\rm inst}\sim
\mathcal K_{\rm UV}\frac{\det M_P^*}{v_H^2}
 e^{-8\pi^2/g_{H,\rm UV}^2(v_H)+i\vartheta_{\rm UV}}.
\label{eq:Z9-instanton-mass}
\end{align}
Here $g_{H,\rm UV}$ is the hidden gauge coupling in the theory
containing the three additional pairs $P_I,\widetilde P_I$.
The dimensionless factor $\mathcal K_{\rm UV}$ denotes the
remaining instanton prefactor after the exponential and
the factor $\det M_P^*/v_H^2$ have been extracted. 
In these equations, $\vartheta_{\rm UV}$ is the coefficient of the
hidden gauge topological term in the theory containing the heavy
pairs. It is not defined as a phase of $M_P$. With the mass and instanton
conventions used here, the coefficient has phase
\begin{align}
\arg\mu_{\rm inst}=\vartheta_{\rm UV}-\arg\det M_P
+\arg\mathcal K_{\rm UV}\pmod{2\pi}.
\label{eq:UV-instanton-phase}
\end{align}
This is a relation between coefficients in a fixed field basis, not
by itself a basis-independent CP observable. An anomalous rephasing
of the heavy fermions changes their mass phases and the gauge theta
angle together. 
After the additional heavy fermions are integrated out,
the phase contribution $-\arg\det M_P$ may be included either
in the effective theta angle or in the instanton prefactor, but not in both.
The coefficient $\zeta_H$ is defined after this threshold matching. 
For the interactions in
Eq.~\eqref{eq:general-messenger-matching}, the overall phase of
$\mu_H$ cancels from $Y_\nu Y_\nu^\dagger$ and does not give an
independent PMNS phase.

This contribution involves one instanton and three mass insertions.
The mass-insertion expansion is valid only when each heavy-fermion
mass multiplied by $\rho$ is much smaller than one. For masses of
order $v_H$, the expression gives only the dimensional scaling,
and the full mass dependence must be included. For $M_P\gg v_H$,
threshold matching replaces the small-mass expansion. In the
effective theory with two hidden Weyl doublets, $\mu_H=v_H\zeta_H$
and $\mathcal K_H$ includes the heavy-fermion mass dependence. With the scalar
content used in the main text, the one-loop coefficient is $25/6$
at energies above the masses of the three additional pairs and
$37/6$ below them, as in Eq.~\eqref{eq:hidden-beta-coefficient}.

\subsection{Independent ultraviolet contributions}

The instanton calculation determines one contribution to the mixing.
To test whether it dominates, one must also examine local operators
allowed by the UV gauge symmetries. Examples are
\begin{align}
\Delta\mathcal L_{\rm UV}\supset{}&
\frac{c_6}{\Lambda_{\rm UV}^2}(S^\dagger)^3(\eta N_D^c)\nonumber\\
&+\frac{c_{7,i}}{\Lambda_{\rm UV}^3}(S^\dagger)^3
\left(\frac{\varphi}{\Lambda_{\rm FN}}\right)^{\ell_i+1}L_iHN_S^c
+\mathrm{h.c.}
\label{eq:Z9-independent-operators}
\end{align}
The powers of $\varphi/\Lambda_{\rm FN}$ in the second term follow
from the separate FN flavor assignment; these powers are not required by $\mathbb Z_9$. 
After $S$ condenses, the
second term gives an independent $N_S^c$ Yukawa column, without the
instanton factor in Eq.~\eqref{eq:rank-lifting-estimate}. Its
contribution must be subleading under the assumptions of
Secs.~\ref{subsec:instanton-stability} and
\ref{subsec:singlet-corrections}.
The first term induces
\begin{align}
\delta\mu_{\rm UV}=c_6\frac{(f^*)^3}{\Lambda_{\rm UV}^2}.
\label{eq:Z9-independent-mass}
\end{align}
This coefficient is allowed by the gauge symmetries. Requiring
$|\delta\mu_{\rm UV}|\ll|\mu_H|$ gives
\begin{align}
\Lambda_{\rm UV}\gg
\left(\frac{|c_6||f|^3}{|\mu_H|}\right)^{1/2}.
\label{eq:Z9-quality-bound}
\end{align}
Taking $|f|=v_H=10^{10}\,\mathrm{GeV}$ and
$|\mu_H|\simeq0.204\,\mathrm{MeV}$ makes the right-hand side of
Eq.~\eqref{eq:Z9-quality-bound} equal to
$7.0\times10^{16}\sqrt{|c_6|}\,\mathrm{GeV}$. At $|c_6|=1$ and
$\Lambda_{\rm UV}=2.4\times10^{18}\,\mathrm{GeV}$,
$|\delta\mu_{\rm UV}|\simeq0.174\,\mathrm{keV}$, about
$8.5\times10^{-4}$ of the benchmark value of $|\mu_H|$. Other independent
operators must be checked separately. If an independent coefficient
is generated at a heavy-particle threshold below the assumed cutoff,
its size must be evaluated at that threshold rather than by using
the larger cutoff as its suppression scale.

The complete scalar and neutral-fermion spectra are not determined
by this construction. 
Any additional scalar or neutral fermion whose mass is not large
compared with the energies under consideration must be included
explicitly in the effective theory. 
In this extension, all gauge-invariant renormalizable interactions
preserve the accidental chiral charge $p$. They therefore cannot
generate an independent $(\eta N_D^c)$ mass term in perturbation
theory. Higher-dimensional interactions that violate $p$ remain
allowed and must be sufficiently suppressed for the instanton
contribution to dominate.
We do not determine all flavon-dependent couplings, the scalar
mass hierarchies, or the cosmological evolution of the additional
fields.

\section{Detailed finite-group and modular anomaly analysis}
\label{app:flavor-completions}

This appendix gives the detailed anomaly and fusion analysis for the
finite-group and modular extensions summarized in Sec.~\ref{sec:family}.
We use the chiral-defect construction reviewed in
Sec.~\ref{sec:noninv-review}. The fermion measure determines an anomaly
character or cocycle. We then attach a three-dimensional TQFT whose
anomaly cancels the corresponding phase. The original flavor or modular
transformation is invertible, while the TQFT-dressed defect can be
non-invertible. Under the assumptions stated below, we obtain
associative TQFT-valued fusion data.
These examples concern the non-invertible completion of anomalous
finite-group and modular transformations, separately from the
interactions of the FN mass model.

\subsection{Anomaly}

Consider a set of fields $\phi_r$ transforming in an irreducible
representation $r$ of a family symmetry group $G_f$. Under
$g\in G_f$,
\begin{align}
\phi_r(x)
\longrightarrow
\rho_r(g)\phi_r(x),
\end{align}
where $\rho_r(g)$ is the representation matrix. If $G_f$ is
continuous, the classical theory has a Noether current $J_F$. If
$G_f$ is discrete, the finite transformations are implemented by
codimension-one symmetry walls.

For a continuous $U(1)_F$ symmetry, we define the mixed
$U(1)_F$--$G_a^2$ anomaly coefficient for a non-Abelian gauge factor
$G_a$ by
\begin{align}
\mathcal A_{Faa}
\equiv
\sum_f
F_f\,T_a(r_f)\,d_f^{(a)},
\label{eq:general-nonabelian-anomaly}
\end{align}
where $F_f$ is the $U(1)_F$ charge of the left-handed Weyl fermion
$f$, $T_a(r_f)$ is the Dynkin index of its representation under
$G_a$, and $d_f^{(a)}$ counts the multiplicity from the remaining
quantum numbers of $f$, apart from its $G_a$ representation. We use the convention
\begin{align}
T_a(\mathbf{fund})=\frac12.
\end{align}
It is useful to define
\begin{align}
\kappa_{Faa}
\equiv
2\mathcal A_{Faa}.
\label{eq:kappa-nonabelian}
\end{align}
With the instanton number normalized as
\begin{align}
\nu_a
=
\frac{1}{8\pi^2}
\int
\operatorname{tr}(F_a\wedge F_a)
\in\mathbb Z,
\end{align}
the anomaly equation takes the form
\begin{align}
d\star J_F
=
\sum_a
\frac{\kappa_{Faa}}{8\pi^2}
\operatorname{tr}(F_a\wedge F_a)
+
\frac{\kappa_{F{\rm grav}}}{192\pi^2}
\operatorname{tr}(R\wedge R).
\label{eq:general-anomaly-equation}
\end{align}
A unit $G_a$ instanton therefore violates the $U(1)_F$ charge by
\begin{align}
\Delta F_a
=
\kappa_{Faa}
=
2\mathcal A_{Faa}.
\label{eq:general-deltaF}
\end{align}
We are interested in an ABJ-type anomaly involving a dynamical gauge field. Gauge configurations in the path integral then violate the
corresponding flavor transformation, and it does not remain an exact
invertible symmetry of the quantum theory. This is different from an
't Hooft anomaly of an exact symmetry coupled to background fields,
and from an anomaly cancelled by a Green-Schwarz term or a local
counterterm. The construction in Sec.~\ref{sec:noninv-review} applies
when the anomalous phase can instead be cancelled by a wall TQFT
coupled to an exact higher-form symmetry of the gauge sector.

Note that Eq.~\eqref{eq:general-anomaly-equation} also contains a possible
mixed flavor-gravitational contribution. In this work, we consider
flat spacetime, where the curvature contribution to the Fujikawa
phase vanishes. More generally, the same argument applies to wall
deformations for which this contribution vanishes throughout the
four-dimensional region between the original and deformed walls.
A trivial gravitational phase on a closed four-manifold 
does not ensure invariance under local wall deformations.
An extension to general curved backgrounds would require the
mixed gravitational contribution to be absent or cancelled by
an additional inflow or wall theory. We do not consider such
an extension here.

Finite non-Abelian groups and modular symmetries do not in general
admit a Noether-current description of the form in
Eq.~\eqref{eq:general-anomaly-equation}. Their anomaly data are instead
encoded in the fermion measure and, for modular transformations, in a
modular anomaly cocycle. We focus on the case in which the relevant
gauge-field-dependent anomaly contains a rational phase and the gauge
sector has the higher-form symmetry required for the TQFT dressing.

\subsection{Finite non-Abelian flavor symmetries}
\label{subsec:finite-nonabelian-flavor}

Let $G_f$ be a finite flavor group and let $\psi_f$ denote a
left-handed Weyl multiplet transforming as
\begin{align}
\psi_f
\longmapsto
\rho_f(g)\,\psi_f,
\end{align}
with $g\in G_f$. 
For an element $g$ of order $N_g$, its determinant can be written as
\begin{align}
\det \rho_f(g)
=
\exp\!\left[
\frac{2\pi i}{N_g}\,q_f(g)
\right],
\label{eq:finite-group-determinant}
\end{align}
with $q_f(g)\in\mathbb Z_{N_g}$. 

Consider a non-Abelian gauge factor $G_a$, with the instanton
number normalized as above. In a self-dual instanton background
with topological charge $\nu_a=1$, the index theorem gives
the number of chiral zero modes per flavor component as
\begin{align}
m_f^{(a)}
=
2T_a(r_f)\,d_f^{(a)}.
\end{align}
Here $d_f^{(a)}$ counts the multiplicity from the other gauge
factors. The flavor multiplicity is already included in
$\det\rho_f(g)$ and is not included in $d_f^{(a)}$.
The fermion measure transforms as
\begin{align}
\mathcal D\psi\,\mathcal D\bar\psi
\longmapsto
J_a(g;\nu_a)\,
\mathcal D\psi\,\mathcal D\bar\psi,
\end{align}
with
\begin{align}
J_a(g;\nu_a)
&=
\prod_f
\left[\det\rho_f(g)\right]^{m_f^{(a)}\nu_a}
=
\exp\!\left[
\frac{2\pi i}{N_g}\,
\nu_a\,\mathfrak A_a(g)
\right],
\label{eq:finite-group-anomaly}
\end{align}
and 
\begin{align}
\mathfrak A_a(g)
=
\sum_f m_f^{(a)}q_f(g)
\quad (\operatorname{mod}N_g).    
\end{align}
This is the Fujikawa form of the mixed $G_f$--$G_a^2$ anomaly
\cite{Araki:2008ek}. The local Jacobian depends on
$\det\rho_f(g)$ and therefore probes a one-dimensional character of
$G_f$, rather than the full non-Abelian representation matrix.

Evaluating the Jacobian in the sector $\nu_a=1$, we define
\begin{align}
\eta_a(g)
=
J_a(g;1).
\end{align}
Since determinants are multiplicative,
\begin{align}
\eta_a(g_1g_2)
=
\eta_a(g_1)\eta_a(g_2),
\end{align}
and hence $\eta_a:G_f\rightarrow U(1)$ is a group homomorphism. Its
kernel,
\begin{align}
G_{0,a}
=
\ker\eta_a,
\end{align}
is a normal subgroup, and
\begin{align}
G_f/G_{0,a}
\simeq
\operatorname{Im}\eta_a
\end{align}
is cyclic because every finite subgroup of $U(1)$ is cyclic.

When several independent dynamical gauge sectors are present, the
transformations that are anomaly free in all sectors form
\begin{align}
G_0
=
\bigcap_a G_{0,a}.
\label{eq:combined-anomaly-free-subgroup}
\end{align}
The combined character map is
\begin{align}
\eta:G_f\longrightarrow\prod_a\operatorname{Im}\eta_a
\qquad
\mathrm{with}
\qquad
\eta(g)=\bigl(\eta_a(g)\bigr)_a,
\end{align}
so that
\begin{align}
G_f/G_0
\simeq
\operatorname{Im}\eta
\end{align}
is finite Abelian. It need not be cyclic when several independent
anomaly characters are present. If the anomalous phases are generated
by a single character, the quotient reduces to
\begin{align}
G_f/G_0
\simeq
\mathbb Z_{N_{\rm an}}.
\label{eq:anomalous-cyclic-quotient}
\end{align}
The normal-subgroup structure associated with a single anomaly
character was analyzed systematically in Ref.~\cite{Kobayashi:2021xfs}.

Every determinant character equals one on the commutator subgroup,
\begin{align}
[G_f,G_f]
\subseteq
G_{0,a}.
\label{eq:derived-subgroup-anomaly-free}
\end{align}
Thus a local Fujikawa anomaly can detect only the Abelianization
$G_f^{\rm ab}=G_f/[G_f,G_f]$.

\subsection{Completion of anomalous flavor symmetries by non-invertible symmetries}

We now construct the dressed flavor walls described in
Sec.~\ref{subsec:finite-flavor-summary}. Consider a compact Abelian
gauge sector $U(1)_V$ with a fixed global form and an exact magnetic
one-form symmetry. Here $U(1)_V$ is the gauge sector entering the
mixed flavor anomaly in this example. It is distinct from $U(1)_X$,
which is used to embed $\mathbb Z_8^D$ in a continuous gauge theory
in Appendix~\ref{app:continuous-Dirac-lift}.
We write the Fujikawa phase of a flavor transformation $g$ as
\begin{align}
J_V(g)
=
\exp\!\left[
2\pi i\,\alpha_V(g)\,\mathcal I_V
\right],
\label{eq:finite-anomaly-character}
\end{align}
with $\alpha_V(g)\in\mathbb Q/\mathbb Z$, 
where $\mathcal I_V$ is the properly quantized Abelian topological
number. Since the Jacobian is multiplicative,
\begin{align}
\alpha_V(g_1g_2)
=
\alpha_V(g_1)+\alpha_V(g_2)
\qquad(\operatorname{mod}1).
\label{eq:finite-anomaly-additivity}
\end{align}
Thus $\alpha_V$ is a rational character of $G_f$. Its kernel is the
anomaly-free normal subgroup $G_0$, while its image is a finite cyclic
group.

For $\alpha_V(g)=p/n$ in lowest terms, the Abelian wall theory
$\mathcal A^{n,p}$ cancels the anomalous phase when coupled to the
magnetic one-form symmetry of $U(1)_V$
\cite{Cordova:2022ieu,Choi:2021kmx,Choi:2022zal}. More generally, let
$\mathcal B_g$ be a three-dimensional TQFT with the required one-form
anomaly. We define the dressed flavor wall by
\begin{align}
\mathfrak D(g,\mathcal B_g;M_3)
=
U_g(M_3)\,
Z_{\mathcal B_g}[M_3,F_V].
\label{eq:finite-complete-defect}
\end{align}
The wall anomaly of $\mathcal B_g$ cancels the Fujikawa phase of
$U_g$. Therefore $\mathfrak D(g,\mathcal B_g)$ is gauge invariant and
topological as long as the magnetic one-form symmetry is exact.

The fusion rule follows from group multiplication and TQFT stacking,
\begin{align}
\mathfrak D(g,\mathcal B_g)
\times
\mathfrak D(h,\mathcal B_h)
=
\mathfrak D\!\left(
gh,
\mathcal B_g\boxtimes\mathcal B_h
\right).
\label{eq:finite-complete-fusion}
\end{align}
Here $\boxtimes$ means that the two wall theories are placed on the
same interface and coupled to the same bulk gauge field.
Eq.~\eqref{eq:finite-anomaly-additivity} shows that this combined
wall theory has the anomaly required for $gh$. Since group
multiplication and TQFT stacking are associative,
Eq.~\eqref{eq:finite-complete-fusion} gives associative TQFT-valued
fusion data. No splitting of
\begin{align}
1\longrightarrow G_0\longrightarrow G_f
\longrightarrow\operatorname{Im}\alpha_V\longrightarrow1
\end{align}
is required. 
In other words, the construction does not require
$G_f$ to be a semidirect product of $G_0$ and the anomalous quotient.

For $h\in G_0$, the anomalous phase is trivial and the wall theory can
be taken to be trivial. We then recover the ordinary invertible defect
$U_h$. We assume that $G_f$ acts trivially on the wall TQFT, and
we obtain
\begin{align}
\mathfrak D(g,\mathcal B_g)U_h
=
U_{ghg^{-1}}\mathfrak D(g,\mathcal B_g).
\label{eq:finite-crossed-relation}
\end{align}

For an element with a nontrivial rational anomaly,
$\alpha_V(g)=p/n$, the minimal dressed defect satisfies
\begin{align}
\mathfrak D(g,\mathcal A^{n,p})
\times
\overline{\mathfrak D(g,\mathcal A^{n,p})}
=
\mathcal C_{n,p}^{(1)},
\qquad
\mathcal C_{n,p}^{(1)}\neq\mathbf1.
\label{eq:finite-noninvertible-fusion}
\end{align}
Here $\mathcal C_{n,p}^{(1)}$ is the magnetic condensation defect.
Although $g$ itself has the inverse $g^{-1}$ in $G_f$, the dressed
defect is non-invertible because the wall TQFT leaves
$\mathcal C_{n,p}^{(1)}$ after fusion with the orientation-reversed
defect. Thus non-invertibility is a property of the TQFT-dressed
defect, not of the group element itself.

If several independent gauge sectors contribute rational anomaly
characters, we stack the corresponding wall theories
$\boxtimes_a\mathcal B_{g,a}$. 
The following quotients illustrate the one-dimensional characters
that can appear in a local Fujikawa anomaly:
\begin{align}
S_3/A_3&\simeq\mathbb Z_2,
&
A_4/(\mathbb Z_2\times\mathbb Z_2)&\simeq\mathbb Z_3,
\nonumber\\
S_4/A_4&\simeq\mathbb Z_2,
&
\Delta(6n^2)/\Delta(3n^2)&\simeq\mathbb Z_2.
\label{eq:finite-group-anomaly-examples}
\end{align}

\subsection{An $A_4$ example}

As an example, consider a flavor symmetry based on
\begin{align}
A_4
&=
\langle S,T\mid S^2=T^3=(ST)^3=1\rangle .
\end{align}
We use the convention:
\begin{align}
\rho_{\mathbf{1}'}(S)
&=
\rho_{\mathbf{1}''}(S)
=1,
\qquad
\rho_{\mathbf{1}'}(T)
=
\omega,
\qquad
\rho_{\mathbf{1}''}(T)
=\omega^2,
\end{align}
with $\omega=e^{2\pi i/3}$. 
Consider a compact $U(1)_V$ gauge theory with an exact magnetic one-form symmetry. 
The fields in this example are separate from those in
Table~\ref{tab:matter-content}; in particular, $\eta$ below is not
the hidden doublet of the FN model.
We introduce three left-handed Weyl fermions $\psi$, $\eta$, and $\lambda$, with
\begin{align}
\bigl(
q_V(\psi),
q_V(\eta),
q_V(\lambda)
\bigr)
&=
(+1,-1,0),
\nonumber\\
\bigl(
r_{A_4}(\psi),
r_{A_4}(\eta),
r_{A_4}(\lambda)
\bigr)
&=
\bigl(
\mathbf{1}',
\mathbf{1},
\mathbf{1}''
\bigr).
\end{align}
The $U(1)_V^3$ and mixed $U(1)_V$-gravitational gauge anomalies cancel because
\begin{align}
\sum_f q_{V,f}^3
&=
1-1
=0,
\qquad
\sum_f q_{V,f}
=1-1=0.
\end{align}
The gauge-neutral fermion $\lambda$ does not contribute to the mixed $A_4$-$[U(1)_V]^2$ anomaly. It is included to cancel the determinant character of $\psi$ in the purely gravitational sector. Indeed,
\begin{align}
\det\rho_{\mathbf{1}'}(T)\,
\det\rho_{\mathbf{1}''}(T)
&=
\omega^3
=1.
\end{align}
Since both one-dimensional representations are trivial under $S$, the purely gravitational determinant character is trivial for every element of $A_4$.

Let
\begin{align}
I_V
&=
\frac{1}{8\pi^2}
\int F_V\wedge F_V
\in
\mathbb Z
\end{align}
be the properly quantized Abelian topological number on a spin four-manifold. Under the order-three transformation $T$, the charged fermion $\psi$ produces the Fujikawa phase:
\begin{align}
J_V(T;I_V)
&=
\left[
\det\rho_{\mathbf{1}'}(T)
\right]^{q_{V,\psi}^2{I_V}}
=
\exp\left(
\frac{2\pi i}{3}I_V
\right).
\end{align}
The corresponding rational anomaly character is
\begin{align}
\bigl(
\alpha_V(S),
\alpha_V(T),
\alpha_V(T^2)
\bigr)
&=
\left(
0,
\frac{1}{3},
\frac{2}{3}
\right)
\qquad
(\mathrm{mod}\ 1).
\end{align}
Its image is $\mathbb Z_3$, and its kernel is the normal Klein subgroup
\begin{align}
G_0
\equiv
\ker\alpha_V
&=
V_4
=\{1,S,TST^2,T^2ST\}.
\end{align}
Therefore,
\begin{align}
A_4/V_4
&\simeq
\mathbb Z_3,
\end{align}
in agreement with the Abelianization of $A_4$.
The wall associated with $T$ is dressed by the minimal Abelian wall theory $\mathcal A^{3,1}$, whereas the wall associated with $T^2$ is dressed by $\mathcal A^{3,2}$:
\begin{align}
D_T(M_3)
&=
U_T(M_3)
Z_{\mathcal A^{3,1}}[M_3,F_V],
\nonumber\\
D_{T^2}(M_3)
&=
U_{T^2}(M_3)
Z_{\mathcal A^{3,2}}[M_3,F_V].
\end{align}
The elements of $V_4$ have a trivial anomaly character and are represented by ordinary invertible walls. By contrast, fusion of $D_T$ with its orientation-reversed defect gives
\begin{align}
D_T
\times
\overline{D_T}
&=
\mathcal C_{3,1}^{(1)}
\neq
1.
\end{align}
This example gives the mixed $A_4$--$[U(1)_V]^2$ anomaly
and the corresponding TQFT-dressed flavor defects. 
Since all three fermions are in one-dimensional representations, $V_4$ acts trivially on these fermions: this fermion sector realizes only the quotient $A_4/V_4\simeq\mathbb Z_3$. A faithful $A_4$ model requires additional fields on which $V_4$
acts nontrivially, for example a gauge-neutral scalar transforming
as an $A_4$ triplet, with an $A_4$-invariant potential and zero vacuum
expectation value. Such a scalar does not change the chiral anomaly.
In this extension, the flavor walls are labeled by elements of the
full group $A_4$, and group multiplication determines the flavor
label of their fusion product. The local Fujikawa anomaly still
depends only on the Abelian quotient.

More generally, the quotients in
Eq.~\eqref{eq:finite-group-anomaly-examples} specify which
one-dimensional characters can appear in a local Fujikawa anomaly.
Which character occurs depends on the matter representations. For example, an $S_4$ chiral anomaly can distinguish the two cosets of $A_4$, but it is trivial on the commutator subgroup $[S_4,S_4]=A_4$.

Under the assumptions stated above,
Eqs.~\eqref{eq:finite-complete-defect}--\eqref{eq:finite-noninvertible-fusion}
give the TQFT-valued fusion data for the anomalous flavor
transformations. The ordinary representation-theoretic selection
rules remain valid in perturbation theory, while instantons or
monopoles can break the anomalous flavor transformations once the
required higher-form symmetry is broken.

\subsection[Modular flavor symmetries and residual defects]{Modular flavor symmetries and residual defects}
\label{subsec:modular-flavor}

A modular flavor symmetry acts on the modulus, matter fields,
and coupling functions~\cite{Feruglio:2017spp}. Let $\tau$ be a dimensionless complex
modulus in the upper half-plane,
\begin{align}
\mathbb H
=
\{\tau\in\mathbb C\mid \operatorname{Im}\tau>0\}.
\end{align}
We denote the homogeneous modular group by
$\Gamma=SL(2,\mathbb Z)$. An element of $\Gamma$ is written as
\begin{align}
\gamma
=
\begin{pmatrix}
a&b\\
c&d
\end{pmatrix},
\qquad
a,b,c,d\in\mathbb Z,
\qquad
ad-bc=1.
\end{align}
The matrices $\gamma$ and $-\gamma$ give the same fractional
linear transformation of $\tau$. The group acting on the modulus is therefore
\begin{align}
\overline\Gamma
=
\Gamma/\{\pm\mathbf 1_2\}
=
PSL(2,\mathbb Z).
\end{align}
We keep the distinction between $\Gamma$ and $\overline\Gamma$,
since the central element $-\mathbf 1_2$ can act nontrivially
on the matter fields.

For an integer $N\geq2$, the principal congruence subgroup of
level $N$ is
\begin{align}
\Gamma(N)
=
\{\gamma\in\Gamma\mid
\gamma\equiv\mathbf 1_2\pmod N\}.
\end{align}
Writing $\overline\Gamma(N)$ for its image in
$\overline\Gamma$, we define the finite modular groups by
\begin{align}
\Gamma_N
=
\overline\Gamma/\overline\Gamma(N),
\qquad
\Gamma_N'
=
\Gamma/\Gamma(N).
\end{align}
For $N>2$, $\Gamma_N'$ is a double cover of $\Gamma_N$.

With the automorphy factor
\begin{align}
j(\gamma,\tau)=c\tau+d,
\end{align}
the modulus and matter fields respectively transform as
\begin{align}
\tau
&\longmapsto
\gamma\tau
=
\frac{a\tau+b}{c\tau+d},
\nonumber\\
\psi_i
&\longmapsto
j(\gamma,\tau)^{-k_i}
\rho_i(\gamma)\,\psi_i.
\label{eq:modular-transformation-general}
\end{align}
Here $k_i$ determines the modular weight of $\psi_i$, and
$\rho_i(\gamma)$ is a representation matrix of the chosen
finite modular group, $\Gamma_N$ or $\Gamma_N'$.
The matrix $\rho_i(\gamma)$ is evaluated on the image of $\gamma$
in the chosen finite modular group.
A vector-valued modular
form in a Yukawa coupling transforms as
\begin{align}
Y_I(\gamma\tau)
=
j(\gamma,\tau)^{k_Y}
\rho_Y(\gamma)_{IJ}Y_J(\tau).
\label{eq:modular-form-transformation}
\end{align}
Here $k_Y$ is the modular weight of $Y_I(\tau)$,
$\rho_Y$ is its representation of the finite modular group,
and $I,J$ label the components of the modular multiplet. 
The full action is invariant when the modulus and the matter fields
are transformed simultaneously. Classical modular invariance then
constrains both the modular weights and the tensor products of the
finite modular representations in each interaction.

We next consider the modular anomaly. Let $d_i=\dim\rho_i$ and let
$m_i^{(a)}$ denote the number of $G_a$ instanton zero modes associated
with one flavor component of the multiplet $\psi_i$. 
The flavor multiplicity $d_i$ is written explicitly below. In a holomorphic field basis, we denote the matter-fermion contribution to the Jacobian in an instanton sector of charge $\nu_a$ by
\begin{align}
\mathcal J_a^{\rm mat}(\gamma,\tau;\nu_a)
=
\left[\Xi_a^{\rm mat}(\gamma,\tau)\right]^{\nu_a}.
\label{eq:modular-fujikawa-jacobian}
\end{align}
The matter-fermion factor has the schematic form
\begin{align}
\Xi_a^{\rm mat}(\gamma,\tau)
=
\prod_i
\left[
j(\gamma,\tau)^{-k_i d_i}
\det\rho_i(\gamma)
\right]^{m_i^{(a)}}.
\end{align}
This expression does not give the full anomaly in a general
supergravity effective theory. Gaugino, K\"ahler-connection,
field-redefinition, and supergravity contributions must also
be included. The form of the automorphy-factor contribution can also change when the fields are canonically normalized. 
We use the expression above to distinguish the two contributions
to the matter-fermion factor,
\begin{align}
\Xi_a^{\rm mat}(\gamma,\tau)
=
\Xi_a^{\rm wt}(\gamma,\tau)
\Xi_a^{\rm rep}(\gamma),
\end{align}
where $\Xi_a^{\rm wt}$ arises from the automorphy factors and
$\Xi_a^{\rm rep}$ from the determinants of the finite modular
representations. The latter is a root of unity, whereas the former
is generally modulus dependent. This modulus-dependent contribution
is the field-theory counterpart of the target-space duality anomaly
controlled by modular weights in string compactifications
\cite{Ibanez:1992hc}.

The automorphy factors satisfy
\begin{align}
j(\gamma_1\gamma_2,\tau)
=
j(\gamma_1,\gamma_2\tau)
j(\gamma_2,\tau).
\end{align}
Together with the group law of the representation matrices,
this gives the composition law for the matter-fermion factor,
\begin{align}
\Xi_a^{\rm mat}(\gamma_1\gamma_2,\tau)
=
\Xi_a^{\rm mat}(\gamma_1,\gamma_2\tau)
\Xi_a^{\rm mat}(\gamma_2,\tau).
\label{eq:modular-anomaly-cocycle}
\end{align}
This relation follows from the matter-field transformation law.
It does not determine the remaining contributions to the anomaly.
After these contributions are included, the complete
gauge-field-dependent anomaly must satisfy the same modular
composition law, as required by the Wess--Zumino consistency
condition~\cite{WessZumino}.
Counterterm and Green-Schwarz contributions, as well as the
subsequent TQFT dressing, must be compatible with this composition
law.

In string-derived effective theories, modular anomalies are often
cancelled by the Green-Schwarz mechanism, possibly together with
moduli-dependent threshold corrections~\cite{Ibanez:1992hc}.
When the cancellation is complete, the modular transformation
remains an invertible symmetry of the quantum theory.
We consider instead the case in which a residual rational
anomalous phase is cancelled by a three-dimensional TQFT on the
modular wall. This is also different from explicit nonperturbative
breaking. For example, D-brane instantons in magnetized orbifold
models can break a cyclic part of a modular flavor group while
leaving a normal subgroup unbroken~\cite{Kikuchi:2022bkn}.

For the Abelian gauge sector $U(1)_V$, let
$\Xi_V^{\rm tot}(\gamma,\tau)$ denote the complete
gauge-field-dependent anomaly factor in a unit topological sector.
It includes the matter-fermion contribution and all other
applicable fermionic, gaugino, K\"ahler, field-redefinition,
supergravity, counterterm, Green-Schwarz, and threshold
contributions. It is defined before the wall TQFT is attached
and need not equal $\Xi_V^{\rm mat}(\gamma,\tau)$.

We restrict ourselves to modular transformations for which
the corresponding anomaly takes the rational form:
\begin{align}
\mathcal J_V(\gamma,\tau;\mathcal I_V)
=
\left[\Xi_V^{\rm tot}(\gamma,\tau)\right]^{\mathcal I_V}
=
\exp\!\left[
2\pi i\,\alpha_V(\gamma,\tau)\,\mathcal I_V
\right],
\label{eq:modular-rational-cocycle}
\end{align}
with
\begin{align}
\alpha_V(\gamma,\tau)\in\mathbb Q/\mathbb Z.
\end{align}
This condition is not satisfied for a generic modular transformation.
The construction below applies only to transformations satisfying
this condition. Here $\alpha_V$ is determined by the complete
anomaly factor $\Xi_V^{\rm tot}$ after all relevant contributions have been included. The Wess--Zumino consistency condition for the complete
anomaly then gives
\begin{align}
\alpha_V(\gamma_1\gamma_2,\tau)
=
\alpha_V(\gamma_1,\gamma_2\tau)
+
\alpha_V(\gamma_2,\tau)
\qquad(\operatorname{mod}1).
\label{eq:modular-additive-cocycle}
\end{align}

We assume that the transformations considered here are closed under
composition and inversion, and that their rational cocycle is defined
on a common domain. For a dynamical $\tau$, rationality at isolated
points is not sufficient. A continuous rational-valued cocycle is
locally constant on a connected domain, so the wall TQFT data must also
be fixed there. After modulus stabilization, we can instead apply the
construction to the finite subgroup that leaves the full vacuum
invariant, using the anomaly character at $\tau_\star$. This does not require a continuation to
generic values of the modulus.

For each $(\gamma,\tau)$ satisfying these assumptions, we choose a wall
TQFT $\mathcal B_{\gamma,\tau}$ whose one-form anomaly cancels
$\alpha_V(\gamma,\tau)$. We denote the modular transformation wall by
$U_\gamma(\tau)$ and define the dressed wall by
\begin{align}
\mathfrak D(\gamma,\tau;\mathcal B_{\gamma,\tau})
=
U_\gamma(\tau)\,
Z_{\mathcal B_{\gamma,\tau}}[F_V].
\label{eq:modular-complete-interface}
\end{align}
The composition law is
\begin{align}
&\mathfrak D(\gamma_1,\gamma_2\tau;
\mathcal B_{\gamma_1,\gamma_2\tau})
\times
\mathfrak D(\gamma_2,\tau;
\mathcal B_{\gamma_2,\tau})
=
\mathfrak D\!\left(
\gamma_1\gamma_2,
\tau;
\mathcal B_{\gamma_1,\gamma_2\tau}
\boxtimes
\mathcal B_{\gamma_2,\tau}
\right).
\label{eq:modular-complete-composition}
\end{align}
The cocycle condition guarantees anomaly cancellation for the
composite transformation. Since the modular group law and TQFT
stacking are associative,
Eq.~\eqref{eq:modular-complete-composition} gives associative
TQFT-valued composition data for the dressed modular transformations.

We now consider a stabilized vacuum. Let
\begin{align}
\langle\tau\rangle=\tau_\star.
\end{align}
A modular transformation generally maps this vacuum to another
vacuum related by the same transformation,
\begin{align}
\tau_\star
\longrightarrow
\gamma\tau_\star.
\end{align}
The transformations that leave the stabilized value invariant form
the subgroup
\begin{align}
H_{\tau_\star}
=
\{\gamma\in\Gamma\mid\gamma\tau_\star=\tau_\star\}.
\end{align}
Taking also the vacuum expectation values of the other fields into
account, we denote the unbroken subgroup by
\begin{align}
H_{\rm vac}
\subseteq
H_{\tau_\star}.
\label{eq:modular-vacuum-stabilizer}
\end{align}
Since the projective modular group is generated by the transformations:
\begin{align}
S:\tau\longmapsto-\frac{1}{\tau},
\qquad
T:\tau\longmapsto\tau+1,
\end{align}
the elliptic fixed points are
\begin{align}
S:i\longmapsto i,
\qquad
ST:\omega\longmapsto\omega,
\label{eq:modular-fixed-points}
\end{align}
with $\omega=e^{2\pi i/3}$. 
For $\gamma\in H_{\rm vac}$, the modular transformation acts within
the chosen vacuum and gives a residual symmetry transformation.

If
\begin{align}
\alpha_V(\gamma,\tau_\star)
=
0
\qquad(\operatorname{mod}1),
\end{align}
no nontrivial wall TQFT is required, and the corresponding residual
modular defect is invertible. For a nontrivial rational anomaly,
\begin{align}
\alpha_V(\gamma,\tau_\star)
=
\frac{p}{n}
\neq0
\qquad(\operatorname{mod}1),
\end{align}
we take the minimal wall theory $\mathcal A^{n,p}$. The dressed
residual defect is then
\begin{align}
\mathfrak D_\gamma
=
U_\gamma(\tau_\star)\,
Z_{\mathcal A^{n,p}}[F_V].
\end{align}
The modular transformation itself is invertible and has the inverse
$\gamma^{-1}$. However, fusion with the orientation-reversed dressed
defect gives
\begin{align}
\mathfrak D_\gamma
\times
\overline{\mathfrak D_\gamma}
=
\mathcal C_{n,p}^{(1)}
\neq
\mathbf1.
\label{eq:modular-noninvertible-fusion}
\end{align}
Thus the TQFT-dressed residual defect is non-invertible, although the
original modular transformation is invertible. The condition
$\gamma\in H_{\rm vac}$ identifies the transformations that act within
the chosen vacuum; the non-invertibility follows from the nontrivial
TQFT dressing.

For $\gamma\notin H_{\rm vac}$, the modular transformation maps the
chosen vacuum to a different vacuum. It is therefore not a residual
symmetry acting within the chosen vacuum, and we do not treat its
wall as an internal symmetry defect of the theory about that vacuum.

\paragraph{Summary}

The construction above requires the following data:
\begin{enumerate}[leftmargin=8mm,itemsep=1mm]
\item the rational anomaly character $\alpha_V(g)$, or the
      rational (torsion) part of the modular anomaly cocycle
      $\alpha_V(\gamma,\tau)$;
\item a compact gauge sector with an exact higher-form symmetry and a
      fixed global form;
\item cancellation or triviality of anomaly phases from all other
      gauge sectors;
\item a wall TQFT with the required one-form anomaly.
\end{enumerate}
For a finite flavor group, an anomaly-free element gives an ordinary
invertible defect, while an element with a nontrivial rational anomaly
can give a non-invertible dressed defect. For a modular symmetry,
$H_{\rm vac}$ gives the transformations that remain unbroken in the
chosen vacuum. An element of $H_{\rm vac}$ with a trivial anomaly gives
an invertible residual defect, while a nontrivial rational anomaly can
give a non-invertible TQFT-dressed residual defect. Dynamical monopoles
or other magnetic objects break the higher-form symmetry required for
this construction and make the corresponding defects non-topological.

The finite-group and modular constructions have the same basic
structure. The anomalous phase fixes the anomaly that the wall TQFT
must cancel. Group multiplication or modular composition fixes the
transformation carried by the fused defect, while the wall theories
are combined by stacking. The Abelian realization in Secs.~\ref{sec:application}--\ref{sec:messenger-z4}
shows how this construction can enter an FN mass model.

\section{A numerical example for lepton masses and mixing}
\label{app:lepton-benchmark}

We use the leading Yukawa matrix in Eq.~\eqref{eq:factorized-Ynu}
and the charge assignment of Eq.~\eqref{eq:benchmark-charges}. 
In Eq.~\eqref{eq:orientation-independent-matching},
we set $\widehat Z_R^{-1/2}=\mathbf1_3$, omitting the
$\mathcal O(|\mu_H|^2/w_H^2)$ correction to this factor.
The additional $N_S^c$ interactions discussed in
Sec.~\ref{subsec:singlet-corrections} are also neglected.

This appendix gives a numerical example for the model with the
two singlet messengers $X_1$ and $X_2$. We take the coefficients
$C_e,A,B,a,b,c,d$ to be real and of order unity.
We use the normal-ordering NuFIT 6.0 values in
Table~\ref{tab:NuFIT} as the experimental reference
\cite{Esteban:2024eli}. The chosen coefficients reproduce the
charged-lepton masses, the neutrino mass ratio $m_2/m_3$ and
the three lepton mixing angles.

\renewcommand{\arraystretch}{1.25}
\begin{table}[H]
\caption{Neutrino oscillation parameters for normal ordering used in our numerical analysis. The values are taken from the NuFIT 6.0 global analysis including the Super-Kamiokande atmospheric-neutrino data~\cite{Esteban:2024eli}. We use
$\Delta m_{31}^{2}=m_3^2-m_1^2>0$.}
\label{tab:NuFIT}
\centering
\small
\begin{tabular}{|c||c|c|}
\hline
\multirow{2}{*}{Observables}
& \multicolumn{2}{c|}{Normal Ordering (NO)}
\\
\cline{2-3}
& best fit $\pm 1\sigma$
& $3\sigma$ range
\\
\hline
$\sin^{2}\theta_{12}$
& $0.308_{-0.011}^{+0.012}$
& $0.275\rightarrow 0.345$
\\
$\theta_{12}/^\circ$
& $33.68_{-0.70}^{+0.73}$
& $31.63\rightarrow 35.95$
\\
\hline
$\sin^{2}\theta_{13}$
& $0.02215_{-0.00058}^{+0.00056}$
& $0.02030\rightarrow 0.02388$
\\
$\theta_{13}/^\circ$
& $8.56_{-0.11}^{+0.11}$
& $8.19\rightarrow 8.89$
\\
\hline
$\sin^{2}\theta_{23}$
& $0.470_{-0.013}^{+0.017}$
& $0.435\rightarrow 0.585$
\\
$\theta_{23}/^\circ$
& $43.3_{-0.8}^{+1.0}$
& $41.3\rightarrow 49.9$
\\
\hline
$\delta_{\text{CP}}/\pi$
& $1.18_{-0.23}^{+0.14}$
& $0.69\rightarrow 2.02$
\\
$\delta_{\text{CP}}/^\circ$
& $212_{-41}^{+26}$
& $124\rightarrow 364$
\\
\hline
$\Delta m_{21}^{2}/10^{-5}\,\eV^{2}$
& $7.49_{-0.19}^{+0.19}$
& $6.92\rightarrow 8.05$
\\
\hline
$\Delta m_{31}^{2}/10^{-3}\,\eV^{2}$
& $+2.513_{-0.019}^{+0.021}$
& $+2.451\rightarrow +2.578$
\\
\hline
\end{tabular}
\renewcommand{\arraystretch}{1}
\end{table}

For the charged leptons, we write the Yukawa matrix in
Eq.~\eqref{eq:charged-textures} as
\begin{align}
Y_e=D_LC_eD_e
\qquad\mathrm{with}\qquad
D_e=\operatorname{diag}(\epsilon^7,\epsilon^5,\epsilon^3),
\end{align}
and choose
\begin{align}
C_e=\begin{pmatrix}
1.160974&-1.169743&2.019520\\
-0.767913&-1.074960&-1.087150\\
1.335447&2.403772&0.527311
\end{pmatrix}.
\label{eq:Ce-benchmark}
\end{align}
Thus $C_e$ is the coefficient matrix of the charged-lepton interaction
in Eq.~\eqref{eq:FN-perturbative-lagrangian}.

For the neutrino sector, we specify the heavy-fermion mass matrix
$\widehat M_X$ in Eq.~\eqref{eq:orientation-independent-mass}
and the coefficients $A_i,B_i$ multiplying
$\epsilon^{\ell_i}L_iHX_1$ and
$\epsilon^{\ell_i+1}L_iHX_2$, respectively, in
Eq.~\eqref{eq:portal}. The original mass matrix is
$M_X=\widehat M_XU_\Omega$, with its dependence on $z_1,z_2$
contained in $U_\Omega$. In the same basis for $L_i$ as in
the charged-lepton Yukawa matrix, we take
\begin{align}
\widehat M_X=w_H\begin{pmatrix}1&\epsilon\\-\epsilon&1\end{pmatrix}
\qquad\mathrm{with}\qquad a=b=d=1,\quad c=-1,
\label{eq:heavy-benchmark}
\end{align}
and
\begin{align}
A=\begin{pmatrix}
0.938022\\0.454625\\-0.852711
\end{pmatrix},\qquad
B=\begin{pmatrix}
1.371603\\1.411913\\-0.700177
\end{pmatrix}.
\label{eq:exact-rank-two-benchmark}
\end{align}
The magnitudes of the entries lie in the ranges
$0.455\lesssim|A_i|\lesssim0.938$ and
$0.700\lesssim|B_i|\lesssim1.412$.
Together with $0.527\lesssim|(C_e)_{ij}|\lesssim2.404$,
all input coefficients are real and of order unity.

Substituting $a=b=d=1,c=-1$ into the coefficient definitions
in Eq.~\eqref{eq:effective-flavor-vectors} gives
\begin{align}
C_1=\frac{A+B}{1+\epsilon^2},\qquad
C_2=\frac{\epsilon^2B-A}{1+\epsilon^2}.
\label{eq:benchmark-C-columns}
\end{align}
This substitution fixes both coefficient vectors. With
Eq.~\eqref{eq:factorized-Ynu}, it determines the neutrino Yukawa matrix up
to the common factor
$\zeta_\nu$. 
At $\epsilon=0.2$, biunitary diagonalization of $Y_\nu$ gives
\begin{align}
Y_\nu^{\rm diag}
&=|\zeta_\nu|\operatorname{diag}(0,s_2,s_3)\nonumber\\
&\simeq|\zeta_\nu|\operatorname{diag}(0,0.174,1.00).
\label{eq:benchmark-dimensionless-diagonal}
\end{align}
Here $s_i=y_i/|\zeta_\nu|$ for $i=2,3$, where $y_i$ are the
nonzero diagonal Yukawa couplings. The values of $s_2,s_3$ are
fixed by the FN factors and the chosen input coefficients,
independently of $|\zeta_\nu|$. Thus $m_2/m_3=s_2/s_3$, while
the absolute masses also depend on $|\zeta_\nu|$, which is
determined in Sec.~\ref{subsec:neutrino-normalization}.
The $N_S^c$ column of the Yukawa matrix vanishes.
Since the other two columns are independent, one neutrino
remains massless at this order, without a cancellation among
the numerical coefficients.

Diagonalizing $Y_e$ and $Y_\nu$ with the convention in
Eq.~\eqref{eq:yukawa-diagonal-form}, and using
Eq.~\eqref{eq:PMNS-minimal}, gives
\begin{align}
(m_e,m_\mu,m_\tau)\simeq
(0.510\,\mathrm{MeV},106\,\mathrm{MeV},1.78\,\mathrm{GeV}),
\label{eq:charged-lepton-masses-benchmark}
\end{align}
\begin{align}
(\sin^2\theta_{12},\sin^2\theta_{13},\sin^2\theta_{23})
\simeq(0.308,0.0225,0.470),
\label{eq:recomputed-lepton-benchmark}
\end{align}
\begin{align}
m_2/m_3\simeq0.174.
\label{eq:benchmark-neutrino-mass-ratio}
\end{align}
For these real lepton and messenger coefficients, $Y_e$ and
$\widehat Y_\nu/\zeta_\nu$ are real matrices. The common phase
of $\zeta_\nu$ cancels from $Y_\nu Y_\nu^\dagger$, and the
dependence on $z_1,z_2$ enters only through a unitary rotation
of the right-handed neutrinos.
The left-handed diagonalization matrices can therefore be chosen
real, so this example conserves CP in lepton mixing. We obtain
\begin{align}
J_{\rm CP}=0,
\qquad
\delta_{\rm CP}=\pi=180^\circ.
\label{eq:benchmark-deltaCP}
\end{align}
Here $J_{\rm CP}=\operatorname{Im}(U_{e1}U_{\mu2}U_{e2}^*U_{\mu1}^*)$.
Real coefficients are chosen only for this numerical example.
The general model allows complex coefficients and does not require
CP conservation. The three mixing angles and the CP-conserving
value $\delta_{\rm CP}=\pi$ lie within the $3\sigma$ intervals
of Table~\ref{tab:NuFIT}. 
The mass ratio is consistent with
Eq.~\eqref{eq:neutrino-mass-scaling} and is close to
$\sqrt{\Delta m_{21}^2/\Delta m_{31}^2}\simeq0.173$,
obtained from the best-fit oscillation parameters in
Table~\ref{tab:NuFIT}.

The charged-lepton and neutrino Yukawa matrices are written
in the same lepton-doublet basis, so that their left-handed
rotations can be combined to obtain the PMNS matrix.
The comparison with the lepton data uses the tree-level matrices
specified above and does not include their renormalization-group
evolution. Appendix~\ref{app:heavy-fermion-stability}
shows that the chosen $\widehat M_X$ is realized by order-one
couplings for arbitrary $z_1,z_2$. The two heavy Dirac masses are
$w_H\sqrt{1+\epsilon^2}$ at $\mu_H=0$, so a small messenger mass is
not responsible for the neutrino hierarchy.

\subsection{Neutrino mass scale}
\label{app:lepton-benchmark-normalization}

We use $\Delta m_{31}^2$ to fix the common factor $|\zeta_\nu|$.
Since $m_1=0$ and $m_3=(v/\sqrt2)|\zeta_\nu|s_3$, we find
\begin{align}
|\zeta_\nu|=\frac{\sqrt{2\Delta m_{31}^2}}{v\,s_3}
\simeq2.88\times10^{-13}.
\label{eq:benchmark-zeta-normalization}
\end{align}
Here $s_3$
is the value obtained in
Eq.~\eqref{eq:benchmark-dimensionless-diagonal}. 
Once $\Delta m_{31}^2$ is fixed, the ratio $s_2/s_3$ obtained
from the chosen coefficients determines the solar mass splitting
through $\Delta m_{21}^2=(s_2/s_3)^2\Delta m_{31}^2$.
We obtain
\begin{align}
Y_\nu^{\rm diag}\simeq
\operatorname{diag}(0,5.00\times10^{-14},2.88\times10^{-13}).
\label{eq:normalized-Yukawa-benchmark}
\end{align}
This gives
\begin{align}
m_2\simeq8.70\times10^{-3}\,\mathrm{eV},
\label{eq:coefficient-benchmark-masses}
\end{align}
and hence
\begin{align}
\Delta m_{21}^2\simeq7.57\times10^{-5}\,\mathrm{eV}^2.
\label{eq:solar-splitting-benchmark}
\end{align}
The small difference from the best-fit solar mass-squared difference
in Table~\ref{tab:NuFIT} follows from the chosen coefficients and
lies within the quoted $1\sigma$ interval.

\bibliography{ref}
\bibliographystyle{JHEP}

\end{document}